\documentclass[%
prb,
superscriptaddress,    
twocolumn,showpacs,floats,aps
]{revtex4-1}

\usepackage{graphicx}
\usepackage{amsmath}

\usepackage{color}

\def\ket#1{\left |#1\right\rangle }
\def\bra#1{\left \langle#1 \right | }

\def\expect#1{\left\langle#1 \right \rangle}

\def\non{\nonumber \\ }
\def\w{\omega}

\def\Tr#1{\textrm{Tr}\left[#1\right]}

\newcommand{\la}{\left\langle}
\newcommand{\lv}{\left\lvert}
\newcommand{\ra}{\right\rangle}
\newcommand{\rv}{\right\rvert}

\begin{document}

\title{Spin-noise in the anisotropic central spin model}

\author{Johannes Hackmann and Frithjof B.~Anders}
\affiliation{Lehrstuhl f\"ur Theoretische Physik II, Technische Universit\"at Dortmund, 44221 Dortmund,Germany}

\date{\today}

\begin{abstract}
Spin-noise measurements can serve as direct probe for the microscopic decoherence 
mechanism of an electronic spin in semiconductor
quantum dots (QD). We have calculated the spin-noise 
spectrum in the anisotropic central spin model using a Chebyshev expansion technique 
which exactly accounts for the dynamics up to an arbitrary  
long but fixed time in a finite size system.
In the isotropic case, describing QD charge with  a single electron, 
the short-time dynamics is in 
good agreement with a quasi-static approximations for 
the thermodynamic limit. The spin-noise spectrum, however, shows strong deviations at 
low frequencies with a power-law %
behavior of $\w^{-3/4}$ corresponding to a $t^{-1/4}$ decay at intermediate and
long times. In the Ising limit, applicable to  QDs
with heavy-hole spins, 
the spin-noise spectrum exhibits a threshold behavior of $(\w-\w_L)^{-1/2}$ above 
the Larmor frequency $\w_L=g\mu_B B $. 
In the generic anisotropic central spin model we have found a crossover from a 
Gaussian type of spin-noise spectrum 
to a more Ising-type spectrum with increasing anisotropy in a finite magnetic field. 
In order to make contact with experiments, we 
present ensemble averaged spin-noise spectra for 
QD ensembles charged with single electrons or holes. 
The Gaussian-type noise spectrum evolves to a more Lorentzian shape 
spectrum with increasing spread of characteristic time-scales and $g$-factors of the individual QDs.
\end{abstract}





\maketitle

\section{Introduction}

Single electron or hole spins  confined in semiconductor quantum dots
(QDs)  have been suggested\cite{SchliemannKhaetskiiLoos2003,
  HansonSpinQdotsRMP2007} as prime candidates for the realization of
solid-state qubits.  Single-shot readout of the
electron spin has been demonstrated in gate controlled\cite{Elzerman04} QDs, and a very
high degree coherent control of spins has been achieved in
self-assembled
QDs.\cite{Bonadeo1998,GreilichBayer2007,FokinaBayer2010,SpatzekGreilichBayer2011}  
While the strong confinement of the electronic wave function in QDs
reduces the interaction with the  substrate and,  therefore,
suppresses electronic decoherence mechanisms, it simultaneously
enhances the hyperfine interaction between the confined electronic
spin and the nuclear spin bath formed by the underlying lattice.
Although spin-lattice relaxation processes might contribute to the
spin dephasing, it is
believed\cite{SchliemannKhaetskiiLoos2003,HansonSpinQdotsRMP2007,Merkulov2002,KhaetskiiLoss2003,CoishLoss2004}
that the hyperfine interaction dominates the spin relaxation in such
systems.

The dynamics of a single-electron spin coupled to a nuclear spin
bath of non-interacting
spins\cite{HansonSpinQdotsRMP2007,FischerLoss2008,Testelin2009} is
described by the  Gaudin's  central spin model\cite{Gaudin1976}
(CSM). Even though the CSM is  exactly solvable\cite{Gaudin1976} using a Bethe ansatz
(BA), up to now  there exist no thermodynamic  Bethe
ansatz equations for this model  as for the Kondo
model.\cite{AndreiFuruyaLowenstein83} The explicit  solution of the BA
equations are restricted to a finite size system of $N<25$  bath
spins,\cite{BortzStolze2007,Bortz2010} while larger spin bath sizes
require  stochastical
techniques\cite{FaribautSchuricht2013a,FaribautSchuricht2013b} to
extract the spin dynamics of the  central spin and are still limited to 
a small number of nuclear spins of $N<50$.

Recent  spin-noise measurements performed on 
quantum dots\cite{CrookerBayerSpinNoise2010,Dahbashi2012,LiBayer2012,*ZapasskiiGreilichBayer2013}
charged with single electrons or holes
aimed to  directly reveal   the intrinsic dynamics of central spins
interacting with a nuclear spin bath.  The  spin-noise
spectrum  measured in $z$-direction shifts upon increasing the transversal magnetic field $B_x$  to
higher frequencies and traces the Larmor frequency $\w_L$, while the
low-frequency range crosses over from a nearly Lorentz shape to a
$1/f$
noise.\cite{CrookerBayerSpinNoise2010,LiBayer2012,*ZapasskiiGreilichBayer2013}

In this paper, we investigate the spin-noise spectra for the
anisotropic CSM\cite{Gaudin1976,FischerLoss2008,Testelin2009}   using
the Chebyshev expansion technique (CET).  The CET has been developed
30 years ago\cite{TalEzer-Kosloff-84,*Kosloff-94,Fehske-RMP2006}  and
offers an accurate way to calculate the time evolution of a single
initial state $|\psi_0 \rangle$ under the influence of a general
time-independent Hamiltonian ${\cal H}$ operating on a finite
dimensional Hilbert space.  This  approach has been
proposed\cite{Dobrovitski2003} as an efficient scheme for numerical
simulations of the spin-bath decoherence and applied to the  isotropic
CSM\cite{ZangHarmon2006} as well as two coupled spins 1/2 in contact
with a spin bath.\cite{Dobrovitski2003,Yuan2008} 
The original application\cite{Dobrovitski2003,ZangHarmon2006,Yuan2008}
of the CET was restricted to the propagation of a single wave
function. We have extended the approach (i) to thermodynamic
ensembles and also (ii) have averaged over many randomly generated
hyperfine coupling constant configurations. While (i) significantly  
increases the accuracy of the CET for incoherent spin baths 
in the high temperature limit relevant to  the
experiments,\cite{CrookerBayerSpinNoise2010,Dahbashi2012,LiBayer2012,*ZapasskiiGreilichBayer2013}
(ii) turns out  to be crucial for obtaining a smooth noise
spectrum. In any finite-size system, the  exact spectral functions are
given by a finite number of $\delta$-functions. Since the
eigenenergies depend on the configuration of hyperfine couplings,
averaging over many configurations mimics a much larger system and
smoothens the superposition  of $\delta$-functions to a continuous
function when introducing a very small,  but finite broadening
similar to the $z$-averaging\cite{YoshidaWithakerOliveira1990}
used in the time-dependent numerical renormalization group
approach\cite{BullaCostiPruschke2008,AndersSchiller2005,*AndersSchiller2006} 
to the non-equilibrium dynamics.

For a rigorous
solution\cite{Gaudin1976,BortzStolze2007,Bortz2010,FaribautSchuricht2013a,FaribautSchuricht2013b}
of the central spin dynamics  the surrounding spin bath must be taken
into account exactly.  Spin baths differ fundamentally from  bosonic
baths\cite{Leggett1987} due to their  degeneracies and their finite
dimensional Hilbert space.  Over the last decade, very intuitive
pictures for the qualitative  understanding of the  decoherence
induced by a spin bath have emerged. The separation of time scales
\cite{Merkulov2002}  -- a fast electronic precession around an
effective  nuclear magnetic field, and  slow nuclear spin precessions
around the fluctuating electronic spin -- has motivated various
quasi-static
approximations\cite{Merkulov2002,KhaetskiiLoss2003,ErlingssonNazarov2004,Cucchietti2005,Merkulov2010}
(QSA)  and semiclassical
approximations\cite{Al-Hassanieh2006,ChenBalents2007,Sinitsyn2012} which describe
very well the short-time dynamics but predict a non-decaying
fraction of the central spin polarization.  Early on, it became
clear\cite{Merkulov2002} that non-Markovian
corrections\cite{KhaetskiiLoss2003,CoishLoss2004} caused by  slowly
fluctuating nuclear bath configurations generate corrections to this
non-decaying fraction as well as the long-time decay of spin
polarization. The functional form is non-universal and depends on the
details of the distribution function of the hyperfine coupling
constants.\cite{KhaetskiiLoss2003,CoishLoss2004}  A crossover from
$\propto 1/\log^\alpha(t)$ in the absence of an external magnetic
field to a $\propto 1/t \log^2(t)$ in a finite field has been
predicted\cite{KhaetskiiLoss2003,CoishLoss2004}  where the exponent
$\alpha$  is non-universal and depends on the distribution function.

Semiclassical
approaches\cite{Merkulov2002,KhaetskiiLoss2003,ErlingssonNazarov2004,Cucchietti2005}
have been employed to access the long-time dynamics of the
spin-dynamics using either a spin-coherent-state P
representation\cite{Al-Hassanieh2006,Sinitsyn2012} or a path integral
formulation,\cite{ChenBalents2007} both based on coherent spin
states. Spin fluctuations in quantum dot ensembles have been addressed
by a semi-classical Langevin term in the Bloch
equation.\cite{Glazov2012,StanekRaasUhrig2013} 

All quasi-statical and semiclassical approximations or truncation
schemes in quantum-master equations allow to access the thermodynamic
limit by neglecting higher order correlations in the spin bath.  While
such an approximation is very useful for tracing the spin-decay of
an initially polarized central spin coupled to an incoherent spin
bath, these approximations become questionable in coherent control
experiments\cite{GreilichBayer2007,FokinaBayer2010,SpatzekGreilichBayer2011}
where a pulsed pump laser induces  frequency focusing of electron spin
coherence\cite{GreilichBayer2007} by a non-equilibrium nuclear spin
polarization.

Numerical methods, however, accurately  include the
entanglement between the central spin and the spin bath but are
limited to finite spin bath size. Recently, the  time-dependent  density matrix
renormalization group
(TD-DMRG)\cite{White92,*SchollwoeckDMRG2005,*Schollwoeck2011} has been
adapted  to the central spin
model\cite{FriedrichPhD2006,StanekRaasUhrig2013} and has been able to
push this limit to up to  $N\approx 1000$ bath
spins\cite{StanekRaasUhrig2013}  in calculation for the short time
dynamics.

While most of the theoretical literature has focused on the  dynamics 
of the central spin in a single QD, only recently a semi-classical 
approach\cite{Glazov2012} has
been applied to the calculation of spin-noise in an ensemble of 
QDs charged with single electrons or holes
as investigated in experiments.\cite{CrookerBayerSpinNoise2010,Dahbashi2012,LiBayer2012,ZapasskiiGreilichBayer2013} 
We have extended the CET approach to ensembles of QDs and present spin-noise 
spectra for this case as well. 
We find an evolution of a Gaussian-type spin-noise spectrum of
a single QD  to a more Lorentzian
shape spectrum  at finite transversal magnetic field in a QD ensemble 
similar to the observed experimental
data on electron spins \cite{CrookerBayerSpinNoise2010} 
or hole spins.\cite{LiBayer2012}

Most of those approaches predict a non-decaying fraction of the central spin polarization
and a very slow, non-exponential decay of the spin in the long-time limit
which  has not been observed in the experiments. 
Therefore, it has been suggested  that  taking into account additional nuclear quadrupole
couplings\cite{Sinitsyn2012} can lead to an exponential decay when these quadrupole 
coupling constants exceed the hyperfine coupling strength. 
In this paper, however, we do not consider such an additional quadrupole 
term. We restrict ourselves to the minimal   anisotropic CSM 
and focus on  an exact finite size calculation  
including all correlations between the electronic spin and the nuclear spins.

\subsection{Preliminaries}

We have calculated the spin-correlation function 
$S(t)=[\expect{S^z(t)  S^z} +\expect{S^z S^z(t) }]/2$  and its Fourier transformation, the spin-noise spectrum
$S(\w)$, for the anisotropic CSM.  Its isotropic limit  is relevant
to  QDs charged with a single electron,  while the maximally anisotropic case, the
exactly solvable Ising limit,  can be applied to dot-confined heavy-hole 
spins.\cite{LeeEtAl2005,HansonSpinQdotsRMP2007,Testelin2009}  The
generic anisotropic regime  interpolates between these  two extreme
cases   and accounts for dot-confined hole spin of 
arbitrary mixtures of light-hole and heavy-hole contributions.  
We have studied $S(\w)$  as a function of the
external transversal magnetic  field $B_x$: while an external
longitudinal field $B_z$  suppresses the spin decay of a spin
initially polarized in longitudinal  direction, a transverse field
$B_x$ induces a Larmor precession with the  Larmor frequency
$\w_L\propto |\vec{B}|$ of the electronic spin. 

The following qualitative picture has emerged for the spin-noise
spectrum $S(\w)$. In addition to a $\delta$-peak at zero frequency
whose spectral weight is given by the non-decaying fraction of the
spin polarization, we  find a Gaussian type  noise spectrum plus
corrections in the isotropic CSM.  The width and center  of the
spectrum  are given by the  intrinsic energy scale of the fluctuating
nuclear hyperfine field $1/T^*$. A finite transversal magnetic field
destroys the $\delta$-peak and the center of the spectrum is shifted
to larger frequencies which is given by the Larmor frequency in the
limit $\w_L T^*\gg 1$. 

In the Ising limit, the quasi-static approximation becomes exact. In
zero-field, $S_z$ cannot decay at all,  and at any finite magnetic
field  a  finite non-decaying fraction of the spin polarization
remains after averaging over all randomly precessing  configurations
in the long-time limit. In the thermodynamic limit, the finite
frequency part of the noise spectrum shows a threshold behavior, where
the threshold frequency $\w_{\rm th}$ is given by  $\w_L$. Above the
threshold, we find $S(\w) \propto (\w-\w_{\rm th})^{-\alpha}$ for
$\w>0$  where the fits to our numerics are consistent with the
predictions\cite{KhaetskiiLoss2003,CoishLoss2004} of $\alpha=1/2$. Far
away from the threshold, the noise spectrum contains non-universal
parts and is cutoff sharply at the largest eigenenergy difference
$\w_{\rm max}<\sqrt{\w^2_L + A^2_s/4}$, where  $A_s$ is determined by
the details of the electronic wave function of the confined  
hole in the QD.

In the anisotropic CSM, the spin-noise spectrum depends strongly on the
anisotropy parameter $\lambda$ and the external magnetic field. We
find a crossover from a more threshold-like noise spectrum for
transversal fields that are small compared to $\lambda$ to a more
Gaussian type shape but with a renormalized width $1/T^*_\lambda$
which depends on the asymmetry parameter.

\subsection{Plan of the Paper}
As outlined above, the main objective of the paper is the discussion
of the electronic spin noise  in the anisotropic  CSM. Since the two
extreme limits, the isotropic CSM and the Ising limit, show two distinct
spectral properties, we divide the part on the results  in three sections.

But first we begin with an introduction of the model 
in Sec.\ \ref{sec:model} and
the discussion of a realistic distribution of hyperfine coupling
constants in Sec.\ \ref{sec distribution}. That  distribution depends
not only on the envelope of the electronic wave-function but also on
the finite volume $V\propto r_0^3$ that encloses the QDs since with
increasing volume the number of nuclear spins which couple
exponentially weak is increasing. We  briefly review the CET in Sec.\
\ref{Sec:Chebyshev} before we state the expansion of $S(\w)$ in terms
of Chebychev polynomials in Sec.\ \ref{sec:CET-spin-noise}. 

Sec.\ \ref{sec:results:isotropic-CSM} is devoted to the results for
the isotropic CSM  while Sec.\ \ref{sec:results-ising-limit} focuses
on the Ising limit.  In Sec.\ \ref{sec:results-anisotropic-CSM} we
present our data  for the fully anisotropic case  and investigate the
crossover from small to large transversal magnetic fields. 

In order to establish the accuracy of the CET approach, we  compare
the CET results  with exact diagonalization (ED) for small bath sizes
in Sec.\ \ref{sec:benchmark}; we also augmented our data with the
prediction of QSA for the short time dynamics  that has been reviewed
in Sec.\ \ref{sec mf}. Sec.\ \ref{sec:influence-distribution-function}
is devoted to an investigation of the influence of the distribution function
on the real-time dynamics while we extract the cutoff dependence of
the non-decaying fraction of spin polarization in Sec.\ \ref{sec:influence-of-r0}.

In Sec.\ \ref{sec:QD-ensemble-average} we present results for ensemble averaged
spin-noise spectra for parameters which closely resemble the recent 
experiments.\cite{CrookerBayerSpinNoise2010,Dahbashi2012,LiBayer2012,*ZapasskiiGreilichBayer2013}
We explicitly demonstrate that a distribution of characteristic time scales of the quantum dots
modifies the spectral properties from a more Gaussian like shape to an ensemble averaged spectrum
which can be fitted with a Lorentzian. We will discuss the $g$-factor induced and 
hyperfine interaction induced broadening of the single QD spectra. We summarize our findings in
Sec.\ \ref{sec:discussion-outlook} and give a brief outlook.

\section{Theory}
\label{sec:theory}

\subsection{Modelling of the quantum dots}
\label{sec:model}

For the spin decoherence in semiconductor QDs various interactions
play a  role. As main contributions three sources  have been
identified for relativistic electrons confined in a semiconductor QD:
the (i) Fermi contact hyperfine interaction, (ii) the dipole-dipole
interaction and  (iii)   the coupling of the orbital angular momentum
to the  nuclear spin.\cite{FischerLoss2008}  The Fermi contact
hyperfine interaction provides the largest energy scale of the three
contributions.\cite{FischerLoss2008}

Since the atomic contribution\cite{LeeEtAl2005,FischerLoss2008} to
the electron wave function stems mainly from 4s-orbitals  in Ga and
As, the Fermi-contact hyperfine-interaction dominates. The wave
functions for light and heavy holes, however, are dominated by
4p-orbitals which vanish at the nuclei. Therefore, the sources (ii)
and (iii)  govern the coupling for light and heavy holes to the
nuclear spins. 

Fischer et al.\cite{FischerLoss2008} have shown that all cases can be
casted into an anisotropic CSM\cite{Testelin2009} given
by the Hamiltonian
\begin{eqnarray} H &= &\w_L \vec{S} \vec{n}_B + \sum_{k=1}^{N} A_k
\left( S^z I_k^z + \frac{1}{\lambda} \left( S^x I_k^x + S^y I_k^y
\right) \right) 
\label{eq hamiltonian}
\end{eqnarray} 
where $ \vec{S}$ denotes the electron spin operator, $
\vec{I}_k$ the nuclear spin of the $k$-th nucleus, $N$ the number of
nuclear spins, and $\vec{n}_B =\vec{B}/|\vec{B}|$ is
the unit vector of the external magnetic field direction.
We include the electron or hole $g$-factor as well as the external
magnetic field strength $B= |\vec{B}|$ into the Larmor frequency
$\w_L=g \mu_B B$.  The anisotropy parameter $\lambda$ distinguishes
the three different cases: $\lambda = 1$ for electrons,
$\lambda= 1/2$ for light holes and $\lambda \rightarrow \infty$ for
heavy holes. For mixed heavy and light hole states $1<\lambda <\infty$
holds.  We will review a realistic distribution of the $A_k$ and an
estimate of the orders of magnitude in the next section below. Since
$\lambda$ introduces an anisotropy axis, we use the term longitudinal
for external magnetic fields in $z$-direction and call $B_x$ a
transverse field.

For $\lambda = 1$, we recover the standard isotropic central-spin
model\cite{Gaudin1976} which conserves the total spin $\vec{J} =
\vec{S}+\sum_k \vec{I}_k$ of the coupled system in absence of an
external field, and the spin component of the total spin in the
direction of the applied field.  For $\lambda \neq 1$ only the
component $J^z = S^z+\sum_k I^z_k$ of the total spin commutes with the
Hamiltonian for the absence of a transversal external
field. Throughout the paper we will use the convention $\hbar=1$,
$k_B=1$ unless otherwise stated.

Recently, the effect of additional nuclear quadrupole couplings
in the Hamiltonian on the central spin dynamics
 have been investigated.\cite{Sinitsyn2012} 
Such  nuclear quadrupole terms  significantly 
change the bath characteristics.  They lift the very large degeneracies of 
the nuclear spin bath and  lead to an exponential
spin decay\cite{Sinitsyn2012} once the nuclear quadrupole coupling strength exceeds 
the hyperfine interaction. In this paper, however, we do not include  these additional 
nuclear quadrupole couplings and present an exact finite size calculation that avoids any
truncations or factorization of correlations which typically changes
the type of long-time dynamics.

{\em Definition of a time scale:} In addition to the Larmor frequency
$\w_L$, the fluctuations of the transversal and longitudinal component
of an unpolarized nuclear spin bath in the absence of an external field
defines the time scale $T^*_\lambda$
\begin{eqnarray} [T_\lambda^*]^{-2} &=& \frac1{\lambda^2}
\sum_{k=1}^{N} A_k^2
\end{eqnarray} and $T^*=T^*_{\lambda=1}$ respectively.  These scales
govern the short-time spin decay of the electronic spin polarized
along the $z$-axis. We use the transversal time scale to define the
dimensionless hyperfine couplings $a_k = A_k T_\lambda^*$ which enters
the dimensionless Hamiltonian $\tilde H =  H T^*_\lambda$
\begin{eqnarray}
\label{eq hamiltonian-dim-less} 
\tilde H &=&  \lambda b \vec{S} \vec{n}_B 
+
\sum_{k=1}^{N} a_k \left( S^z I_k^z + \frac{1}{\lambda} \left( S^x
I_k^x + S^y I_k^y \right) \right) 
\,.
\end{eqnarray} The longitudinal scale $T^*$  has been used to define
the dimensionless magnetic field $b= \w_L T^*$.

\subsection{Distributions of the coupling constants $A_k$} 
\label{sec distribution}

In numerical simulations of the CSM either a
model\cite{BortzStolze2007,Bortz2010,StanekRaasUhrig2013,FaribautSchuricht2013a,FaribautSchuricht2013b}
distribution function $P(A)$ for the hyperfine coupling constants
$A_k$, or a more realistic\cite{SchliemannKhaetskiiLoos2003}
distribution based on the envelope function (\ref{eqn:psi-R}) have
been used.
In materials the coupling constants $A_k$ are given
by\cite{AbragamNMR1961,HansonSpinQdotsRMP2007,
LeeEtAl2005,FischerLoss2008}
\begin{align} 
A_k &= \frac{16\mu_B \mu_N \gamma_k}{I_k} \left\lvert \Psi
(\vec{R}_k) \right\rvert^2 \eta_k 
& = A_s v_0 \eta_k \left\lvert \Psi (\vec{R}_k) \right\rvert^2,
  \label{eq akdef}
\end{align} 
where $\mu_B$ is the Bohr magneton, and $\mu_N$ the  nuclear magneton, 
and $\gamma_k$ the gyro-magnetic factor of the $k$-th nuclei. 
 $I_k$ denotes the spin of
the $k$-th nucleus. $v_0$ is the average volume
occupied by a single nucleus within the crystal, and
$A_s = 16\mu_B \mu_k/v_0 I_k$.

The electron (or hole) wave function $\psi
(\vec{r}) = \Psi (\vec{r}) u(\vec{r})$ is divided into a slowly
varying envelope function $\Psi (\vec{r})$, that appears in (\ref{eq
akdef}), and a fast varying dimensionless Bloch factor $u (\vec{r})$
describing the wave function in the individual unit cells at the
nuclei $k$ and determining $\eta_k$.  The factor $\eta_k$
encodes the symmetries of the Bloch factor and differs for electrons
and holes:\cite{Testelin2009} 
\begin{align} 
\eta^e_k &= \frac\pi3 \left\lvert u(\vec{R}_k)
\right\rvert^2\\ \eta^h_k &= \frac85 v_0 \left\langle
\frac{1}{\left\lvert \vec{r} - \vec{R}_k \right\rvert^3} \right\rangle
   \label{eq etaholes} \, .
\end{align} 
(For details on the definition of
Eq.\ (\ref{eq etaholes}) see appendix A of Ref.\
\onlinecite{Testelin2009}.)
In case of a simple Bloch wave of a free electron,
$|u(\vec{R}_k)|^2=1$, and different factors $\eta^e_k$ and $\eta^h_k$
are discussed in the
literature\cite{SchliemannKhaetskiiLoos2003,Testelin2009} to account
for the different values of the electronic wave function at Ga and As
nuclear sites in the unit cell.

In this paper, however, we neglect these differences and restrict our
investigation to a generic spin bath with $j=1/2$. We set $\eta^{e(h)}_k=1$
and absorb the value into the definition of $A_s$.

Since $\Psi (\vec{R}_k)$ varies slowly over the volume of a single
nucleus, $|\Psi (\vec{R}_k) |^2$ is taken as constant over the volume
$v_0$, and the normalization integral can be approximated by a
discrete sum over all nuclei
\begin{align} 
1 &= \int \text{d}^3 R \left\lvert \Psi (\vec{R}_k)
\right\rvert^2 \approx \sum_k v_0 \left\lvert \Psi (\vec{R}_k)
\right\rvert^2
\end{align} from which we conclude:
\begin{eqnarray} 
A_s &=& \sum_k A_k 
\end{eqnarray} 
and is constant independent of the details of the wave function
as a consequence of the wave function normalization.
$A_s$ is
typically\cite{LeeEtAl2005,HansonSpinQdotsRMP2007,Testelin2009} of the
order $O(10^{-5}\,\text{eV})$ for electrons and predicted\cite{Testelin2009} about a factor
$10-1000$ times smaller for holes yielding a much larger decoherence
time $T^*$  for hole-spins.

A typical distribution of $A_k$ for an (InGa)As self-assembled quantum
dot with base diameter of $15\,\text{nm}$ is depicted in Fig.\ 2 of
Ref.~\onlinecite{LeeEtAl2005}. This suggests a normalized envelope
function of
\begin{eqnarray}
\label{eqn:psi-R} \Psi(\vec{r})&=& (\sqrt{\pi} L_0)^{-3/2}
e^{-\frac{r^2}{2L_0^2}}
\end{eqnarray} varying on a length scale of $L_0\approx 5\,\text{nm}$. 
Using\cite{LeeEtAl2005} the lattice constant of GaAs of $\approx
0.5\,\text{nm}$, $L_0=5\,\text{nm}$ and the ${\rm max}\{ A_k\}\approx 8\,\text{neV}$,
we obtain a realistic  estimate\cite{LeeEtAl2005,HansonSpinQdotsRMP2007,Testelin2009}  
for $A_s\approx 20\,\mu \text{eV}$.

\begin{figure}[t]
\begin{center}

\includegraphics[width=80mm]{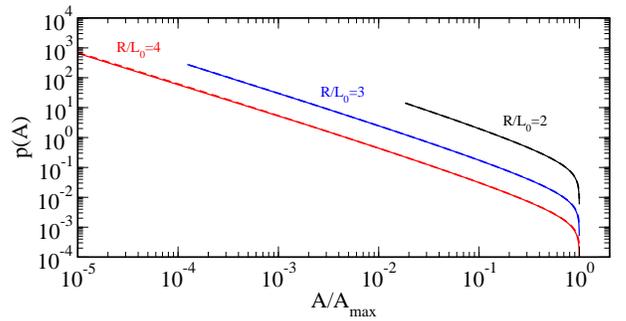}

\caption{Probability density $P(A)$ vs $A_s/A_{max}$ 
for three different radii $R/L_0=2,3,4$. The dashed lines
of the same color depict the histogram of the distribution generated by $10^{8}$ random picks
for $A_k$.}
\label{fig:probability-pa}
\end{center}
\end{figure}

Apparently, the probability to find a nucleus with coupling $A(r)$
increases quadratically with radius $r$: the distribution function
$P(A)$ diverges for $A\to 0$ and requires a finite cutoff radius
$R>L_0$ for which the envelope wave function has almost vanished and
$\Psi(\vec{r})$ is approximately normalized in that sphere of radius
$R$.  Although the physics must be independent of this artificial
cutoff $R$, only those nuclei with a significant coupling constant
contribute to the dynamics of the central spin.  Using the probability
density for point of fixed radius $r$, $P(r)= 3r^2/R^3$ and
$P(A)dA=P(r)dr$, we derive the probability distribution
\begin{eqnarray}
\label{eqn:p-a} P(A,r_0) &=& \frac{3}{2} \frac{1}{r_0^3 A}
\sqrt{\log(A_{max}/A)}
\end{eqnarray} 
where $A_{max}=A(0) =A_s v_0/(\sqrt{\pi} L_0)^3$ and
the ratio $r_0$ is defined as $r_0=R/L_0$.  This distribution is shown
in Fig. \ref{fig:probability-pa} for three different ratios $r_0 =
R/L_0=2,3,4$. We also added the histogram from $10^{8}$ randomly
generated $A_k$ values as dashed line in the same color. They are
nearly indistinguishable from the analytic function stated in
(\ref{eqn:p-a}). Note that this distribution function is almost identical to the one
used by Coish and  Loss.\cite{CoishLoss2004}

Using this probability distribution $P(A)$ it is straight forward to
calculate the average $\bar A = \int dA P(A) A \approx A_s/N(R)$ where
$N(R)= n_0 4\pi R^3/3$ denotes the number of nuclei in the sphere of
radius $R$.  The square average $\bar A^2$ is approximately given by
$\bar A^2 \approx \sqrt {2} A_s^2 [3\sqrt{\pi} N(R)N(L_0)]^{-1}$.
Both approximations become exact for $R\to\infty$.

Since the choice of the radius $R$ should be arbitrary, as long as
$R>L_0$ and $|\Psi(R)|^2\approx 0$, the physical properties of the
central spin model must not depend on $R$. This is clearly the case
for $A_s=\sum_k A_k = N(R) \bar A$ since we have $N(R)$ different
coupling constants.  For the time scale $T^*$, the inverse rms of the Overhauser
field, which governs the short
time dynamics, we obtain
\begin{eqnarray} 
[T^*]^{-2} &=& \sum_{k=1}^{N} A_k^2 = N(R) \bar A^2 =
A_s^2 \frac{1}{3\sqrt{\pi}N(L_0)}
\label{eqn-11}
\end{eqnarray} 
which is also independent of the radius $R$. It is only
dependent on $A_s$ and the number of nuclei which are located within
the sphere defined by the length scale $L_0$ of the electronic
envelope function. Using the parameters from above an estimate for
$T^* \approx 30$ ns and a characteristic frequency
$f=1/T^*\approx 32$ Mhz for electrons and a factor
$10^{-1}-10^{-3}$ times smaller value for hole spins.

Although  the cutoff $r_0$ controls the width of the distribution function 
$P(A,r_0)$, the physics must remain invariant in the thermodynamic limit,
when sending $N\to\infty$ first and then $r_0\to\infty$. 
In a finite size calculation, however,  each random configuration
of hyperfine couplings  $\{ A_k \}$ generated by $P(A,r_0)$
yields a slightly different dynamics. 
To bridge between the typically $N(L_0)\approx 10^5$ nuclear spins in
real QDs and the numerical CET simulations of a spin bath with $N=20$,
each configuration $\{ A_k\}$  is normalized to a fixed $1/T^* = \sqrt{\sum_k A_k^2}$ 
which is the  energy unit used in all calculations. 
Therefore, each configuration is
characterized by exactly the same short time dynamics, and by
averaging over typically $n=50-100$ different configurations we mimic
a much larger spin-bath.

A word is in order about varying the cutoff $r_0$. For very large $r_0$, the ratio $a$ 
between the largest and the smallest hyperfine coupling is exponentially large and
the probability of generating exponentially low coupling constants is large. In this case, we will end
up with one or two large couplings $a_k=A_K T^*\to 1$, while all other are exponentially small for
a fixed $N$. The resulting unphysical dynamics  will be discussed  in Sec.\ \ref{sec:benchmark} below.

In order to avoid this effect one could demand that the largest dimensionless coupling 
$a_{max} = {\rm max}\{a_k\}$ must be a constant when varying $r_0$. This requires 
a simultaneous change of the bath size $N$ when varying $r_0$.
Choosing $a_{max}=0.5$ ensures a reasonable distribution  of dimensionless coupling constants $\{a_k\}$
since the sum of all other couplings squares 
must be $3/4$. For $r_0=1.5$ we can fulfill this condition with 
$N=18$ while for $r_0=1$ only $N=8$ would be sufficient. For $r_0=2$, however, we would
need $N=42$ nuclear spins which is beyond the reach of the CET.
For $r_0=2$ and $N=18$ we find $a_{max}=0.64$: this is only slightly larger than $0.5$
and  implies that all other hyperfine constants still contribute $\approx 60\%$  to
$T^*$. In our simulations, we typically us $r_0=1.5$ and $N=18$.

\subsection{Definition of the spin-noise function}
 
Experimentally the spin-noise is measured via fluctuations of the
Faraday rotation angle using a linearly  polarized probe laser in
$z$-direction of the sample. The auto-correlation function of the
Faraday rotation angle is equivalent to the symmetrized fluctuation
function
\begin{eqnarray}
\label{eqn:def-spin-noise} S (t) &=& \frac{1}{2}\left[ \expect{ S^z(t)
S^z} +\expect{ S^z S^z(t)} \right] - \expect{S^z}^2
\end{eqnarray} where $\expect{S^z}$ denotes the average
spin-polarization which vanishes in the absence of an external
magnetic field.  The probe laser only weakly perturbs the system, and
all expectation values are calculated using the equilibrium density
operator.  Since $S(t)$ is symmetric in time, the spin-noise spectrum
\begin{eqnarray} 
S (\w)&=& \int_{-\infty}^\infty S (t) e^{-i\w t} dt =
\int_{-\infty}^\infty S (t)\cos(\w t) dt \, . \label{s_alpha}
\end{eqnarray} 
From these definitions, we obtain the obvious sum-rule
\begin{eqnarray} 
\int_{-\infty}^\infty \frac{d\w}{2\pi} \, S (\w) &=&
S(0) = \expect{(S^z)^2} -\expect{S^z}^2 
\label{eq sum-rule}
\end{eqnarray} 
for the spin-noise spectrum. In the absence of an
external magnetic field, its value is fixed to $1/4$ for  a QD
filled with a single electron or hole spin. This sum-rule is useful
to test the accuracy of any numerical spin-noise calculation.


\subsection{Connection between the spin-noise function and the
real-time dynamics}
\label{sec:spin-noise-real-time-dynamics}

The spin-noise measurements are performed in thermal
equilibrium\cite{CrookerBayerSpinNoise2010} at $T \approx 5\,\text{K}
\rightarrow k_B T \approx 4\cdot 10^{-4}\,\text{eV}$.  Since the
intrinsic energy scale $A_s$ of the system is of the order
$O(10^{-5}\,\text{eV})$ for electrons and even one order of magnitude
smaller for holes, $\beta A_s\ll 1$ holds, and we can consider the
coupled system consisting of the nuclear spin bath and the central
spin in the limit of high temperature and being characterized by the
initial density operator
\begin{align} \rho_0 &= \frac1D \underline{\underline{1}} \, ,
\end{align} where $D$ is the dimension of the Hilbert space and
$\underline{\underline{1}}$ is the identity matrix.  Using the
commutator $\left[ \rho_0, S^z \right] = 0$, we conclude:
$\expect{S^z(t) S^z} = \expect{S^z S^z(t)}$.

If we prepare an initially fully polarized electron (hole) spin along
the $z$-direction coupled to an incoherent nuclear spin bath, the
density operator for such a system is given by
\begin{align} 
\rho_p &= \rho_0 (\underline{\underline{1}} + 2 S^z)
\,\, .
\end{align} 
The time evolution $\expect{S_z (t)}$ for this initial
condition
\begin{eqnarray} 
\expect{S^z (t)}_{\rho_p} &= & \Tr{\rho_pS^z (t)} =
\left( \expect{S^z} + 2 \expect{S^z (t) S^z} \right) \non &=& 2
\expect{S^z (t) S^z} = 2 S(t)
   \label{eq:sz-s-vs-t}
\end{eqnarray} 
is equivalent to twice the correlation function $S (t)$
where the expectation values are calculated with respect to $\rho_0$.
Therefore, $S(t)$ can also be interpreted as the dynamics of an
initially fully polarized spin coupled to a bath at high
temperature. In this limit, we still can neglect the spin polarization
in Eq.\ (\ref{eqn:def-spin-noise}) in a magnetic field, since the
large field limit discussed below implies a large magnetic field in
comparison to the hyperfine energy scale but still small compared to
the temperature.


\subsection{Quasi-static approximation for $\vec{B} =0$}
\label{sec mf}

Merkulov\cite{Merkulov2002} et al.\ proposed a quasi-static approximation (QSA) to
calculate the short-time dynamics of an initially polarized electron
spin which later has been extended to dot-confined hole spins by Testelin et al.\cite{Testelin2009} 
It is based on a separation of energy scales
and, therefore, time scales.  While a single nuclear spin just is
exposed to the field generated by the single central spin whose
magnitude is proportional to $A_k\ll A_s$, the electron spin precesses
in a constant effective magnetic field $\vec{B}_{\text{eff}}$ provided
by a frozen nuclear spin bath configuration $\ket{\psi_{\rm bath}}$
\begin{eqnarray} 
\vec{B}_{\text{eff}} &=& \frac{1}{\mu_B g_e} \sum_k
A_k \bra{\psi_{\rm bath}}\vec{I}_k\ket{\psi_{\rm bath}} =
B_{\text{eff}}\vec{n}
  \label{glg beff}
\end{eqnarray} 
for the time scale defined by the effective Larmor
frequency $\omega_{\text{eff}} = g_e \mu_B |\vec{B}_{\text{eff}}|$
which is of the order of $O(A_s)$.
In this momentarily frozen field in the direction 
$\vec{n} = \vec{B}_{\text{eff}} /|\vec{B}_{\text{eff}} |$,
the Bloch equations for the electronic spin dynamics have the simple
solution
\begin{eqnarray}
\label{eq eom} 
\expect{\vec{S}(t)} &=& \left( \vec{S}_0 \cdot \vec{n} \right)
\vec{n} + \left( \vec{S}_0 - \left( \vec{S}_0 \cdot \vec{n} \right)
\vec{n} \right) \text{cos}(\omega_{\text{eff}} t) 
\nonumber
\\ 
&& + \left[
\vec{n} \times \left( \vec{S}_0 - \left( \vec{S}_0 \cdot \vec{n}
\right) \vec{n} \right) \right] \text{sin}(\omega_{\text{eff}} t),
\end{eqnarray} 
with initial polarization of the electron spin
$\vec{S}_0$.

The effective magnetic field is generated by a large number of small
contributions from randomly oriented nuclear spins. Therefore, the
direction is isotropically distributed over a unit-sphere, and, in the
limit of large $N$, the magnitude of the effective field is described
by the Gaussian probability distribution
\begin{align} 
W (\vec{B}_{\text{eff}}) &= \frac{1}{\pi^{3 / 2}
\Delta_B^3} \text{exp}\left( -
\frac{\vec{B}_{\text{eff}}^2}{\Delta_B^2} \right), \label{glg
feldverteilung} \\ \Delta_B^2 &= \frac{1}{2(\mu_B g_e)^2} \sum_k A_k^2
= \frac{1}{2(\mu_B g_e T^{*})^2}
\end{align} 
whose width is defined by the fluctuation time scale
$T^*$.
Averaging the central spin dynamics (\ref{eq eom}) over the
distribution function $W (\vec{B}_{\text{eff}}) $
\begin{eqnarray} 
\la \vec{S} (t) \ra &= & \int d\Omega \int_0^{\infty}
B^2 \text{d}B \, W(\vec{B}_{\text{eff}})\vec{S}(t)
\end{eqnarray} 
the QSA result\cite{Merkulov2002} for $\vec{S}(t) =
S^z(t) \vec{e}_z$
\begin{eqnarray} 
\left\langle S^z (t) \right \rangle &=& \frac{S^z_0}{3}
\left[ 1 + 2 \left(1 - \left(\frac{t}{2T^{*}} \right)^2 \right)
e^{-\frac{1}{2}\left(\frac{t}{2T^{*}} \right)^2} \right] \non &=& 2M(t)
 \label{glg merkulov}
\end{eqnarray} 
has been obtained. It is straight forward to calculate
its Fourier transformation $M(\w)$
\begin{eqnarray}
\label{eq:m-w} M(\omega) &=& \int_{-\infty}^\infty \text{d}t\,
\text{e}^{-i\omega t} M(t) \\ &=& \frac{S_0}{3} \left[ 2\pi
\delta(\omega) + \omega^2 (\sqrt{8}T^*)^3 \sqrt{\pi}
\text{e}^{-2(\omega T^*)^2} \right] \, ,  \nonumber
\end{eqnarray}
where $S_0 = S^z_0 / 2$, because $S_0$ refers to the correlation
function $S(t)$ instead of $\left\langle S^z (0) \right\rangle$.
Since the QSA generates decoherence only by angular averaging, it
lacks long-time decay and contains a large non-decaying contribution
of $1/3$ of the initial spin-polarization $S_0$. This large
non-decaying contribution defines the weight of the spin-noise
$\delta$-function at $\w=0$.

\subsection{The Chebyshev expansion technique}
\label{Sec:Chebyshev}

\subsubsection{Expansion of the time evolution operator}

Since all our results have been obtained using the
CET,\cite{TalEzer-Kosloff-84,Kosloff-94,Fehske-RMP2006} we briefly
review the CET, in order to introduce the notation used below.

The CET\cite{TalEzer-Kosloff-84,Kosloff-94,Fehske-RMP2006} has been
developed 30 years ago and offers an accurate way to calculate the
time evolution of a single initial state $|\psi_0 \rangle$ under the
influence of a general time-independent and finite-dimensional
Hamiltonian ${\cal H}$:
\begin{eqnarray} |\psi(t) \rangle &=& e^{-i{\cal H}t}|\psi_0\rangle .
\label{stationarySolution}
\end{eqnarray} The main idea of the method is to construct a stable
numerical approximation for the time-evolution operator $e^{-i{\cal
H}t}$ that is independent of the initial state $|\psi_0\rangle$ and
whose error can be reduced to machine precision for any given time
$t$. Its limitation lies in the need to explicitly store certain
states in the course of the calculation, which limits the size of the
Hilbert space that can be handled.

There are different ways to expand the time-evolution operator. The
most direct one is the conventional expansion of the exponent in
powers of ${\cal H}$ using the definition of any operator function.
One would like, however, to use an expansion that converges uniformly,
independent of the initial state $|\psi_0\rangle$.  The Chebyshev
polynomials turned out to be such a suitable
choice\cite{TalEzer-Kosloff-84}. They are defined by the recursion
relation
\begin{equation} T_{n+1}(z) = 2zT_n(z)-T_{n-1}(z) ,
\end{equation} subject to the initial conditions $T_0(z) = 1$ and
$T_1(z) = z$. Those polynomials can be used to expand any function
$f(z)$ on the interval $-1 \leq z \leq 1$. Explicitly, $f(z)$ is
expressed as an infinite series
\begin{eqnarray} 
f(z) &= &\sum_{n=0}^{\infty} b_n T_n(z) ,
\label{f-z-expantion}
\end{eqnarray} 
where the expansion coefficients $b_n$ are given by
\begin{eqnarray} 
b_n &= &\frac{2 - \delta_{n, 0}}{\pi} \int_{-1}^{1}
dx\, \frac{f(x)T_n(x)}{\sqrt{1-x^2}} \, .
\label{b_n_integral}
\end{eqnarray} 
Using the integral
representation\cite{AbramowitzStegun} of the Bessel function
\begin{eqnarray} 
J_n(z) &=& \frac{i^{-n}}{\pi} \int_0^\pi d\vartheta
e^{iz\cos\vartheta} \cos(n\vartheta)
\end{eqnarray} 
and $T_n(\cos \vartheta)= \cos(n\vartheta)$, we
immediately arrive for $z\in[-1:1]$ at
\begin{eqnarray} 
e^{-iz \tau } &=& \sum_{n = 0}^{\infty} b_n T_n(z)
\end{eqnarray} 
with the expansion coefficients $b_{n} =
(2-\delta_{0,n}) i^n J_n (\tau)$.

If the spectrum of the Hamiltonian is bound to $E_{\rm min}\le E\le
E_{\rm max}$, the time-evolution operator $e^{-i {\cal H} t}$ can be
expanded in the same fashion after mapping the Hamiltonian to the
dimensionless $H' = (H-\alpha)/\Delta E$ where we have defined the
center of the energy spectrum $\alpha = (E_{\rm max}+E_{\rm
min})/2$ and its half-width $\Delta E = (E_{\rm max}-E_{\rm
min})/2$. Identifying $\tau=\Delta E t$ we arrive at
\begin{equation} 
e^{-i H t} = \sum_{n = 0}^{\infty} b_n(t) T_n(H') \,
\label{Chebyshev-exp-e^-iH}
\end{equation} with
\begin{eqnarray}
\label{eqn:b_n} b_n(t)&=& (2-\delta_{0,n} ) i^n e^{-i\alpha t} J_n (
\Delta E t).
\end{eqnarray}
Finally, applying Eq.~(\ref{Chebyshev-exp-e^-iH}) to the initial state
$|\psi_0\rangle$ one obtains
\begin{equation} 
|\psi(t) \rangle = \sum_{n = 0}^{\infty} b_n(t)
|\phi_n \rangle ,
\label{CET}
\end{equation} where the infinite set of states $|\phi_n \rangle =
T_n({\cal H}') |\psi_0\rangle$ obey the recursion
relation\cite{Fehske-RMP2006}
\begin{equation} 
|\phi_{n+1} \rangle = 2{\cal H}' |\phi_n \rangle -
|\phi_{n-1} \rangle ,
\label{chebyshev-recursion-relation}
\end{equation} 
subject to the initial condition $|\phi_0 \rangle =
|\psi_0 \rangle$ and $|\phi_1 \rangle = {\cal H}' |\psi_0 \rangle$.

Several comments are in order. First, all time dependence is confined
in Eq.~(\ref{CET}) to the expansion coefficients $b_n(t)$, which are
independent of the initial state $|\psi_0\rangle$.  Second, the
Chebyshev recursion relation of
Eq.~(\ref{chebyshev-recursion-relation}) reveals the iterative nature
of the calculations. Starting from the initial state $|\psi_0\rangle$,
one constructs all subsequent states $|\phi_n\rangle$ using repeated
applications of the ``transformed'' Hamiltonian ${\cal H}'$. Third,
since $J_n(x) \sim (e x/2 n)^n$ for large order $n$, the Chebyshev
expansion converges quickly as $n$ exceeds $\Delta E t$. This allows
to terminate the series (\ref{CET}) after a finite number of elements
$N_C$ guaranteeing an exact result up to a well defined order.
Finally, the Chebyshev expansion has the virtue that numerical errors
are practically independent of $t$, allowing access to very long
times. The main limitation of the approach, as commented above, stems
from the size of the Hilbert space, since each of the states
$|\phi_n\rangle$ must be constructed explicitly.

For the application of the CET to the central spin model, an
estimation of the upper and lower bound of the Hamiltonian is required
entering the center of the energy spectrum $\alpha$ and its half-width
$\Delta E$.  Applying the power iteration method, the series
\begin{eqnarray} \lv \varphi_n \ra = \frac{{\cal H}^{n} \lv \varphi_0
\ra}{\sqrt{\la \varphi_0 \rv {\cal H}^{n} {\cal H}^{n} \lv \varphi_0
\ra}}
\end{eqnarray} converges to the eigenvector associated with the
eigenvalue $\bar E$ the largest absolute value ${\rm max}\{
|E_{\text{min}}|, |E_{\text{max}}| \}$.  For $\lambda=1$, and $A_k>0$,
one can show that the eigenvalue obtained by the power iteration
determines $E_{\text{min}}$ while $E_{\text{max}} =  \w_L/2 +
A_s /4$. For the Ising regime, $\lambda\to\infty$, the largest
eigenvalue $E_+$ and the smallest eigenvalue $E_-$ are exactly known
\begin{eqnarray} E_\pm &=& \pm\sqrt{\left( \frac{\w_L}2 \right)^2 +
\left( \frac{A_s}{4} \right)^2}
\end{eqnarray} and for any finite $\lambda$, we interpolate between
these to limits. Alternatively, one can set $\alpha=0$ and only use
the eigenenergy $\bar E$ to define $\Delta E = 2\bar E$. In either
case, $\alpha$ and $\Delta E$ entering the Chebyshev expansion are
easily obtained.

\subsubsection{Evaluating traces}
\label{sec:traces}

The original application\cite{Dobrovitski2003} of the
CET\cite{TalEzer-Kosloff-84,*Kosloff-94} focused on the dynamics of a
single wave-function. We have extended the approach to thermodynamic
ensembles to incorporate the incoherent spin-bath at high-temperature
relevant to the experiments.

The expectation value of an arbitrary time-dependent observable $O$ is
given by
\begin{eqnarray} 
\left \langle O(t) \right \rangle &= &\text{Tr}\left[
\rho_0 O (t) \right] = \sum_{i = 1}^{D} \left \langle i \right \rvert
\rho_0 \text{e}^{iHt} O \text{e}^{-iHt} \left \lvert i \right \rangle
\non &=& \sum_{i = 1}^{D} \bra{i'(t) }O \ket{i(t)}
\label{glg spurbildung}
\end{eqnarray} 
where $\left \lvert i \right \rangle$ denotes a state
of the complete basis set.  $D= 2^{N+1}$ grows exponentially with the
number of bath spins, and the direct evaluation of the trace cannot be
computed in moderate time for large $N$.  In addition, the CET
provides only the time evolution of a single state $\ket{i(t)}=
\text{e}^{-iHt} \ket{i}$ and $\ket{i'(t)} = \text{e}^{-iHt}\rho_0
\ket{i}$.

Therefore, we employ a stochastical method discussed by Weisse et
al..\cite{Fehske-RMP2006} It is based on the generation of $N_s$
random states $\ket{r}$ of the form
\begin{eqnarray} 
\left \lvert r \right \rangle &= &\sum_{i = 1}^{D}
\xi_{ri} \left \lvert i \right \rangle 
\label{eq ranstate}
\end{eqnarray} 
with the real coefficients $\xi_{si}$ fulfilling the
relations
\begin{eqnarray} 
\left \langle \left \langle \xi_{ri} \right \rangle
\right \rangle &=& 0, \\ 
\left \langle \left \langle \xi_{ri}
\xi_{r'j} \right \rangle \right \rangle &= &\delta_{r,r'} \delta_{i,j}
\end{eqnarray} 
where $\left \langle \left \langle \cdots \right
\rangle \right \rangle$ refers to the statistical average of these
random numbers. Note that $\ket{r}$ is not a normalized state for
$\xi_{si}$ fulfilling those relations.  However, the trace of an
operator $\hat A$ can be evaluated\cite{Fehske-RMP2006} by
\begin{eqnarray} 
\left \langle \left \langle \frac{1}{N_s} \sum_{r =1}^{N_s} 
\bra{r} \hat A \ket{r} \right \rangle \right \rangle &=&
\frac{1}{N_s} \sum_{r = 1}^{N_s} \sum_{i,j = 1}^{D} \left \langle
\left \langle \xi_{ri} \xi_{rj} \right \rangle \right \rangle
\bra{i}\hat A\ket{j} \nonumber\\ 
&=& \sum_{i= 1}^{D} \bra{i}\hat
A\ket{i}
\end{eqnarray} 
by statistical average of the random numbers.

Using the self-averaging properties of $\xi_{ri}$ drawn from a
Gaussian distribution, the trace is approximated\cite{Fehske-RMP2006}
by
\begin{eqnarray} 
\frac{1}{N_s} \sum_{r = 1}^{N_s} \bra{r} \hat A
\ket{r} &=& \sum_{i= 1}^{D} \bra{i}\hat A\ket{i} + 
O\left( \frac{1}{\sqrt{N_s D}} \right) \, .
\end{eqnarray} 
The error is well controlled and scales with $(N_s
D)^{-1/2}$: only a few states $N_s$ are needed for an exponentially
large Hilbert space. In our simulations we typically use $N_s=5$
different randomly generated states for the evaluation of the
traces.

For very small Hilbert-spaces $N<10$, we have the reverse situation:
the number $N_s$ of random states required for a small error might
exceed the dimension of the Hilbert-space $D$.  In such cases, the
trace has been evaluated exactly.

\subsection{Spin-noise spectra obtained from Chebyshev polynomial
expansion}
\label{sec:CET-spin-noise}

Since the time-dependent coefficients of the CET are known and are
stated in Eq.\ (\ref{b_n_integral}), we can analytically perform the
Fourier transformation of $S(\w)$ in Eq.\ (\ref{s_alpha}) and derive
an explicit expression for the spin noise in terms of the momentum
$\mu_{n,m}$ and a convolution of two Chebyshev polynomials
\begin{eqnarray}
\label{eqn-CET-spin-noise-full} 
S(\w)
&=& \frac{2\pi}{\Delta E} \sum_{n,m = 0}^\infty \mu_{n,m}
\int_{-1}^{1-\frac{\omega}{\Delta E}} \text{d}\tilde\omega \,
\nonumber \\ 
&& \times \frac{T_{n}(\tilde\omega) T_m (\tilde\omega +
\frac{\omega}{\Delta E})} {\sqrt{(1 - \tilde\omega^2) (1 -
(\tilde\omega + \frac{\omega}{\Delta E})^2)}}
\label{eq calcspec}
\end{eqnarray} 
for $\omega \geq 0$. While the convolution of two
Chebyshev polynomials only depends on the half-width $\Delta E$ of the
spectrum of $H$ and is independent of the dynamics, the momentum
$\mu_{n,m}$ gather all Hamiltonian dependent information about the
dynamics and are defined as
\begin{eqnarray}
\label{eqn:momenta-nm} 
\mu_{n,m} &= &\frac{2 - \delta_{n,0}}{\pi}
\frac{2 - \delta_{m,0}}{\pi} \nonumber \\ 
&&\times \text{Tr}\left\{
\rho_0 T_n (H') S^z T_m (H') S^z \right\} \, .
\end{eqnarray} 
and evaluated with the method presented in Sec.\
\ref{Sec:Chebyshev}.

The additional prefactors $g_n$
\begin{eqnarray} 
g_n &=& \frac{(N_C - n + 2)\text{cos} \frac{\pi
n}{N_C + 2} + 
\text{sin} \frac{\pi n}{N_C + 2}\text{cot}\frac{\pi}{N_C
+ 2}}{N_C + 2},
\end{eqnarray} 
referring to as Jackson kernel,\cite{Fehske-RMP2006}
considerably reduce the truncation error\cite{Fehske-RMP2006} when
evaluating the truncated series
\begin{eqnarray} 
S(\w) &=& \frac{2\pi}{\Delta E}
\sum_{n,m = 0}^{N_C} g_n g_m \mu_{n,m} 
I_{n,m}(\frac{\omega}{\Delta E})
\label{eqn-CET-spin-noise-NC}
%
\end{eqnarray} 
instead of the true infinite series given by Eq.\
(\ref{eqn-CET-spin-noise-full}).  
Since the function $I_{n,m}(x)$  defined as
\begin{eqnarray} 
I_{n,m}(x) &=& \int_{-1}^{1- x} \text{d}\tilde\omega \,
\frac{T_{n}(\tilde\omega) T_m (\tilde\omega +x)} {\sqrt{(1 - \tilde\omega^2) (1 - (\tilde\omega +
x)^2)}} 
\end{eqnarray} 
is independent of the Hamiltonian,
it can be calculated and stored independently, 
and later used in the summation (\ref{eqn-CET-spin-noise-NC}) of the  momenta.

From the orthogonality relation of the Chebyshev polynomials we can
immediately conclude that only the  momentum $\mu_{0,0}$ contributes to
the spin noise sum-rule (\ref{eq sum-rule}):
\begin{eqnarray} 
S(t = 0) &= &\frac{1}{2\pi} \int_{-\infty}^\infty
\text{d}\omega\, S(\omega) \\ 
&=& \sum_{n,m = 0}^{N_C} \text{Tr}\left[
\rho_0 T_n (H') S^z T_m (H') S^z \right] \delta_{n,0} \delta_{m,0}
\nonumber \\ 
&= &\text{Tr}\left[ \rho_0 (S^z)^2 \right] = \frac{1}{4}.
\end{eqnarray} 
Thus all spectral functions calculated from the CET
exactly fulfill the sum-rule independent of the number $N_C$ of included
Chebyshev polynomials.


\section{Dephasing of an electron spin: the isotropic central spin
model}
\label{sec:results:isotropic-CSM}

We begin with the discussion of the spin dynamics for a single
electron confined in a single quantum dot by investigating the isotropic
CSM with $\lambda = 1$.  For all simulations of the real-time
dynamics, we used the CET for the evolution of the states in a system
of $N$ bath spins for a fixed configuration $\{ A_k\}$ drawn from the
probability distribution $P(A,r_0)$ stated in Eq.\
(\ref{eqn:p-a}). Since the number of bath sites is limited to $N\approx
20$, we average over typically $50$ configurations $\{ A_k\}$. By this
averaging we minimize finite size oscillations and essentially mimic
an effectively larger bath.


\subsection{Benchmarks}
\label{sec:benchmark}

In order to establish the virtue and the limitations of the CET
approach in combination with a statistical  evaluation of the traces,
we have investigated the influence of (i) the number of bath spins
$N$, (ii) the order of the largest polynomial $N_C$, (iii) the number
of random states $N_s$.

\begin{figure} [tb] \centering

 \includegraphics[width=75mm]{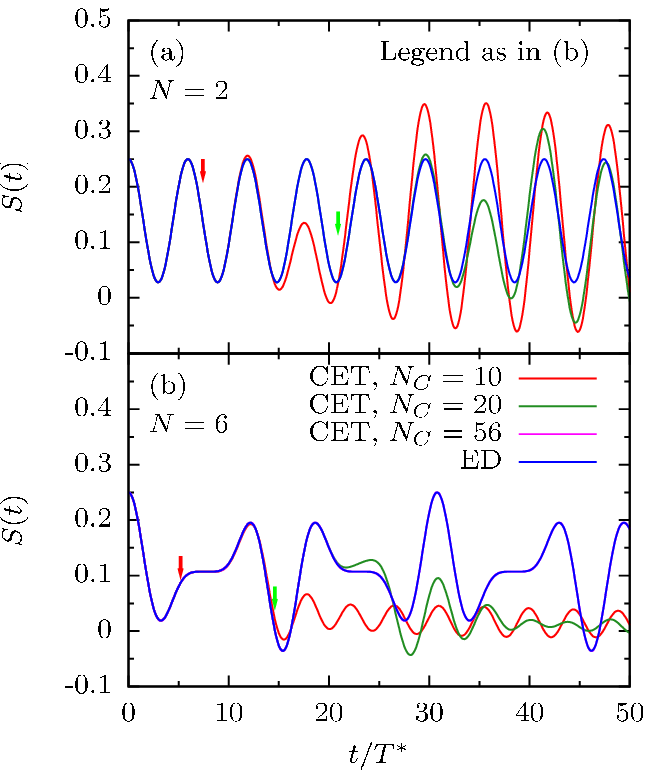}
 
\includegraphics[width=75mm]{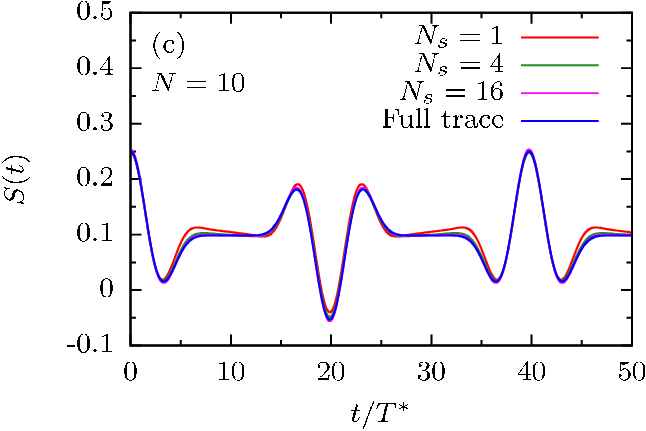}

  \caption{(color online) Benchmark for the CET calculating the time
evolution of the spin correlation function $S(t)$ for uniform coupling
constants $A_k = \frac{A_s}{N}$.  Two different orders $N_C$ of the
expansion are compared to ED for $N = 2$ (a) and $N = 6$ (b) bath
spins. After the time scale indicated by the arrows the CET-error exceeds
$O(10^{-3})$. In (a)-(b) the occurring traces have been calculated
exactly and no error from the statistical evaluation of traces
enters. (c) treats this error for $N=10$, by comparing results for a
varying number $N_s$ of random states to a result taking the full
trace into account with $N_C = 66$. }
  \label{fig benchmark}
\end{figure}

To benchmark the CET in small test systems accessible to exact
diagonalisation (ED), we restrict ourselves to uniform coupling
constants $A_k = A_s/N$ at first, defining $T_1^* = T^* =
\sqrt{N}/A_s$.  We compare results obtained by ED with CET
calculations for $N=2,6$ and three different CET orders $N_C$ in Figs.\
\ref{fig benchmark} (a)-(b). For such small systems, we evaluate the
traces for the momenta $\mu_{n,m}$ in (\ref{eqn:momenta-nm}) exactly,
since the number $N_s$ of randomly generated states needed for an
accurate statistical evaluation of the traces exceeds the dimension of
the Hilbert space. Therefore, the only error of the CET data at longer
times in Figs.\ \ref{fig benchmark} (a)-(b) arises from the finite
$N_C$ while for the short-time dynamics up to $t<t_{\rm max}(N_C)$ an
essentially exact result is obtained.

The convergence of the CET for any given time $t$ is ensured by the
analytic properties of the Bessel functions of large order given by $J_n
(\Delta E t) \sim \left(e\Delta E t/2 n \right)^n$. The vertical
arrows in Figs.\ \ref{fig benchmark} (a)-(b) indicate where this estimate
exceeds the value $10^{-3}$ for $N_C = 10$ and $20$. Apparently the
CET reproduces the exact ED results accurately up to this time. Thus,
the order of the CET for all further calculations is determined by the
smallest $N_C$ fulfilling the condition $\left( e\Delta E t_{\rm max}
/2 N_C \right)^{N_C} \leq 10^{-3}$, where $t_{\rm max}$ is the largest
time of interest. For $N=6$ and $t_{\text{max}}(N_C)/T^*=50$ this estimate yields
$N_C = 56$.  Consequently, the CET renders the ED result exactly up to
$t/T^*=50$.  To illustrate the deviations at larger time for an
insufficiently large $N_C$, $S(t)$ is plotted for the additional two
values $N_C=10,20<56$ in Fig.\ \ref{fig benchmark} (b).

Fig.\ \ref{fig benchmark}(c) illustrates the effect of the error
arising from the statistical evaluation of traces for $N=10$ bath
spins. The CET-result where the traces have been exactly calculated
(blue line) is serving as reference. Since the error of the
statistical evaluation is of the order $O ( (N_s D)^{-1/2})$, each
ascending value for $N_s$ shown in Fig.\ \ref{fig benchmark} (c)
reduces the remaining error by a factor $2$ independent of the time $t$.
This decrease of the statistical error is clearly visible.  For
$N=10$, where the Hilbert space has the dimension of $D=2^{11}=2048$,
our results with $N_s=16$ random states already converged enough to be
optically almost undistinguishable from the exact calculations.
By choosing either a large number of random states $N_s$ or a large
number of nuclei $N$, we are able to obtain an accurate representation
of the exact evaluation of the traces.

The physics of the CSM with uniform coupling constants $A_k = A/N$ is
well understood.\cite{KhaetskiiLoss2003} For a system with only two
bath spins we observe a coherent oscillation as depicted in Fig.\
\ref{fig benchmark}(a) since the central spin effectively only
interacts with the triplet state formed by the two bath spins while
the singlet is decoupled. The oscillation frequency is given by the
full width $2\Delta E = E_{\text{max}} - E_{\text{min}}$ of the
Hamiltonian's spectrum. For larger systems the dynamics is still
coherent and of the form as exemplarily shown for $N=6$ and $N=10$ in
Figs.\ \ref{fig benchmark} (b)-(c). The short time dynamics for $N >
3$ is governed by $T^*$ and the recurrence time $T_{\rm rec}$, where
$S(T_{\rm rec}+t)=S(t)$, increases linearly with the bath size since
the differences of the eigenenergies are commensurable. We find
$T_{\rm rec}\approx 30T^*$ for $N=6$ and $T_{\rm rec}\approx 40T^*$
for $N=10$.

\subsection{Results in the absence of an external magnetic field}

\subsubsection{Influence of the distribution function on the real-time
dynamics}
\label{sec:influence-distribution-function}

\begin{figure} [tb] \centering

\includegraphics[width=80mm]{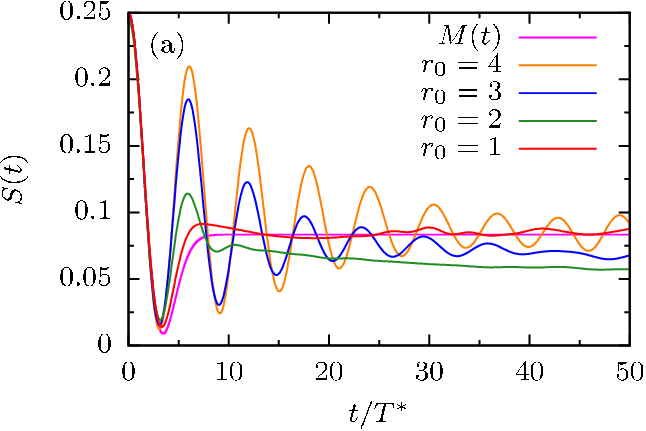}

\includegraphics[width=80mm]{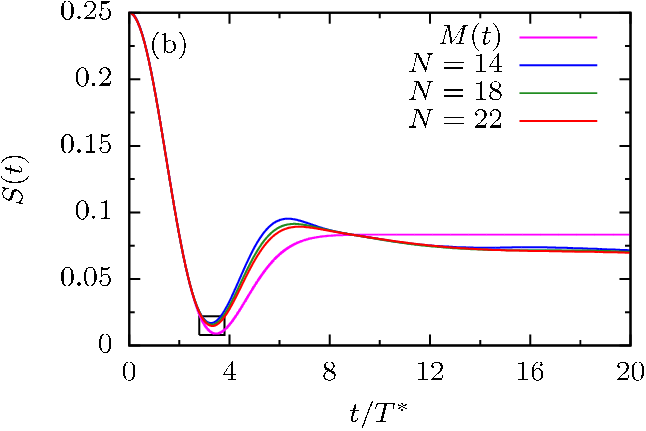}

 \caption{(color online) The spin correlation function $S(t)$ for
randomly generated coupling constants $A_k$ calculated via CET. Each
shown curve has been averaged over $n=50$ different random
realizations of couplings. (a) Short-time evolution of $S(t)$ for
increasing $r_0$ with $N = 18$.  (b) The analytical result $M(t)$ in
comparison to calculated data for increasing bath size $N$ based on a
fixed ratio $r_0 =R/L_0 = 1.5$. The inset shows the area marked
by the box.  }

  \label{fig b0szt}
\end{figure}

Now we discuss the influence of randomly generated coupling constants
$A_k$ onto the time evolution of the spin-correlation function $S(t)$.
In numerically accurate simulations of the real-time dynamics using a
statistical evaluation of the Bethe-ansatz
equations,\cite{FaribautSchuricht2013a,FaribautSchuricht2013b} the
maximum number of bath spins $N\approx 45 $ still remains several
orders of magnitude smaller than the $10^{5}$ nuclear spins present in
experimental samples.  The dynamics of small systems is influenced by
the range of coupling constants defining the ratio
$a=A_{\text{max}}/A_{\text{min}}$ between the largest and the smallest
coupling constant. Increasing $a$ at constant $N$ increases the number
of nuclear spins which are only very weakly coupled to the electronic
spin.  In addition the deviation of the average square, $\expect{A}^2$
and the fluctuation $\expect{A^2}$ increase. The short and intermediate
dynamics is dominated by a decreasing  number of bath spins for a fixed $N$, 
while thevery weakly coupled spins are only contributing significantly at
extremely long times.

In a statistical evaluation of the exact Bethe ansatz
equations\cite{FaribautSchuricht2013a,FaribautSchuricht2013b} the
fixed set of $A_k= A/N \exp[-(k-1)/(N-1)]$ has been chosen, leading to
the ratio $a={\rm e}$.  A recent TD-DMRG\cite{StanekRaasUhrig2013}
study has pushed the limit to up to $N=100-1000$ nuclear bath spins.
In this study, the configurations of $\{A_k\}$ have been drawn from
$P(A)=const$ on the interval $A_0[1/2,1]$, corresponding to a ratio
$a=2$.  For the first distribution, the fluctuation
$u=\expect{A}^2/\expect{A^2}\approx (1+1/e)/(1-1/e)/2\approx 1.082$,
while the distribution $P(A) =const.$ yields $u=28/27$. In both cases
$u\approx 1$ holds which does not differ significantly from
$A_k=const$. Therefore, the non-decaying fraction of the spin
polarization remains close to the QSA result.

In the distribution function $P(A,r_0)$, defined in Eq.\
(\ref{eqn:p-a}), the cutoff ratio $r_0$ directly translates into the
ratio $a = \exp(r_0^2)$. For large $r_0$ the probability $P(A,r_0)$ is
high for adding more and more nuclei to the system whose interaction
with the central spin is negligible, e.~g.\ for $r_0 = 4$ the ratio
between the largest and smallest coupling constant $A_k$ has already
reached $A_{\text{max}}/A_{\text{min}} \approx 10^{7}$. In order to
obtain results faithfully representing a larger system, a set of
several $A_k$ must be taken into account for each order of magnitude
which is impossible for a system size of only $N= 20$ bath spins.

Fig.\ \ref{fig b0szt} (a) illustrates the influence of the cutoff
$r_0$ onto the real-time dynamics.  The results are calculated for $N
= 18$ bath spins and averaged over $n = 50$ different random
configurations $\{ A_k\}$ to reduce the influence of fluctuations and
effectively take more nuclei into account.  We added the QSA result
$M(t)$ stated in Eq.\ (\ref{glg merkulov}) as a guide for the
short-time dynamics obtained from a random nuclear field approximation
in the thermodynamic limit $N\to\infty$.  All CET curves  perfectly coincide 
with $M(t)$ for very short time scales $t/T^*<3$.

For $r_0=1$, the ratio $a= e$, and 
only significant coupling constants of the same order of magnitude are
taken into account. The polarization saturates approximately at
the value $S(0)/3$ as predicted by the QSA.  However, slight deviations
between the CET curve and $M(t)$ are observed for times  $t/T^*>3$.
Nevertheless the CET short-time dynamics and results obtained by other
approaches\cite{FaribautSchuricht2013a,FaribautSchuricht2013b,StanekRaasUhrig2013}
agree remarkably well with each other and with the QSA. 
This indicates that the generic dynamics can already be obtained by a rather
small numbers of bath spins.

For increasing cutoffs $r_0>1$, the short-time dynamics of $S (t)$
evolves from passing through a single minimum as described by the QSA
curve to a damped oscillation as depicted in Fig.\ \ref{fig b0szt}(a).
This behavior is easily understood by the distribution of coupling
constants in any of the random configurations $\{ A_k\}$.  For a large
cutoff $r_0$ and a fixed number of bath spins $N$, the number of nuclei
which couple to the central spin with a coupling constant
$A_k/A_s=O(1)$ becomes very small in $\{ A_k\}$ due to the increasing
probability to find a small coupling. Essentially we see a similar
coherent motion as in Fig.\ \ref{fig benchmark}(a) involving only one
or two bath spins with the strongest coupling constants, while the
slow dephasing is induced by the remaining very weakly coupled nuclear
spins. Therefore, we conclude that such choices of the cutoff $r_0$ do
not render the dynamics for $N\to\infty$ when working with a fixed and
small $N$.

Fig.\ \ref{fig b0szt}(b) focuses on the bath-size dependency of the
short-time evolution of $S(t)$ for $N=14,18,22$ and fixed $r_0 =
1.5$.  This cutoff would correspond to $N(R) \approx 10^5$ 
in a real system, implying a ratio $a \approx 8$. 
The figure compares exact simulations for three
different bath sizes to the QSA result $M(t)$ and demonstrate the
fast convergent with the bath size for $r_0=1.5$. 

The initial decay of
$S(t)$ is well described by $M(t)$ and for increasing $N$ the exact
finite size curves approach the QSA solution for short-time
scales. But after the initial decay the central spin's polarisation
drops below the value predicted by the QSA approach. This deviation
arises from the contribution of the small coupling constants, whose
interaction with the central spin is too weak to have major influence
on the short time behavior of $S(t)$, but on large time scales the
small couplings become dominant. Since the QSA result is based on the
assumption of a static bath, it is not surprising that it is only able
to describe the short-time and intermediate-time evolution of the
central spin that is dominated by the strongly coupling nuclei.

\subsubsection{The influence of $r_0$ onto the long-time limit}
\label{sec:influence-of-r0}

The deviation of the non-decaying fraction of the spin-polarization
from the QSA value of $S(0)/3$ observed in Fig.\ \ref{fig b0szt}(b)
justifies a more detailed analysis.

\begin{figure} [tb] \centering

\hspace*{3mm}\includegraphics[width=75mm]{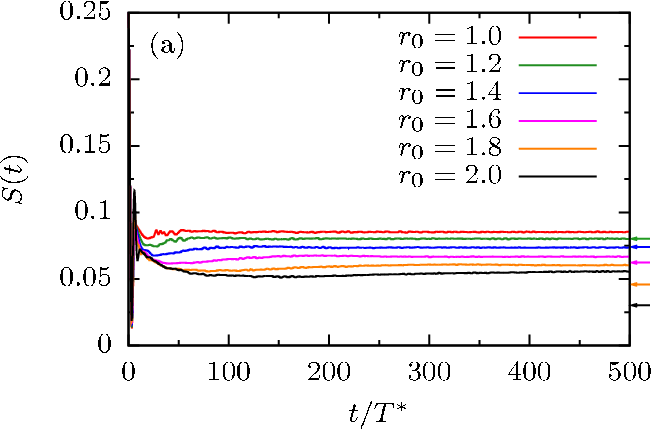}

\includegraphics[width=75mm]{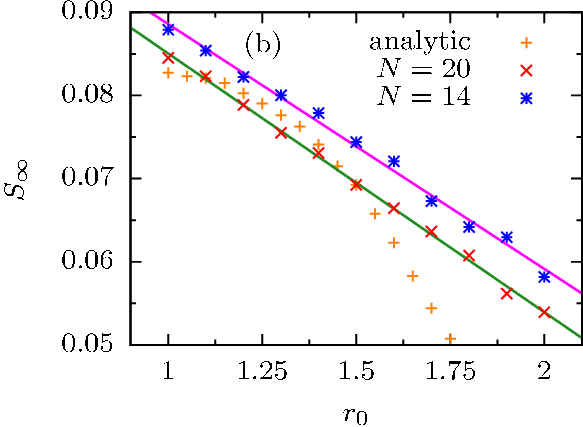}

  \caption{(color online) The behavior of $S(t)$ on long-time scales
for different values of $r_0$ and $N=18$ (a).  Each shown curve has
been averaged over $n=50$ different random configurations $\{ A_k\}$
of couplings.  The colored arrows indicate the long-time values for
$S_\infty$ predicted\cite{Merkulov2002} from the fluctuation ratio
$u=\expect{A^2}/\expect{A}^2$.  (b) The non-decaying fraction
$S_\infty$, obtained by averaging the data over the calculated data
points in the interval $\frac t{T^*} \in [450:500]$, vs. the cutoff
$r_0$ for two different system sizes $N=14,20$. The straight lines are
linear fits. The analytical predictions using the fluctuation ratio $u$
are added for comparison.  }

  \label{fig-4}
\end{figure}

The influence of $r_0$ onto the long-time limit is depicted in Fig.\
\ref{fig-4}(a). With increasing $r_0$ we observe two effects: (i) the
non-decaying part of the polarization $S_\infty=\lim_{t\to\infty}
S(t)$ is decreasing, (ii) the relaxation time from the
pre-equilibrated intermediate state reached after a short transient
time of the order of $O(10T^*)$ into the steady-state is increasing.
Since the weakly-coupled nuclei can only contribute on large-time
scales, the second observation is intuitively clear due to the
increasing number of weakly coupling nuclei with increasing $r_0$ and
fixed $N$.

The first observation can be also understood within a simple argument.
In the QSA approach, no spin polarization transfer between the central
spin and the spin bath can occur since the nuclear magnetic field has
been treated statically.  The spin decay is purely driven through
dephasing by averaging over the random and isotropic effective
magnetic field distribution yielding a finite steady-state value
$S_\infty=S(0)/3$.

Recent Bethe-ansatz
calculations\cite{FaribautSchuricht2013a,FaribautSchuricht2013b} up to
$N=44$ nuclei confirm that the non-decaying fraction of the
spin-polarization depends on the distribution of the coupling
constants.  In any finite size representation of the model with a
small number of coupling constants $\{ A_k\}$ a finite non-decaying
fraction of the spin-polarization is found.  This fraction, however,
decreases when additionally weak coupling nuclear spins have been
added.  Faribault et al.\ gave an analytical
argument\cite{FaribautSchuricht2013b} why there must be a finite
non-decaying fraction of the spin-polarization in any finite-size
system, where the distribution of coupling constants $\{ A_k\}$ is
limited to the same order of magnitude. This agrees perfectly with
our findings for a finite size system.

In contrast to exact evaluations of small systems, approximate
treatments\cite{Merkulov2002,
KhaetskiiGlazman2002,Zhang2004,CoishLoss2004} of the model allows to
access the thermodynamic limit.  Such treatments require the
neglecting of higher order correlation effects and predict a finite
non-decaying fraction. Taking into account non-Markovian contributions
in second order of the transverse coupling,\cite{CoishLoss2004} leads
to a non-exponential correction to the mean-field solution stated in
Eq.\ (\ref{glg merkulov}) of the form $1/\log t$ in the absence of a
magnetic field in the decay to a finite steady-state limit.  Even
though we observe a non-trivial transient behavior with a very slow
decay between $100<t/T^*<1000$ 
the data is not sufficient  to  extrapolate a $[\log t]^{-\alpha}$ correction or a
power-law decay to the non-decaying fraction from the data presented
in Fig.\ \ref{fig-4}(a).
However, we have been able to extract a power-law behavior from the 
low-frequency properties of spin-noise spectra which will be discussed 
in Sec.\ \ref{sec:spin-noise-spectra}.

For bridging to the experiments we can ask the question what is the
asymptotic non-decaying fraction of spin-polarization in the
thermodynamic limit $r_0\to \infty$ and $N\to\infty$.
In order to shed some light on  this question within the framework
of the CET approach, we have investigated the scaling properties of the
non-decaying fraction of the polarization with respect to $r_0$ for
two bath sizes $N=14,20$ in the interval $1\le r_0\le 2$, in which
$S(t)$ has reached a steady-state limit as shown in Fig.\
\ref{fig-4}(a).  The short time dynamics is always governed by the
time-scale $T^*$ and agrees very well with the QSA result.

The long-time limit is plotted as function of the cutoff $r_0$ for two
different bath sizes $N$ in Fig.\ \ref{fig-4}(b). The steady-state
polarisation decreases linearly with $r_0$ for $r_0 \leq 2$.  Since
the CET dynamics does not properly represent the large $N$ limit for
$r_0$ exceeding $r_0>2$, as illustrated in panel (a) and discussed
above, no data is shown for such cutoffs.

The linear scaling of $S_\infty$ as function of $r_0$ suggests
that the non-decaying part of the central-spin polarization should
vanish when the influence of very large numbers of small coupling
constants is taken into account.
Extrapolating our linear fit to the data for $N=20$ indicates that
this is the case for $r_0 \approx 3.7$, corresponding to a ratio
$a\approx 10^6$.  However, with increasing of the number of bath spins
$N$, the predicted cutoff $r_0$ for that $S_\infty$ should vanish
decreases as exemplified by the fit to two different spin bath
dimensions in Fig.\ \ref{fig-4}(b).

Our scaling analysis indicates that the spin correlations will
completely decay at infinitely long times in the thermodynamic limit,
e.g.\ $N\to \infty$ and then $r_0\to \infty$.  This finding is fully consistent with an
extension of the QSA which takes into account the long-time
fluctuations of the nuclear magnetic field.  Averaging Eq.\ (\ref{eq
eom}) over times larger than the electron spin-precession time but
much smaller than the nuclear spin precession time, the spin
precession contribution vanishes and 
only the term  $ (\vec{S}_0 \vec{n}) \vec{n} $ survices.
After inclusion of the explicit
time-dependence of the nuclear field $\vec{B}_{\rm eff} = B_{\rm eff}
\vec{n}(t)$ and spin, the ensemble average is given by Eq.\ (12) in
Ref.\ [\onlinecite{Merkulov2002}]
\begin{eqnarray} 
\expect{\vec{S}(t)} &=& \expect{\vec{n}(t)
[\vec{n}(t)\vec{S}(t)] }
\label{eqn:merkulov-eqn-12}
\end{eqnarray}

Since $[\vec{B}_{\rm eff}(t)\vec{S}(t)]$ accounts for the total energy
of the central-spin model, it is a conserved quantity and time
independent: $[\vec{B}_{\rm eff}(t)\vec{S}(t)] = [\vec{B}_{\rm
eff}(0)\vec{S}(0)]$. Furthermore the nuclear-spin-spin correlation
function is isostropic leading to
\begin{eqnarray} 
\expect{\vec{S}(t)} &=& \gamma(t) \frac{\vec{S}(0)}{3}
\end{eqnarray} 
where $\gamma(t)= \expect{\vec{n}(t)\vec{n}(0)}$ is
defined as correlation function of the nuclear spin orientation.

For the long-time limit, $\gamma(t)$ approaches a stationary
value\cite{Merkulov2002} and is only dependent on the ratio
$u=\expect{A^2}/\expect{A}^2$. We added our estimates for $S_\infty$
using $\gamma(u)$ derived in Ref.\ \onlinecite{Merkulov2002} as
horizontal arrows in Fig.\ \ref{fig-4}(a) as well as crosses labeled
``analytic`` into Fig.\ \ref{fig-4}(b).  
Although our finite size scaling qualitatively agrees  with a
decreasing  $\gamma(u)$ for increasing $r_0$,
the functional form of  $\gamma(u)$ differs
from our linear scaling.   This might be related to
the change of the largest hyperfine coupling when varying $r_0$
for fixed $N$. Keeping both $T^*$ and the  largest hyperfine coupling 
fixed requires the  increase of $N$  when increasing $r_0$. This
would accelerate the decrease of $S_\infty$ when increasing $r_0$
in closer agreement with $\gamma(u)$.

Another argument of why the non-decaying fraction $S_\infty$ must
vanish in the long-time limit for $N\to \infty$ in QDs with smooth
electronic confinement potential was given by Chen et
al.\cite{ChenBalents2007} based on the distribution function of the
$A_k$. Although the total angular momentum in the isotropic CSM is
conserved it will be equally  redistributed onto all nuclear spins at
large times.  Since angular momentum transfer from the central spin to
the nuclear spin $k$ occurs on a time scale $t>1/A_k$, only those
spins within a given radius $R(t) = L_0 [\ln(A_0 t)]^{1/2}$ can
contribute to the spin decay using the Gaussian envelope function
(\ref{eqn:psi-R}) and $A_0=A(R_k=0)$.  Only the strongly coupling
spins in the sphere with radius $R_s=R(T^*)$ contribute to the
short-time dynamics up to the time $t$.  Therefore, the central spin
should decay as
\begin{eqnarray} 
\expect{S^z(t)} &\propto & \frac{N(R_s)}{N(R(t))}
\propto [\ln(A_0 t)]^{-3/2}
\end{eqnarray} in the long time limit $t\gg T^*$ in three dimensions
and its expectation value vanishes for $t\to \infty$.   Hence,
the non-decaying fraction of the central spin-polarization must vanish in
the thermodynamic limit.  In any finite size
calculation,\cite{BortzStolze2007,Bortz2010,StanekRaasUhrig2013,FaribautSchuricht2013a,FaribautSchuricht2013b}
however, there exist a smallest coupling constant $A_{\rm min}$ which
limits the time scale beyond which the exact finite-size calculation
will deviate from the thermodynamic limit.

\subsubsection{Summary}

Before we move on to the discussion of the spin noise spectra, we
briefly summarize the results of this section.  We have demonstrated
the accuracy of the CET by a comparison of the real-time dynamics with
small systems exactly solvable using ED.  While in principle arbitraryly
long times could be reached with the CET approach, it is limited by
the largest polynomial order $N_C$ which has been included in the
calculation. We have established the quality of the statistical
evaluation of the momentum $\mu_{n,m}$ entering the extension of the CET
approach to ensemble averages.  The short-time dynamics is governed by
the time scale $T^*$ and agrees qualitatively well with the QSA
result. However, we observe small deviations which can be traced to
(i) the distribution of the coupling constants $A_k$, to (ii) the
number of bath spins and to (iii) the ratio between the largest and
the smallest $A_k$.  The larger this ratio is for fixed number of bath
spins $N$, the smaller the number of bath spins which couple with an
$A_k=O(1/T^*)$, the less bath spins contribute effectively to the
short time dynamics.  The finite value of $S_\infty$ in a finite-size
system depends on the distribution function $P(A)$.

\subsection{Spin noise spectra}
\label{sec:spin-noise-spectra}

In recent
experiments,\cite{CrookerBayerSpinNoise2010,Dahbashi2012,LiBayer2012,*ZapasskiiGreilichBayer2013}
the spin noise spectra have been measured in QD ensembles.  
Assuming independent QDs, it is sufficient to average the
generic spin-noise spectrum $S(\w)$ over the distribution of time
scales $T^*$ and $g$-factors to make a connection to the experimental
data. Therefore, we focus on calculating the spin-noise spectrum $S(\w)$
for a single QD 
first and postpone the discussion to spin-noise spectra 
for QD ensembles to Sec.\ \ref{sec:QD-ensemble-average}.

\begin{figure} [tb] \centering
 
 \includegraphics[width=75mm]{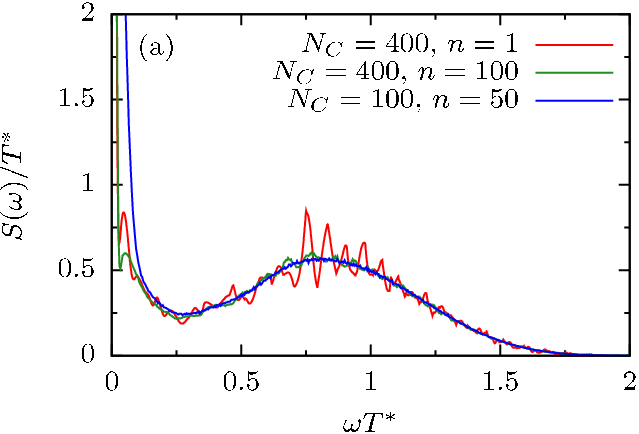}

\includegraphics[width=75mm]{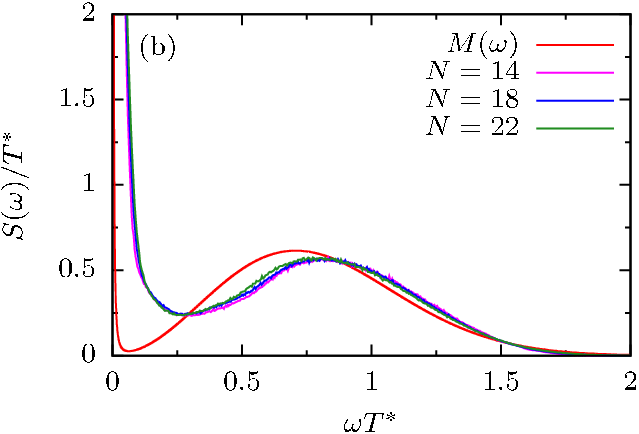}

\caption{(color online) The spin-noise spectrum $S(\omega)$ in the
absence of an external field for $r_0 = 1.5$.  Panel (a) illustrates
the convergence properties of the CET for varying $N_C$ and the effect
of averaging results for several ($n$) configurations of the coupling
constants $A_k$ based on $N=18$ bath spins and $r_0=1.5$.
Panel (b) focuses on the finite bath size effects where we have
averaged over $n = 50$ different configurations $\{A_k\}$. $M
(\omega)$ denotes the Fourier transformation of the semi-classical
result where the $\delta$-peak in $M(\omega)$ at $\omega = 0$ has
been approximated by a Lorentzian.  }
  \label{fig b0spec}
\end{figure}

A realistic modelling of the QD requires the treatment of the order
$O(10^5)$ nuclei. Even on a fine frequency scale, the noise spectrum
will be continuous while $S(\omega)$ obtained from an exact simulation
for $\sim 20$ bath spins with a single distribution $\{ A_k\} $
clearly reflects the spectrum's discrete character.
In order to recover a continuous spectrum from the finite size CET
calculation, we use two different ingredients: (i) averaging over
random distributions $\{ A_k\} $ and (ii) choosing a rather low order
$N_C$ in the CET calculations. The averaging over several random
distributions $\{ A_k\} $ mimics a larger number of nuclei than
contained in a single configuration.  By artificially reducing $N_C$,
we can effectively add a broadening to the individual $\delta$-peaks of
the finite size spectrum which would only be precisely recovered in the
limit $N_C\to \infty$. 
Increasing $N_C$ systematically increases the frequency resolution
which we have employed to reveal a power-law in the low-frequency
spin-noise spectrum.
Note that no spectral weight is lost by this
procedure since the sum-rule (\ref{eq sum-rule}) is exactly fulfilled
for arbitrary $N_C$.

In order to illustrate the effect of those two ingredients, we show a
direct comparison of $S(\w)$ for a single configuration ($n=1$) and
data averaged over $n=100$ random configurations $\{ A_k\}$ with $N_C
= 400$, $r_0 = 1.5$ and $N=18$ in Fig.\ \ref{fig b0spec}(a).
Additionally, the corresponding spectrum for a lower order $N_C = 100$
of the CET and a reduced number of configurations $n=50$ has been
added.  While $S(\w)$ obtained from a single configuration displays a
clear signature of a superposition of discrete peaks a
quasi-continuous spectrum is generated by the configuration averaging.

The calculations with a lower Chebychev order $N_C = 100$ reproduce
the calculations for $N_C = 400$ excellently, except at small
frequencies being consistent with linear scaling of the largest
accessible time scale with $N_C$. Even though it would be sufficient
to use rather small $N_C$ for an accurate description of the
short-time dynamics, the numerical effort increases substantially to
access the low frequency behavior of the spin-noise spectrum.

Based on the largest accessible time discussed in Sec.\
\ref{sec:benchmark}, the smallest accessible frequency of the CET is
given by $\omega_{\text{min}} \geq  10^{3/N_C}\pi \text{e} \Delta E/N_C$.  Note
that there are two limiting factors to the accessibility of small
frequencies in our simulations: the finite $N_C$ and the finite cutoff
$r_0$ which set the boundary to the lowest $A_k$ and therefore, the
lowest non-zero excitation energy of the system.  Choosing a larger $N_C$
than required by the means of this lowest excitation energy does not add
additional information to the finite frequency spectrum but only
sharpens the $\delta(\w)$-peak.

The spectral functions $S(\w)$ depicted in Fig.\ \ref{fig b0spec}(b)
for increasing number of bath spins and $r_0 = 1.5$ correspond to the
time resolved data shown in Fig.\ \ref{fig b0szt}(b). We note the fast
convergence of $S(\w)$ as function of $N$.  We also added the QSA spin
noise $M(\w)$ stated in Eq.\ (\ref{eq:m-w}). The $\delta$-peak in
$M(\omega)$ at $\omega = 0$ has been approximated by a Lorentzian, 
and its spectral weight  is given by the non-decaying fraction of
the spin polarization. 

The
high-frequency part of $S(\w)$ agrees remarkably well with the QSA
spin-noise spectrum $M(\w)$ rendering the excellent agreement in the
short-time dynamics between both approaches.  As expected, the broad
high-frequency peak is centered at $1/T^*$ and its width given by
$1/2T^*$.  However, we notice significant deviations between $M(\w)$
and $S(\w)$ for smaller frequencies. Those differences also reflect
the different transient behavior at times $t\gg T^*$ depending on the
configurations of $\{ A_k \}$.

\begin{figure}[t]
\begin{center}

\includegraphics[width=85mm]{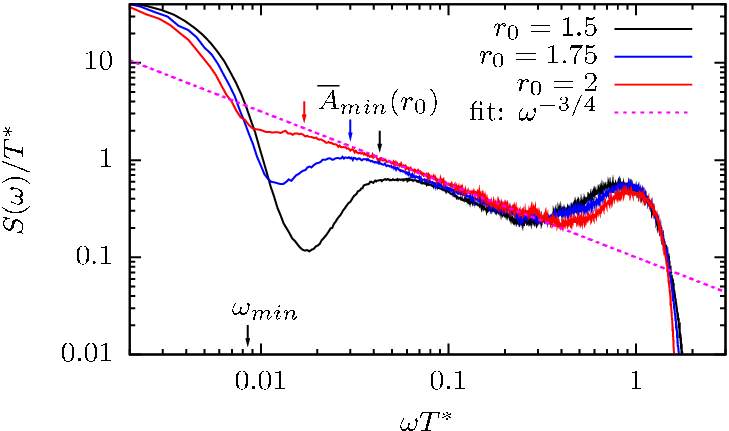}

\caption{(color online)
$S(\w)$ vs $\w T^*$ on a log-log scale for different values of $r_0=1.5,1.75,2$. 
The dashed line is a fit to the low-frequency behavior
above the resolution-broadened $\delta(\w)$-peak. The smallest accessible frequency $\w_{\rm min}$
and the expectation value of the smallest contributing coupling constant $\overline{A}_{\text{min}}$
are indicated by arrows.
Parameters: $N_C=1000,n=100,N=18,b=0$.
}
\label{fig:s-w-isotrop-log}
\end{center}
\end{figure}

At low frequencies a small
shoulder around $\w T^*\approx 0.1$ is observed in $S(\w)$ (green
curve) in Fig.\ \ref{fig b0spec}(a) when using $N_C=400$ and $n=100$
configuration averages. This indicates the existence of an additional
low-frequency feature in $S(\w)$, that is not covered by the QSA
result, located above the resolution-broadened zero-frequency $\delta$-peak
and below the Gaussian type high-energy peak around $\w T^*\approx 1$.

In order to reveal the shape and nature of the low frequency part of
the spin-noise spectrum in greater details, we have pushed the CET
order to $N_C= 1000$ to significantly increase the frequency
resolution to $\w_{\rm min}T^*= 8.5\cdot 10^{-3}$.  As depicted in
Fig.\ \ref{fig:s-w-isotrop-log}, now the resolution-broadened
$\delta(\w)$-peak, whose full-width half maximum can be estimated by
$2\w_{\rm min}$, is well separated from the remaining low-frequency
part of $S(\w)$: the shoulder has evolved into a threshold-type
behavior with a crossover around $\w T^*\approx 0.03$ defined by the
smallest excitation energy of the finite size system.  A power law
$\propto \w^{-3/4}$ can be fitted in this region over approximately
one decade and is indicated as dashed line.
At larger frequencies, the previously discussed Gaussian-like peak
remains visible and is centered around $\w T^*\approx 1$.

The $r_0$-dependency of the crossover scale separating the low-energy
feature in $S(\w)$ from the resolution-broadened $\delta(\w)$-peak is
clearly visible in Fig.\ \ref{fig:s-w-isotrop-log}.  Increasing $r_0$
extends the frequency range which can be fitted by $\propto \w^{-3/4}$
to lower frequencies. The cutoff parameter $r_0$ determines the
smallest hyperfine coupling $A_k$ contained in the configurations $\{
A_k\}$ and, therefore, the smallest excitation energy in the system.
We have indicated the value $\bar A_{\rm min}(r_0)= \expect{{\rm min}[
A_k ]} $ averaged over all configurations $\{ A_k\}$ by an additional
vertical arrow in Fig.\ \ref{fig:s-w-isotrop-log}.  Apparently,
$\bar A_{\rm min}(r_0)$ determines the low-frequency crossover scale
to the power-law behavior.

We have demonstrated in Fig.\ \ref{fig-4} that the non-decaying
fraction of the central spin polarization and, hence the spectral
weight of the $\delta(\w)$ peak decreases with increasing $N$ and
$r_0$ and eventually vanishes in the thermodynamic
limit.\cite{ChenBalents2007} Hence, spectral weight of the
$\delta(\w)$-peak is transferred to the low-frequency part of $S(\w)$
dominating the long-time properties of the spin correlation function
$S(t)$.  We conjecture that the observed finite low-energy crossover
scale is approaching zero-frequency in the limit $r_0\to\infty$ and
$N\to\infty$.

Restricted by the finite size of the spin bath and the frequency
resolution, our numerical data is not accurate enough to predict the
precise analytic form of $S(\w)$ in the thermodynamic limit.  The
guide-to-the-eye fit to our data, however, would suggest $S(\w)\propto
\w^{-3/4}$ implying a $1/t^{1/4}$ decay at long-time scales. This
finding is consistent with the prediction $S(t) \propto
\ln^{-3/2}(t)$, because in the intermediate time regime we can access
via the CET both findings are almost indistinguishable in the time
domain. To predict a deviation a fit over more than one decade is
necessary.
%

\subsection{Magnetic field dependence of the spin noise}

\subsubsection{Transversal magnetic field}

\begin{figure} [tbp]
\centering
  
 \includegraphics[width=85mm]{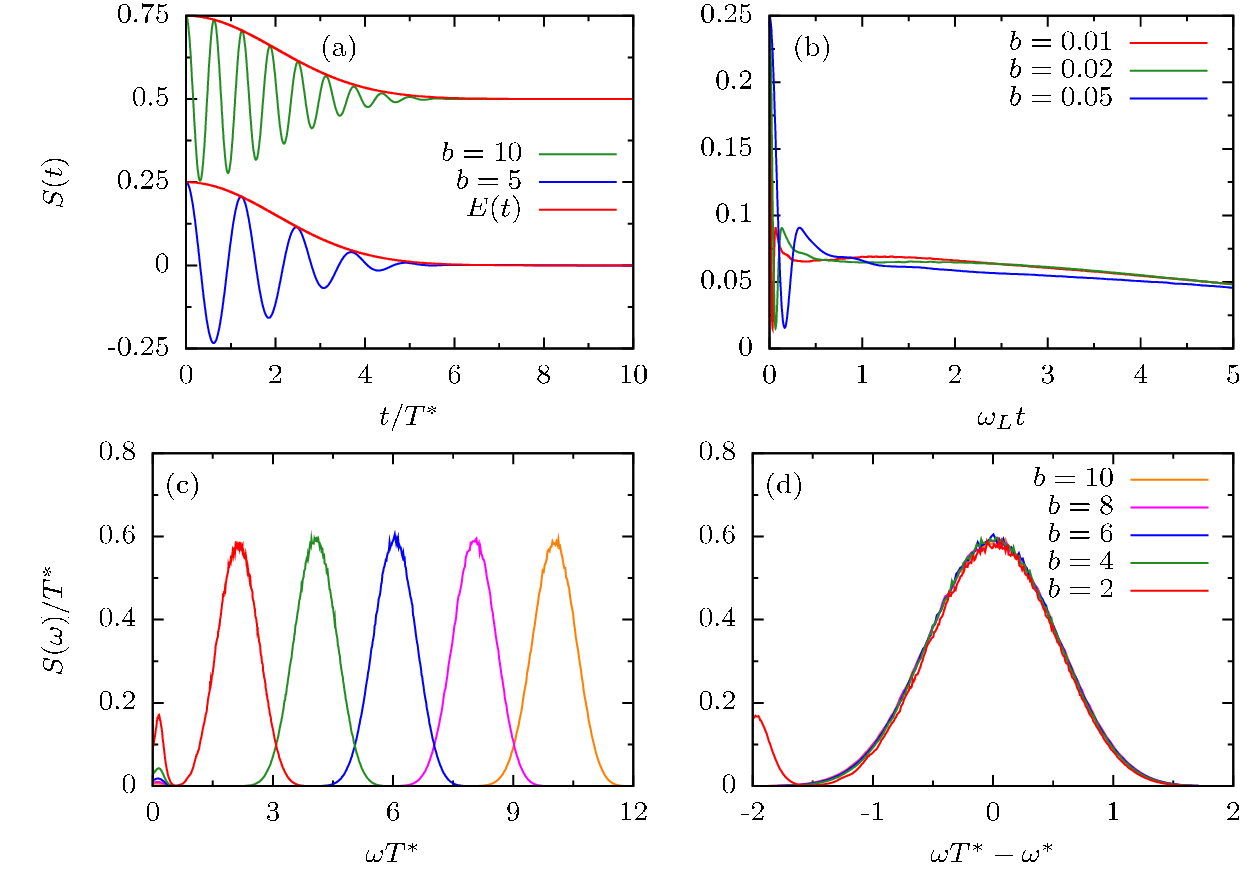}

 \caption{(color online) All shown simulations have been averaged
over $n = 50$ configurations. The spectral functions are all based on
$N_C = 100$.  (a) The correlation function $S(t)$ for two different
values of $b$. To distinguish the results for different $b$ an offset
of $0.5$ has been added to the upper curve. The function $E(t) =
\frac14 \text{exp}\left(-\frac12\left(\frac{t}{2T^*}\right)^2\right)$
approximates the envelope of the signals. (b) $S(t)$ for small
external field strengths. (c) The renormalized spin-noise function
$S(\omega)$ for $b = 2,4,...,10$. (d) The same data as shown in (c)
shifted by $\omega^* = \sqrt{b^2 + \frac12}$, pointing out that the
width of the occuring maxima is given by $T^*$ and that $\omega^*$ is
the exact frequency of the occuring spin precession.  Parameters:
$N=20,r_0=1.5$.}
  \label{fig bxstrong}
\end{figure}

Adding a transversal magnetic field $B_x$ to the system has two
important effects on the time evolution of the correlation function
$S(t)$.  First, it breaks the conservation of the total polarization
along the $z$-axis. Second, the short-time dynamics is now governed by
a shifted time scale $1/T^{'*}(b) = \sqrt{(T^{*})^{-2} +
2\omega^2_L}$ which evolves continuously from $(T^{*})^{-1}$ with
the external magnetic field. This can be either analytically derived
from the von-Neumann equation or can be extracted from the numerical
data for large $b$ as depicted Fig.\ \ref{fig bxstrong}(d).

Applying a large external magnetic field $b=\omega_L T^* \gg 1$ causes
a damped oscillation in $S(t)$ whose dimensionless frequency is given
by $\omega^* = \sqrt{b^2 + 1/2}$.  Since the hyperfine interaction
remains the origin of dephasing, the characteristic time scale of the
decay is governed by the intrinsic time scale $T^*$.  Fig.\ \ref{fig
bxstrong}(a) demonstrates this behavior for two different magnetic
field strengths fulfilling $\omega_L T^* > 1$. Additionally the
function
\begin{eqnarray}  
E(t) &=& \frac{1}{4}
\exp\left[-\frac{1}{2}\left(\frac{t}{2T^*}\right)^2 \right]
\end{eqnarray} 
has been added as an approximation to the envelope of
the real-time dynamics.

Fig.\ \ref{fig bxstrong}(c) shows the spin-noise spectrum $S(\omega)$
for five different magnetic field strengths $b=2,4,6,8,10$.  The
resulting spectrum contains a small contribution near $\omega = 0$
that vanishes quickly for growing magnetic field strength. 
The main contribution of the spectrum, however,  
consists of a central peak around
$\omega^*=\sqrt{b^2+1/2}$ with a field independent width proportional to
$1/T^*$ and is well separated from the low-frequency part
of the spectrum. The universality of $S(\w)$ is revealed by shifting the
dimensionless frequency $\Delta\w = \w T^* -\w^*$. All curves collapse
onto this universal curve independent of $b$ as depicted in Fig.\
\ref{fig bxstrong}(d). This also proves the claim that the envelope
function of the spin-decay is governed by $T^*$ independent of
$b$. Furthermore, we note that $S(\w)$ is independent of the number of
bath spins for a fixed $r_0$.

Fig.\ \ref{fig bxstrong}(b) focuses on the time evolution $S(t)$ for a
weak external field. The short-time evolution of the central spin
remains unaffected by the application of a very small magnetic field,
$b=\omega_L T^*\ll 1$. By symmetry breaking, $S_\infty=0$, so that
the correlation function $S(t)$ approaches zero in the long time
limit.  Thus the $\delta$-peak occurring at $\w=0$ in the spin-noise
function $S(\omega)$ must already vanish for an infinitesimal small
transversal field. Due to the finite-time resolution of the CET, we
only can show the transient behavior, while no change can be resolved
in $S(\w)$ for $\w\to 0$ (not shown) compared to Fig.\ \ref{fig
b0spec}.

The long-time transients are not governed by the nuclear-field
fluctuation time $T^*$ but by $\omega_L$. This is clearly visible in
Fig.\ \ref{fig bxstrong}(b) where we plotted $S(t)$ for three
different weak magnetic field values as function of the dimensionless
time $\w_L t$. We find universality on the intermediate time scale
depicting a very slow decay for $t\to\infty$. This findings agree with
the Bethe-ansatz data of Faribault et
al.\cite{FaribautSchuricht2013a,FaribautSchuricht2013b}

We can ask how does the spectrum evolve from $b=0$ to finite $b>0$.
Since the eigenvalue spectrum evolves adiabatically from $b=0$, we
must observe two effects. (i) the $\delta(\w)$-peak vanishes and its
spectral weight is shifted to finite frequencies, and (ii) all finite
frequency excitation energies will shift with the magnetic field.
Therefore, the center of the Gaussian shaped peak with a width of
$T^*$ gradually evolves to $\sqrt{(T^{*})^{-2} + 2\omega^2_L}$ as
function of the magnetic field as depicted in Fig.\ \ref{fig bxstrong}(d).


\subsubsection{Longitudinal magnetic field $B_z$}

\begin{figure}[tb]
\begin{center}

\includegraphics[width=85mm]{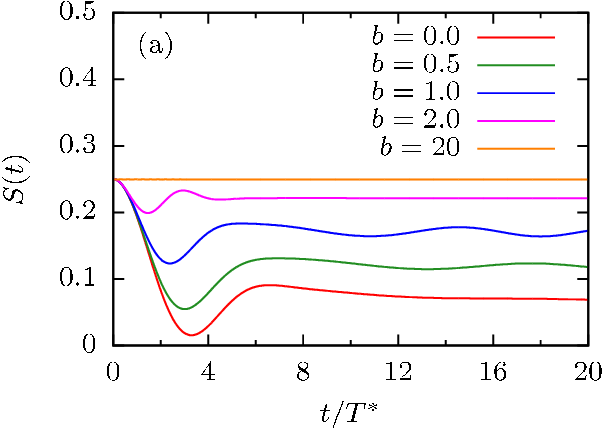}

\includegraphics[width=85mm]{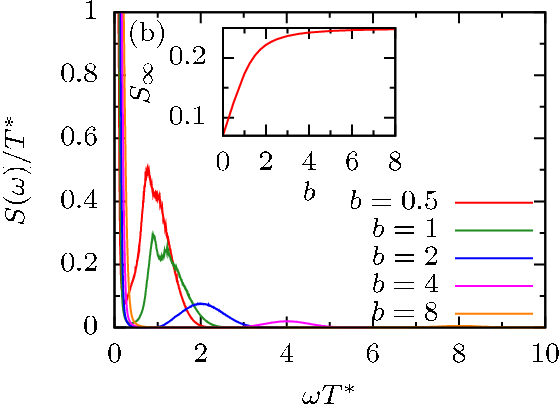}

\caption{(color online) The spin correlation function $S(t)$ for
increasing applied longitudinal fields (a), $N = 18$ and an average
over $n=10$ different configurations. (b) Corresponding calculated
spectral functions $S(\omega)$.  The inset depicts the spectral weight
of the $\delta(\w)$-peak as function of $b$ corresponding to the
non-decaying fraction of $S(t)$.  Parameter: $N_C = 100$, $n = 50$ and
$r_0=1.5$.  
}
\label{fig-7-Sw-for-finite-Bz}
\end{center}
\end{figure}

Applying a magnetic field $B_z$ in longitudinal direction induces a
finite energy difference between the different eigenstates of the
total spin component $J_z$, and the spin-flip processes of the central
spin  are suppressed by this additional energy
barrier.  The initial spin-decay is almost independent of $b$ for
ultra-short time scales until $t\approx 1/b$ where the further decay
of $S(t)$ starts to be oppressed.  Consequently, the non-decaying
fraction of the spin polarization is increasing with $b$: the
spin-decay is completely suppressed for $b\to\infty$ as can be seen in
Fig.\ \ref{fig-7-Sw-for-finite-Bz}(a).

The corresponding spin-noise spectrum is displayed in Fig.\
\ref{fig-7-Sw-for-finite-Bz}(b). The increasing non-decaying fraction
$S_\infty$ in $S(t)$ corresponds to an increasing spectral weight of
the $\delta$-peak located at $\w=0$. Since the total spectral weight
is conserved, there is a spectral weight transfer from the broader
peak centered around $\w^* T^*\approx 1$ to the $\delta(\w)$-peak.
The broad finite-frequency peak is shifted to higher frequencies and
is again centered at $\w^*\approx \sqrt{b^2+1/2}$ at large magnetic
field. The major difference to the application of a transversal
magnetic field is the loss of spectral weight at finite frequencies in
favor of the zero-frequency peak.

As discussed above, the CET is restricted to a finite accessible
$t_{\rm max}\propto N_C$ which defines the lowest frequency resolution
$\Delta \w$.  Therefore, the $\delta(\w)$ cannot be accurately
resolved. We define a low frequency cutoff $\w_{\rm min}$ and
calculate the total spectral weight $S_+$
\begin{eqnarray}  
S_+ &=& \int_{\w_{min}}^\infty \frac{d\w}{2\pi} \, S(\w)
\end{eqnarray} 
of the finite frequency part of the spin-noise
spectrum. Since the CET spin-noise exactly fulfills the spectral
sum-rule, we obtain the spectral weight $S_0=2\pi S_\infty$ of the
$\delta(\w)$-peak from the difference $S_0 = 1/4-S_+$. The resulting
$S_\infty$ as function of $b$ is plotted as inset in Fig.\
\ref{fig-7-Sw-for-finite-Bz}(b) and reveals the increase of the weight
with increasing longitudinal magnetic field.


\section{Spin dynamics of the anisotropic CSM}

\subsection{Spin dynamics in the Ising limit}

\label{sec:results-ising-limit}

Up until now, the results were restricted to the isotropic case,
$\lambda=1$.  In order to set the stage for the anisotropic model with
finite $\lambda<\infty$, we focus on the opposite limit $\lambda
\rightarrow \infty$, the Ising limit, in this section.  Then, the
Hamiltonian (\ref{eq hamiltonian}) of the CSM reduces to an Ising
interaction between the central spin and the nuclear spins
\begin{align} H &= \w_L \vec{S} \vec{n}_B + \sum_k A_k S^z I_k^z
\end{align} and can be solved in a closed analytical form in some
limiting cases.\cite{Koppens2007,FischerLoss2008,Testelin2009} This
limit describes a pure heavy-hole spin.\cite{Koppens2007,FischerLoss2008,Testelin2009}

Obviously, all nuclear spin operators $I_k^z$ commute with $H$ and are
conserved. Therefore, the spin-bath is static and fully determined by
the eigenvalue configuration $\{ m_k \}$ of all $I_k^z$.  In the
absence of an external magnetic field, the eigenenergies are given by
$E^*(\sigma, \{ m_k \})$
\begin{eqnarray} 
E^*(\sigma, \{ m_k \}) &=& 
\sigma \sum_k A_k m_k = \sigma E^* \{ m_k \})
\end{eqnarray} 
where $\sigma$ is the eigenvalue of $S^z$.

\begin{figure} [tb] 

\centering

\includegraphics[width=85mm]{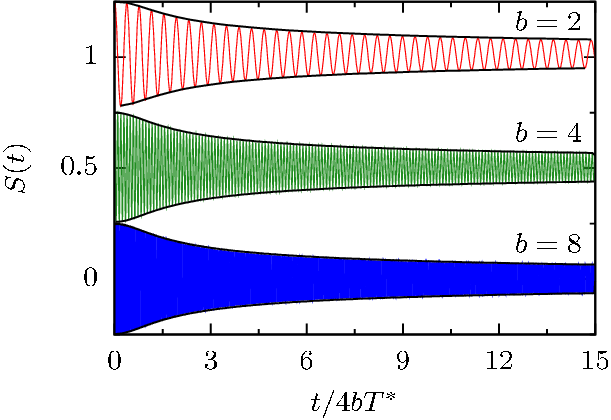}

  \caption{(color online) The spin noise function $S(t)$ for the Ising
limit $\lambda \rightarrow \infty$ of the anisotropic central spin
model plotted vs the dimensionless time $\tau = t/4bT^*$ for three
values of the external magnetic field $b=2,4,8$. An offset of $0.5$
has been added to distinguish the three individual curves.  The
envelope of $E_z (\tau)$ given by Eq.\ (\ref{eq eble}) has been added
as a black line to each simulation.  All curves have been calculated
for $r_0 = 1.5$, $n = 20$ and $N = 18$.  }
  \label{fig an0}
\end{figure}

In a finite external magnetic field, the Hamilton matrix decomposes in
$2\times 2$ subblocks for each fixed nuclear configuration $\{ m_k
\}$. For an external magnetic field in $x$-direction, we obtain the
two eigenenergies $E_\pm( \{ m_k \})) = \pm\sqrt{\w_L^2 +[E^* \{ m_k
\})]^2}/2$. The electronic spin precesses around the resulting
effective magnetic field $\vec{B}_{\rm eff}(\{ m_k \}) = ( \w_L, 0,
E^* (\{ m_k \}) )^T$ which depends on the bath
configuration.\cite{Koppens2007}

Averaging over the Larmor oscillations, the projection of the spin
component $\vec{n} [\vec{n}\vec{S}(0)]$ onto the magnetic field
direction $\vec{n}$ survives as already discussed above in the context
of Eq.\ (\ref{eqn:merkulov-eqn-12}).  For a spin initially polarized
in $z$-direction, the $z$-component of the non precessing contribution
$(\vec{e}_z \vec{n})^2 |\vec{S}(0)|$ has to be averaged over all
nuclear configurations $\{ m_k \}$
\begin{eqnarray}
\label{eqn:non-decaying-fraction-ising} 
\expect{ (\vec{e}_z
\vec{n})^2}_{ \{ m_k \} } &=& 
\expect{\frac{[E^* (\{ m_k \})
]^2}{\w_L^2 +[E^* \{ m_k \})]^2}}_{ \{ m_k \} }
\end{eqnarray} 
where $\vec{e}_z $ is the unit vector in $z$-direction.
For large magnetic fields, this average yields $\expect{ (\vec{e}_z
\vec{n})^2} = 1/4b^2$ while in zero magnetic field $\expect{
(\vec{e}_z \vec{n})^2}=1$.

When applying a transversal external magnetic field $B_x$, the fast
dynamics of a spin initially polarized in $z$-direction is determined
by the Larmor frequency $\w_L\propto b$.  The spin-decay, however, is
governed by a slowly varying envelope function.  Fig.\ \ref{fig an0}
depicts the real-time dynamics of $S(t)$ for three values of a large
external magnetic field for the Ising limit of the Hamiltonian.  An
offset of $0.5$ has been added to distinguish the different curves.

Testelin and collaborators\cite{Testelin2009} have extended the
semi-classical approach of Merkulov et.\ al.\cite{Merkulov2002} to the
Ising limit of the CSM and derived the analytic decay function
$E_z(t)$
\begin{eqnarray} 
E_z (\tau) &=& \frac14
\left[\frac{\text{cos}(4b^2\tau + \frac12 \text{arctan}(\tau))}{(1 +
\tau^2)^{1/4}} \right . \non && \left .  + \frac1{4b^2} \left( 1 -
\frac{\text{cos}(4b^2\tau + \frac32 \text{arctan}(\tau))}{(1 +
\tau^2)^{3/4}} \right)\right],
\label{eq eble}
\end{eqnarray} 
for the spin-component in $z$-direction in the limit of
large magnetic fields $b\gg 1$ where the new dimensionless time scale
$\tau = t/4bT^*$ has been introduced.  It consists of two oscillatory
terms governed by the Larmor frequency $4b^2\tau= \w_L t$ with an
additional phase shift term and two decaying envelope functions $\propto
(1+\tau^2)^{1/4}$ and $\propto (1+\tau^2)^{3/4}$.  The functional form
of this non-exponential decay has been derived\cite{Koppens2007} by
averaging the coherent spin-precession of the central spin for a given
nuclear configuration $\{ m_k \}$ over all nuclear configurations
using a Gaussian distribution of the nuclear field.  The value of the
non-decaying fraction $1/4b^2$ agrees exactly with the prediction of
Eq.\ (\ref{eqn:non-decaying-fraction-ising}).  The increase of the
dephasing time $T_{\rm deph} = 4b T^*$ with increasing field strength
can be understood by the suppression of the effective field fluctuation.

We have added the envelope of the function $E_z(\tau)$ to Fig.\ \ref{fig an0}
for the three different magnetic field strengths. The semi-classical
approach excellently reproduces the exact simulations for large
magnetic fields $b \gg 1$. Our numerical results confirm the slow
decay of the longitudinal spin component $\propto 1/\sqrt{t}$ for
large times in this limit.

\begin{figure} [tb] 
\centering

\includegraphics[width=80mm]{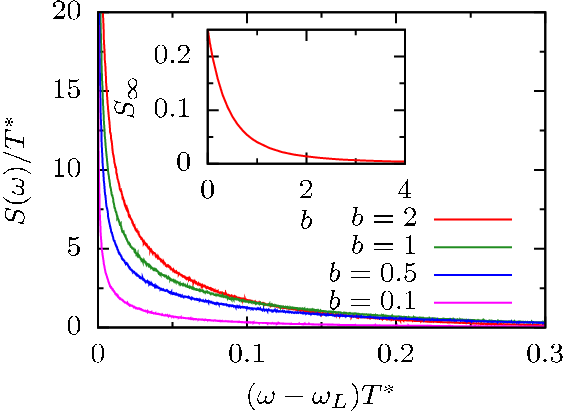}

  \caption{(color online) The spin noise spectra $S(\w)$ for
$(\w-\w_L)\geq 0$ in the Ising limit calculated by exact Fourier
transformation of (\ref{eqn:exact-ising}) for different external
magnetic fields $b$.  The inset shows the non-decaying fraction
$S_\infty$ which defines the spectral weight of the $\delta$-peak at
$\w=0$.  Parameters: $N=18$, $r_0=1.5$.  }
  \label{fig-10-ising-s-w-exact}
\end{figure}

While the power-law decay for large magnetic fields is well
established\cite{Koppens2007,FischerLoss2008,Testelin2009} it is not
obvious whether the analytic form of the long-time envelope prevails
in the crossover regime to small fields.  In order to reveal the
envelope function for the long time decay, one can subtract the
non-decaying fraction from $S(t)$ and multiply the remaining
oscillatory part with the dominating long-time decay. We find that the
amplitude of the oscillatory function
$(S(t)-S_\infty)(1+\tau^2)^{1/4}$ reaches a time-independent long time
limit -- not shown here.

To substantiate these findings, we expand the initial
spin polarized state $\ket{\uparrow, \{ m_k \}} $
\begin{eqnarray} 
\ket{\uparrow, \{ m_k \}} &=& c_+( \{ m_k \}) \ket{+,
\{ m_k \}} + c_-( \{ m_k \}) \ket{-, \{ m_k \}} \nonumber
\end{eqnarray} 
using the exact
eigenstates $\ket{\pm, \{ m_k \}}$ for a given configuration $ \{ m_k \}$ and exactly calculate the spin-noise function $S(t)=
\expect{S^z(t)}_{\rho_p}/2$ by averaging over all excitation energies
$\Delta E( \{ m_k \})= \sqrt{\w_L^2 +[E^* \{ m_k \})]^2}$:
\begin{eqnarray}
\label{eqn:exact-ising} S(t) &= & S_\infty + \sum_{ \{ m_k \}} c_+^2
c_- ^2\cos(\Delta E( \{ m_k \}) t) \\ S_\infty &=& \sum_{ \{ m_k \}}
\frac{1}{4}\left(c_+^2-c_-^2\right)^2
\,\, .
\end{eqnarray} 
While the time-independent part $S_\infty$ is
equivalent to (\ref{eqn:non-decaying-fraction-ising}) and defines the
spectral weight of $S(\w)$ at zero frequency, the Fourier
transformation of the second part can be calculated analytically and
yields the exact finite frequency contribution to the noise spectra.

For the frequency distribution of $\sqrt{\w_L^2 +[E^* \{ m_k \})]^2}$
we conclude that (i) there exists a finite threshold frequency $\w_{\rm
th}$ below which $S(0<\w<\w_{\rm th})=0$ and (ii)
$\w_L=\lim_{N\to\infty} \w_{\rm th}$.  Consequently, the spectral gap
in $S(\w)$ below $\w_{\rm th}$ will prevail in the thermodynamic
limit.  Since the largest frequency in (\ref{eqn:exact-ising}) is
limited by the two fully polarized configurations $\{ m_k =\uparrow\},
\{ m_k =\downarrow\}$, the spin noise spectrum $S(\w)$ also must vanish
for $|\w|> \sqrt{\w_L^2 +[E^*_{\rm max}]^2}$.

Fig.\ \ref{fig-10-ising-s-w-exact} depicts the exact $S(\w)$ as
function of $\w-\w_L$ for different magnetic fields.  $S(\w)$ can be
fitted by a power law $(\w-\w_L)^{-\alpha}$ very close to the Larmor
frequency.  We have extracted a universal exponent $\alpha\approx 1/2$
independent of $b$.  For small $b$, however, the non-decaying fraction
of the spin dominates and the finite frequency spectral function is
very small. Therefore, the fitting accuracy decreases for $b\to 0$.

Approximating $S(\w)$ by $C (\w-\w_L)^{-\alpha}$ close to the threshold
frequency up to some finite cutoff frequency $\w_{\rm max}$ and a
normalization constant $C$, we can analytically extrapolate the
long-time approach to $S_\infty$ from
\begin{eqnarray} 
\Delta S(t) &= &S(t)-S_\infty = 
2\Re e
\left[\int_{\w_L}^{\w_{\rm max}} \frac{d\w}{2\pi}
\frac{C}{(\w-\w_L)^\alpha} e^{i\w t}\right] \non &=& \frac{C}{\pi
t^{1-\alpha}} \Re e \left[ e^{i\w_L t} \int_0^{(\w_{\rm max} -\w_L)t} d
u \frac{e^{iu}}{u^\alpha} \right]
\end{eqnarray} 
which can be approximated for $t\to \infty$ and
$\alpha=1/2$ to
\begin{eqnarray} \Delta S(t) &\approx & \frac{a C}{\pi}
\frac{\cos(\w_L t+\pi/4) }{\sqrt{t}}
\end{eqnarray} where $a\approx 1.25331$. This result recovers the
long-term limit of $E_z(t)$ stated in Eq.\ (\ref{eq eble}) independent
from the $b\gg 1$ limit.


\subsection{Spin dynamics in the generic anisotropic CSM}
\label{sec:results-anisotropic-CSM}

\begin{figure} [t] 

\begin{center}
\includegraphics[width=84mm]{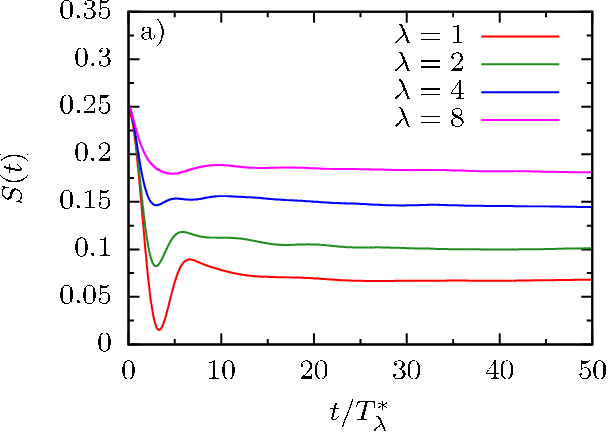}

\hspace*{5mm}\includegraphics[width=78mm]{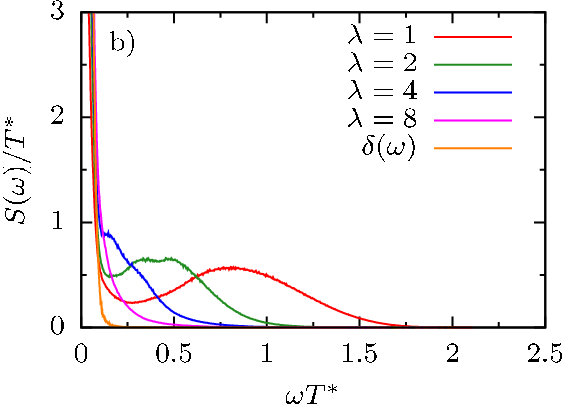}

\end{center}

  \caption{(color online) (a) $S(t)$ vs $t/T^*_\lambda$ for different
values of the anisotropy factor $\lambda=1,2,4,8$.  (b) Spin-noise
spectra vs $\w T^*$ for the same parameters as in panel (a) and $N_C = 100$.
A $\delta$-peak at zero frequency resolved by the CET with the same number
of moments has been added as reference.
Parameters: $N=18, n = 50, r_0=1.5$.  }
  \label{fig-11-b=0-anisotropic-CSM}
\end{figure}

\begin{figure} [bt] 

\centering

\includegraphics[width=75mm]{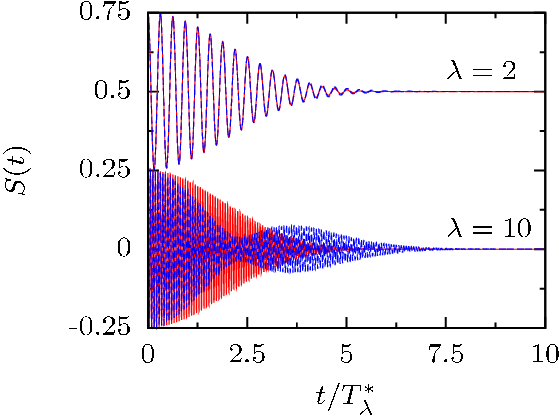}

 \caption{(color online) Comparison between the analytical prediction
(blue) of Eq (26c) of Ref.\ [\onlinecite{Testelin2009}] using a QSA
and the CET approach for $b/\lambda>1$. All curves have been calculated
for $b = 10$ and $\lambda=2$ or $\lambda=10$. We have added an offset
of $0.5$ to the $\lambda=2$-curves for better comparison. Parameters:
$N=18,r_0=1.5$.  }

  \label{fig-CET-vs-QSA-strong-field}
\end{figure}

\begin{figure} [bt]

\begin{center}

\includegraphics[width=70mm]{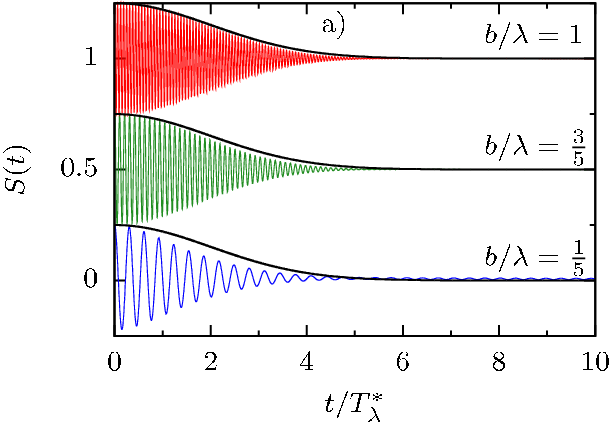}

\includegraphics[width=70mm]{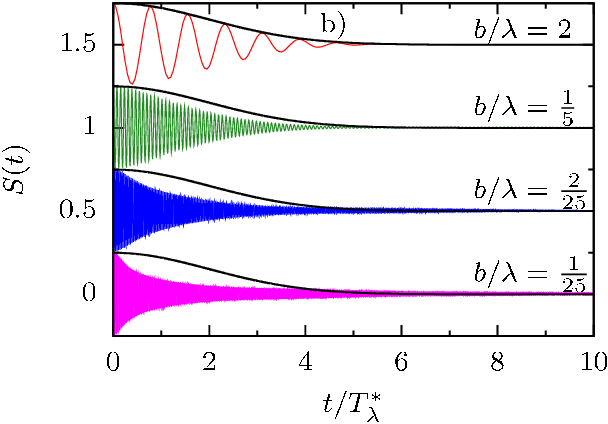}

\includegraphics[width=70mm]{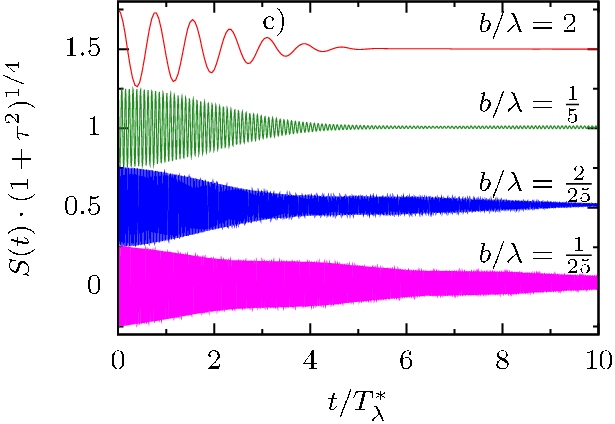}
\end{center}

 \caption{(color online) $S (t)$ in a transversal magnetic field: (a)
for a fixed value $\lambda=10$ and $b=2,6,10$, and (b) for a fixed
value $b=4$ and four different values of $\lambda=2,20,50,100$.  For
all cases we have added the envelope function
$0.25\exp[-(t/T^*_\lambda)^2/8]$ to reveal the difference to the
short-time dynamics of the isotropic CSM.  (c) Rescaled spin
correlation function 
$S(t)(1+\tau^2)^{1/4}$ vs
$t/T^*_\lambda$ for the same parameters as in panel (b) with $\tau =
t/4bT^*_1$.
Parameters: $N=18, n=20, r_0=1.5$.  }
 \label{fig-12-anisotrop-CMS-St-finite-bt}
\end{figure}

So far we have discussed the two extreme limits of the Hamiltonian
(\ref{eq hamiltonian}): the isotropic CSM and the Ising limit. Now we
investigate the influence of an arbitrary anisotropy factor $\lambda$
onto the spin-noise spectra as function of a transversal external
magnetic field.

In the absence of an external magnetic field, the decay of the spin
correlation function $S(t)$ is only caused by the transversal
terms in the Hamiltonian whose relative strength is controlled by the factor
$1/\lambda$. Presenting the data as function of the dimensionless
time $t/T^*_\lambda = t/\lambda T^*$ clearly reveals the
$\lambda$-dependence of the intrinsic time scale: the location of the
minimum in the short-time dynamics remains almost independent of $\lambda$
as depicted in Fig.~\ref{fig-11-b=0-anisotropic-CSM}(a).  With increasing
$\lambda$, the non-decaying fraction of $S(t)$ monotonically increases,
and the curves basically interpolate between the isotropic CSM limit
and Ising limit where $S(t)=1/4=const.$ does not decay at all.

The corresponding spin-noise spectrum $S(\w)$ is shown as a function of
the dimensionless frequency $\w T^*_1$ in
Fig.~\ref{fig-11-b=0-anisotropic-CSM}(b).  We have added the result for
the isotropic CSM as reference for comparison. Since the CET approach
has a finite resolution of a $\delta(\w)$-peak for $\w\to 0$, the data
for $\lambda=\infty$ serves as reference and indicates the change of the
spectral function at finite frequencies as function of $\lambda$.  The
height of the Gaussian shaped peak close to $\w \approx 1/T^*_\lambda$
of the isotropic CSM decreases with increasing $\lambda$, and spectral
weight is shifted in parts into the $\delta$-peak at $\w=0$ and to
smaller frequencies $\w>0$. For $\lambda>3$, the peak close to $\w
\approx 1/T^*_\lambda$ has disappeared and the spectra are
monotonically increasing for decreasing $\w$.

Now we proceed to the spin response in a transversal magnetic field.
A word is in order to clarify the definition of a strong external
field in the anisotropic CSM. The Hamiltonian 
(\ref{eq hamiltonian}) contains two terms driving (i) the decoherence
and (ii) the dephasing of the central spin in $z$-direction: (i) the
transversal spin-flip term is governed by the characteristic timescale
$T^*_\lambda$ and (ii) the external transversal field is governed by
$T^*_\lambda / \lambda b$.  If $b\cdot\lambda \gg 1$ holds, the
external magnetic field dominates over the spin-flip term, and,
therefore, is called a strong external field. The central spin
performs several Larmor precessions before the decoherence due to the nuclei
sets in to the spin bath. In the opposite limit, $b\cdot\lambda \ll 1$, the
central spin cannot complete a single Larmor precession on the time
scale $T^*_\lambda$ on which $S(t)$ decays due to the hyperfine
interaction.

We begin to investigate the strong field regime defined by $b\cdot\lambda > 1$. We
compare the analytical predictions\cite{Testelin2009} by the QSA with our
CET results for short-time and intermediate-time scales as depicted in
Fig.~\ref{fig-CET-vs-QSA-strong-field}. While for $\lambda=2,b=10$ good
agreement is found between both methods significant deviations are
observed for an increased asymmetry $\lambda=10,b=10$ already at short
times. While the numerical exact results show a monotonic decay of the
envelope function, the QSA predicts an oscillation of the spin
polarization amplitude. It turns out that this result can be generalized
to the statement, that Eq (26c) of Ref.\ [\onlinecite{Testelin2009}] using
the QSA only describes $S(t)$ in a transversal magnetic field adequately if
the parameter $b / \lambda$ is large
compared to $1$. This indicates the importance of the ratio $b/\lambda$ 
for separating different types of the dynamics in the anisotropic central spin
model.

To gain more information on the influence of the ratio $b / \lambda$,
CET results for $S(t)$ are shown at either a
fixed $\lambda=10$ and different magnetic field strengths $b=2,6,10$ in
panel (a), or at fixed magnetic field strength $b=4$ and several values
of $\lambda=2,10,20,50$ in panel (b) of
Fig.~\ref{fig-12-anisotrop-CMS-St-finite-bt}.
Augmenting the numerical data with the envelope funtion
$0.25\exp[-(t/T^*_\lambda)^2/8]$ demonstrates that the short-time
response is governed by the characteristic time scale $T^*_\lambda$
and the decay is well captured by a Gaussian envelope function for
$b/\lambda\ge 1$ as predicted by the QSA.\cite{Testelin2009}

Deviations from such a Gaussian decay increase with decreasing ratio
$b/\lambda<1$ and are very pronounced for $b/\lambda=1/25$. For these
parameters, the initial decay occurs much faster but on long-time
scales significantly slower than described by a Gaussian envelope
function. The dependency of the crossover on the ratio $b/\lambda$
can be understood from the fact that the hyperfine-field asymmetry
increases with increasing $\lambda$, and a larger transversal external
magnetic field $b$ is required to induce a dynamics similar to the
isotropic CSM. Only at very strong transversal magnetic fields $B_x$,
the relevance of anisotropy of the hyperfine interaction is reduced
and the Gaussian type of decay is expected for $b/\lambda > 1$.

For strong external fields, $b\cdot\lambda > 1$, the ratio $b/\lambda$ 
solely dictates the form of the envelope of $S(t)$: 
for several different combinations of $b$ and $\lambda$, the envelope functions
for $S(t)-S_\infty$  are identical  - being not explicitly shown here.

For further illustration of the crossover from a Gaussian to the Ising
type decay, we plotted the rescaled spin correlation function
$S(t)[1+\tau^2]^{1/4}$ vs $t/T^*_\lambda$ in
Fig.\ \ref{fig-12-anisotrop-CMS-St-finite-bt}(c), where $\tau = t/4b T^*$.
This reveals an Ising type power law decay for long-time scales that is
recovered for $\lambda\to\infty$.
For $b=4$ and $b/\lambda=2$, we clearly observe a Gaussian dominated
decay. For decreasing $b/\lambda$, an increasing intermediate-time
regime develops where the envelope function follows a slower power-law
type decay. Note that $S(t)[1+\tau^2]^{1/4}$ yields a non-decaying
oscillation in the Ising limit, which can not be shown in figure
\ref{fig-12-anisotrop-CMS-St-finite-bt}(c) due to the underlying
time-scale.

\begin{figure}[t] \centering

\includegraphics[width=75mm]{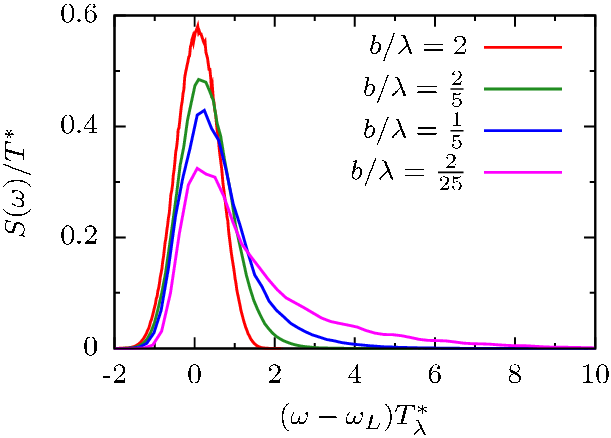}

  \caption{(color online) Spin noise spectra $S(\w)$ for $b = 4$ and
$\lambda = 2,10,20,50$, plotted vs $(\w-\w_L)T^*_\lambda$.  Since the
characteristic time-scale $T^*_\lambda$ increases with increasing
$\lambda$, we adjusted the number of Chebychev momenta to
$N_C(\lambda=2)=100$, $N_C(\lambda=10)=300$, $N_C(\lambda=20)=600$,
and $N_C(\lambda=50)=1500$ to provide an adequate frequency
resolution. The chosen parameters are equivalent to those in
FIG. \ref{fig-12-anisotrop-CMS-St-finite-bt}. }
  \label{fig-spin-noise-rescaled-St}
\end{figure}

We have been able to identify three different regimes for $b\cdot\lambda>1$.
For $b> \lambda$, (i) the decay is of Gaussian type\cite{Testelin2009} similar
to the isotropic CSM.  As long as $\lambda$ is less than one order of
magnitude larger than $b$, (ii) the decay of $S(t)$ deviates from
the Gaussian envelope, but the long-time decay is still governed by
the tail of the envelope function. In the last regime, (iii) where
$\lambda \gg b$ holds, the behavior of the spin-noise function
approaches the Ising regime discussed in the previous section.

The corresponding spin-noise spectrum for this crossover regime
$b/\lambda<1$ is depicted as function of $(\w-\w_L)T_\lambda^*$ in
Fig.\ \ref{fig-spin-noise-rescaled-St}. In leading order, the peak
position of $S(\w)$ is clearly given by the Larmor frequency
$\w_L$, while the peak width is governed by $T^*_\lambda$.  Note that
the smallest and the largest value of $\lambda$ differ by a factor of
$25$.  Therefore, the increase of $T^*_\lambda$ requires a
significant increase of the Chebychev order $N_C$ for a reliable
resolution of the spectra. $S(\w)$ evolves from a Gaussian shape for
$b/\lambda=2$, to a precursor of a threshold behavior for
$b/\lambda=2/25$. While the high frequency tail can be fitted with a
Lorentzian, the low frequency spectrum is rapidly suppressed below the
Larmor frequency for $0<\w<\w_L$ with increasing anisotropy.  In the
limit $\lambda\to \infty$, the Ising limit of the spectrum, as shown
in Fig.~\ref{fig-10-ising-s-w-exact}, must be recovered. Therefore,
the increasing low frequency gap with increasing asymmetry prevails in
the thermodynamic limit.

Let us now turn to small magnetic fields characterized by the
condition $b\cdot\lambda \ll 1$. In this case the central spin
experiences decoherence induced by the spin bath before a whole Larmor
precession can occur.  The short-time dynamics in this regime is
governed by $T^*_\lambda$ and is of the same form as shown for
different values of $\lambda$ in Fig.\
\ref{fig-11-b=0-anisotropic-CSM}(a). After the initial spin decay,
$S(t)$ will approach zero on a large timescale that is dictated by
$\lambda/\omega_L$.  This long-time behavior of $S(t)$ is analogous to
the isotropic CSM as shown in Fig.\ \ref{fig bxstrong}(b). Combining
the information provided by both figures fully describes the long-time
and short-time behavior of $S(t)$ in the weak field limit of the
anisotropic CSM, and we do not show further explicit results for this
regime.

So far we did not discuss the correlation function $S(t)$ for a light hole spin.  
Such QDs are characterized by an anisotropy factor
$\lambda = 1/2< 1$, so that the spin-flip terms dominate over the
Ising contribution.  Since $b/\lambda > 1/\lambda^2>1$ is bound in the
strong field regime, no Ising-type behavior can be observed and $S(t)$
always shows a Gaussian type decay for $b\cdot\lambda > 1$.  However, all other findings
discussed above remain valid for the case $\lambda < 1$: an analogous
behavior for the weak field limit $b\cdot\lambda \ll 1$ is found.
Overall, the spin noise is very similar to the isotropic CSM but with a faster 
characteristic time scale $T^*\to T^*_\lambda<T^*$.

\section{Spin noise spectra in quantum dot ensembles}
\label{sec:QD-ensemble-average}

\begin{figure}[t]
\begin{center}

\includegraphics[width=75mm]{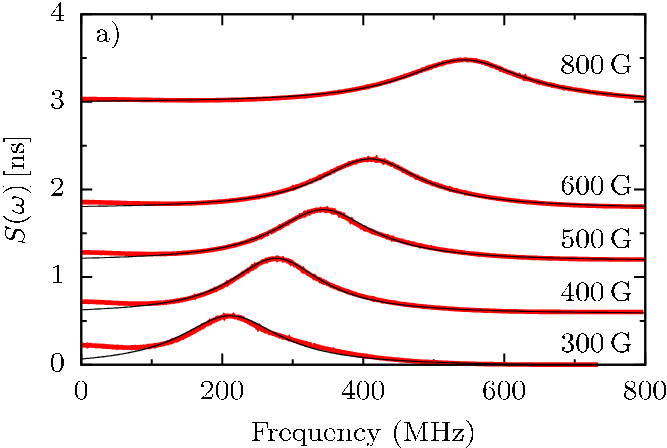}

\includegraphics[width=75mm]{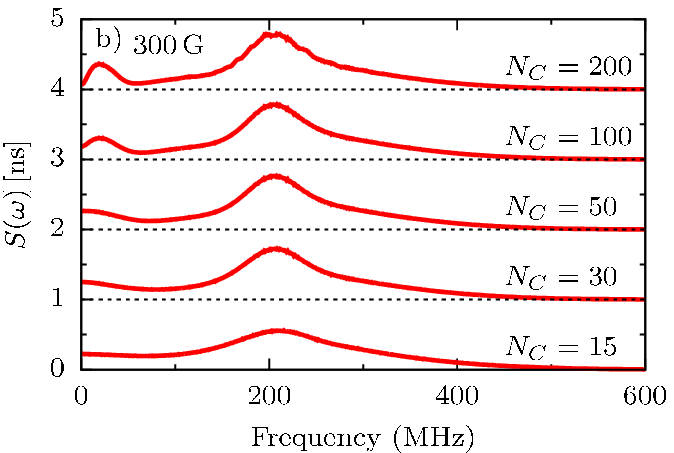}

\caption{(color online)
(a)
Ensemble averaged spin-noise spectra vs $\w$ for five different values of $B$
for  a dot-confined electron spin, i.~e.\ $\lambda=1$.  
Defining the average timescale $\overline{T}^* =\expect{T^*}=1$ ns as reference,
and using $|g_e|=0.54$ as in Ref.~\onlinecite{CrookerBayerSpinNoise2010},
$S(\w)$ has been obtained by averaging over $n=100$ different spectra 
generated from a constant distribution of $T^* \in [0.2\,{\rm ns},1.8\,{\rm ns}]$
of characteristic single QD time scales. 
For better visibility, an offset proportional to $b=\w_L \overline{T}^*$ has been added to $S(\w)$. 
Parameters: $N_C=15,N=18$. The black lines indicate Lorentzian fits to the
individual curves.
(b) Evolution of the ensemble noise spectrum with increasing $N_C$ at fixed $B=300$ G.
}
\label{fig:s-w-ensemble-average-electron}
\end{center}
\end{figure}

The spin noise spectrum has been 
measured\cite{CrookerBayerSpinNoise2010,Dahbashi2012,LiBayer2012,*ZapasskiiGreilichBayer2013}
on  QD ensembles charged with a single electron or hole.  
To bridge between the single QD calculations and the recent experiments,
we have performed an ensemble average of $n$ single independent QD spectra
with different configurations $\{A_k\}$. We have used the average timescale 
$\overline{T^*} =\expect{T^*}$ as reference timescale with
equally distributed individual $T^*\in [0.2\overline{T^*}, 1.8 \overline{T^*}]$  for each individual QD. 

\begin{figure}[tb]
\begin{center}

\includegraphics[width=80mm,height=50mm]{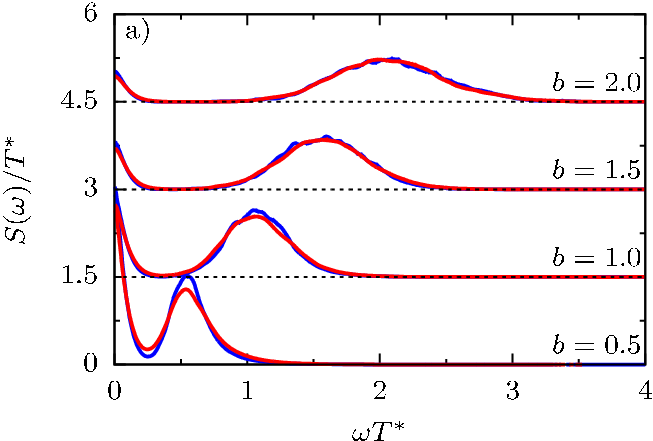}

\includegraphics[width=80mm,height=50mm]{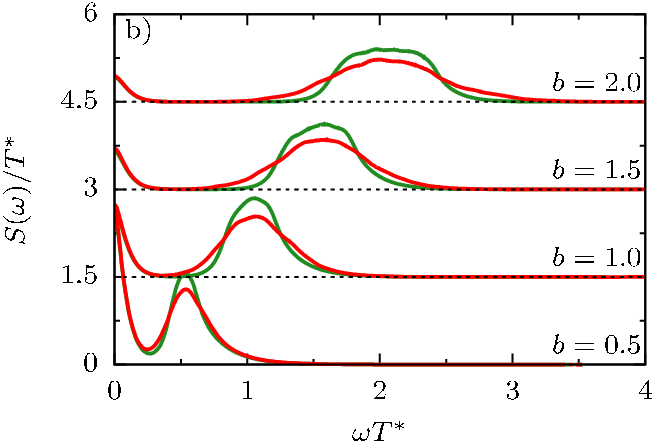}

\caption{(color online)
(a) Ensemble averaged spin-noise  spectra in the asymmetric CSM for $\lambda=5$ (red)
and $\lambda=50$ (blue) for different $b$.
$S(\w)$ has been obtain by averaging over $n=2000$ different single QD
spectra  generated from a constant distribution of $T^*_i\in [0.2 T^*, 1.8
T^*]$, and the individual $g$ factors are taken from a Gaussian distribution
with $ \Delta g / g=0.2$.
(b)  Influence of the $g$-factor distribution function on the shape of the 
spin-noise spectrum: comparing the $\lambda=5$ data of (a) with an 
ensemble averaged $S(\w)$ where the individual $g$ factors $g_i$ 
are taken from a constant  distribution  of $g_i/g \in [0.8,1.2]$ (green curves). 
For better visibility, an offset proportional to $b$ has been added to $S(\w)$.
Parameters: $N = 18, N_C=50, r_0 = 1.5$.
}
\label{fig:s-w-ensemble-average-hole}
\end{center}
\end{figure}

For QD ensembles with  dot-confined electrons
the distribution of $g$-factors is very sharp,\cite{GreilichBayer2007}  and the spread 
of the $g$ factors can be neglected.  The resulting ensemble averaged spin-noise spectra
are depicted in Fig.\ \ref{fig:s-w-ensemble-average-electron}(a). An offset proportional
to the magnetic field $b$ has been added to $S(\w)$ for better
visibility.  The black lines indicate a Lorentzian fit to each individual curve.

For the absolute units in Fig.\
\ref{fig:s-w-ensemble-average-electron}, we have set the average electron
dephasing time $\overline{T^*} = 1\,\text{ns}$ and have used the
experimentally determined\cite{CrookerBayerSpinNoise2010} electron
$g$-factor $|g_e|=0.54$.  Our results demonstrate that phenomenological
Lorentz fits to the experimental spin noise spectra are justified for
strong external magnetic fields $b=\w_L \overline{T^*} >1$. For small
$b$ significant deviations from a simple Lorentzian fit are observed
in the ensemble averaged spectra as being reported in the experimental
data.\cite{CrookerBayerSpinNoise2010} These deviations are related to
the low-frequency features of the spin-noise ignored in such a simple
Lorentzian fit.

Our results are in good agreement with the experimental data of Fig.\
4(a) in Ref.\ \onlinecite{CrookerBayerSpinNoise2010}.  For larger
Chebyshev orders the ensemble averaging still provides peaks around
the Larmor-frequency in $S(\omega)$ that can be approximated by a
Lorentzian, but a clear gap between zero frequency and $\omega_L$
occurs as discussed before.

Note that the spectra presented in Fig.\
\ref{fig:s-w-ensemble-average-electron}(a) have been calculated for a
relatively small Chebyshev order $N_C = 15$
corresponding to a frequency resolution of $\w> 20$ MHz
for a magnetic field of $B=300$ G up to $\w> 60$ MHz for $B=800$ G.
This would mimic a  limited resolution in the experiments. 

The effect of an increase of the frequency resolution (increasing $N_C$)
on the spectral properties
of the ensemble averaged spin noise is  depicted in Fig.\
\ref{fig:s-w-ensemble-average-electron}(b) for a fixed magnetic field $B=300$ G.
While the broadened
peak around $200$ MHz converges very quickly by a small amount of narrowing, 
the additional low-frequency feature  can now be resolved. This would 
apply to all spectra shown in \ref{fig:s-w-ensemble-average-electron}(a): the additional
frequency resolution does not change the qualitative properties of the spin noise in
a transverse magnetic field but only reveals additional spectral information at low frequency.

Now we turn to hole spin  ensembles modeled by the anisotropic CSM. 
In this case the distribution
of the $g$-factors in the QDs is crucial when making connection 
to measurements\cite{LiBayer2012,*ZapasskiiGreilichBayer2013}
 of spin noise in those QD ensembles.  To account for
this spread, we have used a Gaussian probability distribution for the
$g$-factors centered at $g$ with standard deviation $\Delta
g / g = 0.2$ to  include a random value of $g$ for each
individual QD in our simulations. In addition,  
the  average dephasing time $\overline{T^*} =\expect{T^*}$  is set as reference time 
(and reciprocal energy) scale, and
the individual time scales $T^*$ have been generated 
from a constant distribution $T^*\in [0.2 T^*, 1.8
T^*]$ for each of the $n=2000$ individual QD calculations entering the ensemble
average. The resulting averaged $S(\w)$ are depicted in  Fig.\ \ref{fig:s-w-ensemble-average-hole}(a)
for two different values of $\lambda=5,50$.

It is striking that the width of the ensemble averaged $S(\w)$ hardly depends 
of the anisotropy factor $\lambda$, although 
the peak width  is given by $1/T^*_\lambda=1/\lambda T^*$  in a single QD.
In a QD ensemble of hole spins, however, the peak width of the noise spectra is determined by the 
distribution of $g$-factors, and, consequently, the spectral width of $S(\w)$ 
increases with increasing 
external magnetic field.

To illustrate  the influence of the $g$-factor distribution function
onto the shape of ensemble averaged $S(\w)$, 
we have combined 
the $\lambda=5$ spectra of  Fig.\ \ref{fig:s-w-ensemble-average-hole}(a)
with another set of ensemble averaged $S(\w)$ calculated by 
using  a constant 
distribution function $b_i/b \in [0.8,1.2]$ 
but  otherwise identical parameters (green curves)
in  Fig.\ \ref{fig:s-w-ensemble-average-hole}(b). Clearly, the peak shape
of $S(\w)$ obtained from a box-shaped $g$-factor distribution function
resembles strongly this distribution function while the spectra
of Fig.\ \ref{fig:s-w-ensemble-average-hole}(a)
(red curves)  exhibit  a more Gaussian-type  shape.

As  a consequence, the shape of  ensemble averaged spin-noise spectra
is mainly  governed by the two distribution functions for $g$ and $T^*$ and reveal
less information about the dephasing mechanism in a single quantum dot.
Therefore, it would be desirable to be able to measure the spin noise directly on a single
quantum dot in order to obtain additional information on the microscopic dephasing
mechanism of an electron or hole spin in a QD.

\section{Discussion and outlook}
\label{sec:discussion-outlook}

We have presented a comprehensive study of the spin-noise in the minimal
model for dephasing in QDs, the anisotropic CSM. We have investigated
the dependency of $S(t)$ and its Fourier transform $S(\w)$ on the
external magnetic field and the anisotropy parameter $\lambda$ using a
CET approach. This approach provides a time-evolution of an arbitrary
initial state whose error can be reduced to machine precision for any
give time $t$.  We have extended the CET to ensemble averaging of
traces needed to account for a fully incoherent spin-bath.

In the absence of an external magnetic field, $S(t)$ exhibits an
universal time evolution for the isotropic CSM on a very short time
scale governed by the bath-spin fluctuation time $T^*$. At
intermediate times $t/T^*<100$, the dynamics depend on the
distribution function of the hyperfine coupling constants. While a
semi-classical approximation predicts a finite non-decaying fraction
of the spin polarization of $S^{\text{QSA}}_\infty=S(0)/3$ by averaging over
a distribution of static nuclear field configurations, our CET results
are typically below $S^{\text{QSA}}_\infty$.  Fluctuations of the nuclear
spin bath occur on a much longer time scale and can lead to
significant deviations from the quasi-static approximation when the
hyperfine coupling constants cover a large parameter range.  Typically
we find a very slow decay at intermediate and long-time scales driven
by the distribution of the small hyperfine interaction constants
$A_k$. Finite size scaling of the non-decaying fraction of the spin
polarization of $S_\infty$ as function of the cutoff $r_0$ indicates
vanishing $S_\infty$ in the thermodynamic limit in agreement with the
intuitive but not rigorous argument of Chen et
al.\cite{ChenBalents2007}

For the isotropic CSM, we find that the spin-noise spectrum consists
of three parts: (i) a zero-frequency $\delta$-peak whose spectral
weight is given by the non-decaying fraction of the central spin, (ii)
a low-frequency threshold behavior which follows approximately a
power-law and is cut off at the smallest non-zero excitation energy of
the finite-size system determined by $\bar A_{\rm min}(r_0)$ and (iii)
a Gaussian-type high frequency peak around $\w T^*\approx 1$ which
determines the short time dynamics on time scales of $T^*$.  We find a
remarkable agreement with a $\w^{-3/4}$ fit to the low-frequency part
of our high resolution CET spin-noise spectrum which suggests an
asymptotic $1/t^{1/4}$ spin-decay at long times which is not
contradicting the approximative solutions in the literature: on
intermediate time scales $1/t^{1/4}$ and $1/\log(t)^{3/2}$ are nearly
indistinguishable and only deviate significantly in their asymptotic behavior.
We conjecture that in the thermodynamic limit the zero-frequency
$\delta$-peak will disappear, and the low-frequency spectra will show
a pronounced threshold behavior which will govern the long-time
spin-decay. For a detailed analysis of its analytical form, other
approaches have to be employed.

The Ising limit of the CSM is exactly solvable by treating the
$2\times2$ Hamiltonian subspaces for each frozen nuclear spin
configuration leading to a strict threshold behavior of the noise
spectra.  The spectrum consists of a $\delta(0)$ peak whose spectral
weight decreases with increasing $b$ and close to the threshold a
$\w^{-1/2}$ fit can be applied to the data, yielding a $t^{-1/2}$ long
time decay independently of the applied magnetic field. In the Ising
limit pure dephasing occurs and we find that an increase of the
external field yields an increasing coherence time of the central spin
due to the decreasing spread of the eigenenergies.

In the anisotropic CSM without an external magnetic field the
timescale $T^*_\lambda$ determines the short-time evolution of the
central spin. We have demonstrated that in a transversal magnetic
field the value of $b \cdot\lambda$ separates hyperfine interaction
driven spin dynamics from magnetic field dominated spin dynamics.
For $b \cdot\lambda > 1$, the ratio $b / \lambda$ describes the
crossover from Ising-like dynamics ($b / \lambda \ll 1$) to a more
Gaussian type decay ($b / \lambda > 1$.)  For a single QD in a strong
external field we always find a clear energy gap in the spin-noise
between $\omega = 0$ and the broadened peak close to the Larmor
frequency whose width is given by $1/T^*_\lambda$.

Furthermore, we have performed an ensemble average of the
spin-noise spectra to compare our
calculations to recent spin-noise measurements  on QD ensembles. By
using realistic characteristic time scales $T^*$, the experimentally
determined distribution of $g$-factors, we find qualitative 
agreement for QD ensembles of both
hole spins and electron spins. 
Our calculations reveal the shape dependence of 
ensemble averaged noise spectra  on the distribution functions for 
the $g$-factors and the characteristic time scales $T^*$:
the  averaged noise spectrum contains less information  
on the dephasing mechanism than the data for an individual quantum dot.

We acknowledge fruitful discussions with Manfred Bayer, Scott Crooker,
Alex Greilich, Nikolai Sinitsyn, 
Joachim Stolze, G\"otz Uhrig, and Dmitri Yakovlev.



\begin{thebibliography}{51}%
\makeatletter
\providecommand \@ifxundefined [1]{%
 \@ifx{#1\undefined}
}%
\providecommand \@ifnum [1]{%
 \ifnum #1\expandafter \@firstoftwo
 \else \expandafter \@secondoftwo
 \fi
}%
\providecommand \@ifx [1]{%
 \ifx #1\expandafter \@firstoftwo
 \else \expandafter \@secondoftwo
 \fi
}%
\providecommand \natexlab [1]{#1}%
\providecommand \enquote  [1]{``#1''}%
\providecommand \bibnamefont  [1]{#1}%
\providecommand \bibfnamefont [1]{#1}%
\providecommand \citenamefont [1]{#1}%
\providecommand \href@noop [0]{\@secondoftwo}%
\providecommand \href [0]{\begingroup \@sanitize@url \@href}%
\providecommand \@href[1]{\@@startlink{#1}\@@href}%
\providecommand \@@href[1]{\endgroup#1\@@endlink}%
\providecommand \@sanitize@url [0]{\catcode `\\12\catcode `\$12\catcode
  `\&12\catcode `\#12\catcode `\^12\catcode `\_12\catcode `\%12\relax}%
\providecommand \@@startlink[1]{}%
\providecommand \@@endlink[0]{}%
\providecommand \url  [0]{\begingroup\@sanitize@url \@url }%
\providecommand \@url [1]{\endgroup\@href {#1}{\urlprefix }}%
\providecommand \urlprefix  [0]{URL }%
\providecommand \Eprint [0]{\href }%
\providecommand \doibase [0]{http://dx.doi.org/}%
\providecommand \selectlanguage [0]{\@gobble}%
\providecommand \bibinfo  [0]{\@secondoftwo}%
\providecommand \bibfield  [0]{\@secondoftwo}%
\providecommand \translation [1]{[#1]}%
\providecommand \BibitemOpen [0]{}%
\providecommand \bibitemStop [0]{}%
\providecommand \bibitemNoStop [0]{.\EOS\space}%
\providecommand \EOS [0]{\spacefactor3000\relax}%
\providecommand \BibitemShut  [1]{\csname bibitem#1\endcsname}%
\let\auto@bib@innerbib\@empty
\bibitem [{\citenamefont {Schliemann}\ \emph {et~al.}(2003)\citenamefont
  {Schliemann}, \citenamefont {Khaetskii},\ and\ \citenamefont
  {Loss}}]{SchliemannKhaetskiiLoos2003}%
  \BibitemOpen
  \bibfield  {author} {\bibinfo {author} {\bibfnamefont {J.}~\bibnamefont
  {Schliemann}}, \bibinfo {author} {\bibfnamefont {A.}~\bibnamefont
  {Khaetskii}}, \ and\ \bibinfo {author} {\bibfnamefont {D.}~\bibnamefont
  {Loss}},\ }\href@noop {} {\bibfield  {journal} {\bibinfo  {journal} {Journal
  of Physics: Condensed Matter}\ }\textbf {\bibinfo {volume} {15}},\ \bibinfo
  {pages} {R1809} (\bibinfo {year} {2003})}\BibitemShut {NoStop}%
\bibitem [{\citenamefont {Hanson}\ \emph {et~al.}(2007)\citenamefont {Hanson},
  \citenamefont {Kouwenhoven}, \citenamefont {Petta}, \citenamefont {Tarucha},\
  and\ \citenamefont {Vandersypen}}]{HansonSpinQdotsRMP2007}%
  \BibitemOpen
  \bibfield  {author} {\bibinfo {author} {\bibfnamefont {R.}~\bibnamefont
  {Hanson}}, \bibinfo {author} {\bibfnamefont {L.~P.}\ \bibnamefont
  {Kouwenhoven}}, \bibinfo {author} {\bibfnamefont {J.~R.}\ \bibnamefont
  {Petta}}, \bibinfo {author} {\bibfnamefont {S.}~\bibnamefont {Tarucha}}, \
  and\ \bibinfo {author} {\bibfnamefont {L.~M.~K.}\ \bibnamefont
  {Vandersypen}},\ }\href {\doibase 10.1103/RevModPhys.79.1217} {\bibfield
  {journal} {\bibinfo  {journal} {Rev. Mod. Phys.}\ }\textbf {\bibinfo {volume}
  {79}},\ \bibinfo {pages} {1217} (\bibinfo {year} {2007})}\BibitemShut
  {NoStop}%
\bibitem [{\citenamefont {Elzerman}\ \emph {et~al.}(2004)\citenamefont
  {Elzerman}, \citenamefont {Hanson}, \citenamefont {van Beveeren},
  \citenamefont {Witkamp}, \citenamefont {Vandersypen},\ and\ \citenamefont
  {Kouvenhoven}}]{Elzerman04}%
  \BibitemOpen
  \bibfield  {author} {\bibinfo {author} {\bibfnamefont {J.~M.}\ \bibnamefont
  {Elzerman}}, \bibinfo {author} {\bibfnamefont {R.}~\bibnamefont {Hanson}},
  \bibinfo {author} {\bibfnamefont {L.~H.~W.}\ \bibnamefont {van Beveeren}},
  \bibinfo {author} {\bibfnamefont {B.}~\bibnamefont {Witkamp}}, \bibinfo
  {author} {\bibfnamefont {L.~M.~K.}\ \bibnamefont {Vandersypen}}, \ and\
  \bibinfo {author} {\bibfnamefont {L.~P.}\ \bibnamefont {Kouvenhoven}},\
  }\href@noop {} {\bibfield  {journal} {\bibinfo  {journal} {Nature}\ }\textbf
  {\bibinfo {volume} {430}},\ \bibinfo {pages} {431} (\bibinfo {year}
  {2004})}\BibitemShut {NoStop}%
\bibitem [{\citenamefont {Bonadeo}\ \emph {et~al.}(1998)\citenamefont
  {Bonadeo}, \citenamefont {Erland}, \citenamefont {Gammon}, \citenamefont
  {Park}, \citenamefont {Katzer},\ and\ \citenamefont {Steel}}]{Bonadeo1998}%
  \BibitemOpen
  \bibfield  {author} {\bibinfo {author} {\bibfnamefont {N.~H.}\ \bibnamefont
  {Bonadeo}}, \bibinfo {author} {\bibfnamefont {J.}~\bibnamefont {Erland}},
  \bibinfo {author} {\bibfnamefont {D.}~\bibnamefont {Gammon}}, \bibinfo
  {author} {\bibfnamefont {D.}~\bibnamefont {Park}}, \bibinfo {author}
  {\bibfnamefont {D.~S.}\ \bibnamefont {Katzer}}, \ and\ \bibinfo {author}
  {\bibfnamefont {D.~G.}\ \bibnamefont {Steel}},\ }\href {\doibase
  10.1126/science.282.5393.1473} {\bibfield  {journal} {\bibinfo  {journal}
  {Science}\ }\textbf {\bibinfo {volume} {282}},\ \bibinfo {pages} {1473}
  (\bibinfo {year} {1998})}
  \BibitemShut
  {NoStop}%
\bibitem [{\citenamefont {Greilich}\ \emph {et~al.}(2007)\citenamefont
  {Greilich}, \citenamefont {Shabaev}, \citenamefont {Yakovlev}, \citenamefont
  {Efros}, \citenamefont {Yugova}, \citenamefont {Reuter}, \citenamefont
  {Wieck},\ and\ \citenamefont {Bayer}}]{GreilichBayer2007}%
  \BibitemOpen
  \bibfield  {author} {\bibinfo {author} {\bibfnamefont {A.}~\bibnamefont
  {Greilich}}, \bibinfo {author} {\bibfnamefont {A.}~\bibnamefont {Shabaev}},
  \bibinfo {author} {\bibfnamefont {D.~R.}\ \bibnamefont {Yakovlev}}, \bibinfo
  {author} {\bibfnamefont {A.~L.}\ \bibnamefont {Efros}}, \bibinfo {author}
  {\bibfnamefont {I.~A.}\ \bibnamefont {Yugova}}, \bibinfo {author}
  {\bibfnamefont {D.}~\bibnamefont {Reuter}}, \bibinfo {author} {\bibfnamefont
  {A.~D.}\ \bibnamefont {Wieck}}, \ and\ \bibinfo {author} {\bibfnamefont
  {M.}~\bibnamefont {Bayer}},\ }\href@noop {} {\bibfield  {journal} {\bibinfo
  {journal} {Science}\ }\textbf {\bibinfo {volume} {317}},\ \bibinfo {pages}
  {1896} (\bibinfo {year} {2007})}\BibitemShut {NoStop}%
\bibitem [{\citenamefont {Fokina}\ \emph {et~al.}(2010)\citenamefont {Fokina},
  \citenamefont {Yugova}, \citenamefont {Yakovlev}, \citenamefont {Glazov},
  \citenamefont {Akimov}, \citenamefont {Greilich}, \citenamefont {Reuter},
  \citenamefont {Wieck},\ and\ \citenamefont {Bayer}}]{FokinaBayer2010}%
  \BibitemOpen
  \bibfield  {author} {\bibinfo {author} {\bibfnamefont {L.~V.}\ \bibnamefont
  {Fokina}}, \bibinfo {author} {\bibfnamefont {I.~A.}\ \bibnamefont {Yugova}},
  \bibinfo {author} {\bibfnamefont {D.~R.}\ \bibnamefont {Yakovlev}}, \bibinfo
  {author} {\bibfnamefont {M.~M.}\ \bibnamefont {Glazov}}, \bibinfo {author}
  {\bibfnamefont {I.~A.}\ \bibnamefont {Akimov}}, \bibinfo {author}
  {\bibfnamefont {A.}~\bibnamefont {Greilich}}, \bibinfo {author}
  {\bibfnamefont {D.}~\bibnamefont {Reuter}}, \bibinfo {author} {\bibfnamefont
  {A.~D.}\ \bibnamefont {Wieck}}, \ and\ \bibinfo {author} {\bibfnamefont
  {M.}~\bibnamefont {Bayer}},\ }\href {\doibase 10.1103/PhysRevB.81.195304}
  {\bibfield  {journal} {\bibinfo  {journal} {Phys. Rev. B}\ }\textbf {\bibinfo
  {volume} {81}},\ \bibinfo {pages} {195304} (\bibinfo {year}
  {2010})}\BibitemShut {NoStop}%
\bibitem [{\citenamefont {Spatzek}\ \emph {et~al.}(2011)\citenamefont
  {Spatzek}, \citenamefont {Greilich}, \citenamefont {Economou}, \citenamefont
  {Varwig}, \citenamefont {Schwan}, \citenamefont {Yakovlev}, \citenamefont
  {Reuter}, \citenamefont {Wieck}, \citenamefont {Reinecke},\ and\
  \citenamefont {Bayer}}]{SpatzekGreilichBayer2011}%
  \BibitemOpen
  \bibfield  {author} {\bibinfo {author} {\bibfnamefont {S.}~\bibnamefont
  {Spatzek}}, \bibinfo {author} {\bibfnamefont {A.}~\bibnamefont {Greilich}},
  \bibinfo {author} {\bibfnamefont {S.~E.}\ \bibnamefont {Economou}}, \bibinfo
  {author} {\bibfnamefont {S.}~\bibnamefont {Varwig}}, \bibinfo {author}
  {\bibfnamefont {A.}~\bibnamefont {Schwan}}, \bibinfo {author} {\bibfnamefont
  {D.~R.}\ \bibnamefont {Yakovlev}}, \bibinfo {author} {\bibfnamefont
  {D.}~\bibnamefont {Reuter}}, \bibinfo {author} {\bibfnamefont {A.~D.}\
  \bibnamefont {Wieck}}, \bibinfo {author} {\bibfnamefont {T.~L.}\ \bibnamefont
  {Reinecke}}, \ and\ \bibinfo {author} {\bibfnamefont {M.}~\bibnamefont
  {Bayer}},\ }\href {\doibase 10.1103/PhysRevLett.107.137402} {\bibfield
  {journal} {\bibinfo  {journal} {Phys. Rev. Lett.}\ }\textbf {\bibinfo
  {volume} {107}},\ \bibinfo {pages} {137402} (\bibinfo {year}
  {2011})}\BibitemShut {NoStop}%
\bibitem [{\citenamefont {Merkulov}\ \emph {et~al.}(2002)\citenamefont
  {Merkulov}, \citenamefont {Efros},\ and\ \citenamefont
  {Rosen}}]{Merkulov2002}%
  \BibitemOpen
  \bibfield  {author} {\bibinfo {author} {\bibfnamefont {I.~A.}\ \bibnamefont
  {Merkulov}}, \bibinfo {author} {\bibfnamefont {A.~L.}\ \bibnamefont {Efros}},
  \ and\ \bibinfo {author} {\bibfnamefont {M.}~\bibnamefont {Rosen}},\ }\href
  {\doibase 10.1103/PhysRevB.65.205309} {\bibfield  {journal} {\bibinfo
  {journal} {Phys. Rev. B}\ }\textbf {\bibinfo {volume} {65}},\ \bibinfo
  {pages} {205309} (\bibinfo {year} {2002})}\BibitemShut {NoStop}%
\bibitem [{\citenamefont {Khaetskii}\ \emph {et~al.}(2003)\citenamefont
  {Khaetskii}, \citenamefont {Loss},\ and\ \citenamefont
  {Glazman}}]{KhaetskiiLoss2003}%
  \BibitemOpen
  \bibfield  {author} {\bibinfo {author} {\bibfnamefont {A.}~\bibnamefont
  {Khaetskii}}, \bibinfo {author} {\bibfnamefont {D.}~\bibnamefont {Loss}}, \
  and\ \bibinfo {author} {\bibfnamefont {L.}~\bibnamefont {Glazman}},\ }\href
  {\doibase 10.1103/PhysRevB.67.195329} {\bibfield  {journal} {\bibinfo
  {journal} {Phys. Rev. B}\ }\textbf {\bibinfo {volume} {67}},\ \bibinfo
  {pages} {195329} (\bibinfo {year} {2003})}\BibitemShut {NoStop}%
\bibitem [{\citenamefont {Coish}\ and\ \citenamefont
  {Loss}(2004)}]{CoishLoss2004}%
  \BibitemOpen
  \bibfield  {author} {\bibinfo {author} {\bibfnamefont {W.~A.}\ \bibnamefont
  {Coish}}\ and\ \bibinfo {author} {\bibfnamefont {D.}~\bibnamefont {Loss}},\
  }\href@noop {} {\bibfield  {journal} {\bibinfo  {journal} {Phys. Rev. B}\
  }\textbf {\bibinfo {volume} {70}},\ \bibinfo {pages} {195340} (\bibinfo
  {year} {2004})}\BibitemShut {NoStop}%
\bibitem [{\citenamefont {Fischer}\ \emph {et~al.}(2008)\citenamefont
  {Fischer}, \citenamefont {Coish}, \citenamefont {Bulaev},\ and\ \citenamefont
  {Loss}}]{FischerLoss2008}%
  \BibitemOpen
  \bibfield  {author} {\bibinfo {author} {\bibfnamefont {J.}~\bibnamefont
  {Fischer}}, \bibinfo {author} {\bibfnamefont {W.~A.}\ \bibnamefont {Coish}},
  \bibinfo {author} {\bibfnamefont {D.~V.}\ \bibnamefont {Bulaev}}, \ and\
  \bibinfo {author} {\bibfnamefont {D.}~\bibnamefont {Loss}},\ }\href@noop {}
  {\bibfield  {journal} {\bibinfo  {journal} {Phys. Rev. B}\ }\textbf {\bibinfo
  {volume} {78}},\ \bibinfo {pages} {155329} (\bibinfo {year}
  {2008})}\BibitemShut {NoStop}%
\bibitem [{\citenamefont {Testelin}\ \emph {et~al.}(2009)\citenamefont
  {Testelin}, \citenamefont {Bernardot}, \citenamefont {Eble},\ and\
  \citenamefont {Chamarro}}]{Testelin2009}%
  \BibitemOpen
  \bibfield  {author} {\bibinfo {author} {\bibfnamefont {C.}~\bibnamefont
  {Testelin}}, \bibinfo {author} {\bibfnamefont {F.}~\bibnamefont {Bernardot}},
  \bibinfo {author} {\bibfnamefont {B.}~\bibnamefont {Eble}}, \ and\ \bibinfo
  {author} {\bibfnamefont {M.}~\bibnamefont {Chamarro}},\ }\href {\doibase
  10.1103/PhysRevB.79.195440} {\bibfield  {journal} {\bibinfo  {journal} {Phys.
  Rev. B}\ }\textbf {\bibinfo {volume} {79}},\ \bibinfo {pages} {195440}
  (\bibinfo {year} {2009})}\BibitemShut {NoStop}%
\bibitem [{\citenamefont {Gaudin}(1976)}]{Gaudin1976}%
  \BibitemOpen
  \bibfield  {author} {\bibinfo {author} {\bibfnamefont {M.}~\bibnamefont
  {Gaudin}},\ }\href@noop {} {\bibfield  {journal} {\bibinfo  {journal} {J.
  Physique}\ }\textbf {\bibinfo {volume} {37}},\ \bibinfo {pages} {1087}
  (\bibinfo {year} {1976})}\BibitemShut {NoStop}%
\bibitem [{\citenamefont {Andrei}\ \emph {et~al.}(1983)\citenamefont {Andrei},
  \citenamefont {Furuya},\ and\ \citenamefont
  {Lowenstein}}]{AndreiFuruyaLowenstein83}%
  \BibitemOpen
  \bibfield  {author} {\bibinfo {author} {\bibfnamefont {N.}~\bibnamefont
  {Andrei}}, \bibinfo {author} {\bibfnamefont {K.}~\bibnamefont {Furuya}}, \
  and\ \bibinfo {author} {\bibfnamefont {J.~H.}\ \bibnamefont {Lowenstein}},\
  }\href@noop {} {\bibfield  {journal} {\bibinfo  {journal} {Rev. Mod. Phys.}\
  }\textbf {\bibinfo {volume} {55}},\ \bibinfo {pages} {331} (\bibinfo {year}
  {1983})}\BibitemShut {NoStop}%
\bibitem [{\citenamefont {Bortz}\ and\ \citenamefont
  {Stolze}(2007)}]{BortzStolze2007}%
  \BibitemOpen
  \bibfield  {author} {\bibinfo {author} {\bibfnamefont {M.}~\bibnamefont
  {Bortz}}\ and\ \bibinfo {author} {\bibfnamefont {J.}~\bibnamefont {Stolze}},\
  }\href {\doibase 10.1103/PhysRevB.76.014304} {\bibfield  {journal} {\bibinfo
  {journal} {Phys. Rev. B}\ }\textbf {\bibinfo {volume} {76}},\ \bibinfo
  {pages} {014304} (\bibinfo {year} {2007})}\BibitemShut {NoStop}%
\bibitem [{\citenamefont {Bortz}\ \emph {et~al.}(2010)\citenamefont {Bortz},
  \citenamefont {Eggert},\ and\ \citenamefont {Stolze}}]{Bortz2010}%
  \BibitemOpen
  \bibfield  {author} {\bibinfo {author} {\bibfnamefont {M.}~\bibnamefont
  {Bortz}}, \bibinfo {author} {\bibfnamefont {S.}~\bibnamefont {Eggert}}, \
  and\ \bibinfo {author} {\bibfnamefont {J.}~\bibnamefont {Stolze}},\ }\href
  {\doibase 10.1103/PhysRevB.81.035315} {\bibfield  {journal} {\bibinfo
  {journal} {Phys. Rev. B}\ }\textbf {\bibinfo {volume} {81}},\ \bibinfo
  {pages} {035315} (\bibinfo {year} {2010})}\BibitemShut {NoStop}%
\bibitem [{\citenamefont {Faribault}\ and\ \citenamefont
  {Schuricht}(2013{\natexlab{a}})}]{FaribautSchuricht2013a}%
  \BibitemOpen
  \bibfield  {author} {\bibinfo {author} {\bibfnamefont {A.}~\bibnamefont
  {Faribault}}\ and\ \bibinfo {author} {\bibfnamefont {D.}~\bibnamefont
  {Schuricht}},\ }\href {\doibase 10.1103/PhysRevLett.110.040405} {\bibfield
  {journal} {\bibinfo  {journal} {Phys. Rev. Lett.}\ }\textbf {\bibinfo
  {volume} {110}},\ \bibinfo {pages} {040405} (\bibinfo {year}
  {2013}{\natexlab{a}})}\BibitemShut {NoStop}%
\bibitem [{\citenamefont {Faribault}\ and\ \citenamefont
  {Schuricht}(2013{\natexlab{b}})}]{FaribautSchuricht2013b}%
  \BibitemOpen
  \bibfield  {author} {\bibinfo {author} {\bibfnamefont {A.}~\bibnamefont
  {Faribault}}\ and\ \bibinfo {author} {\bibfnamefont {D.}~\bibnamefont
  {Schuricht}},\ }\href {\doibase 10.1103/PhysRevB.88.085323} {\bibfield
  {journal} {\bibinfo  {journal} {Phys. Rev. B}\ }\textbf {\bibinfo {volume}
  {88}},\ \bibinfo {pages} {085323} (\bibinfo {year}
  {2013}{\natexlab{b}})}\BibitemShut {NoStop}%
\bibitem [{\citenamefont {Crooker}\ \emph {et~al.}(2010)\citenamefont
  {Crooker}, \citenamefont {Brandt}, \citenamefont {Sandfort}, \citenamefont
  {Greilich}, \citenamefont {Yakovlev}, \citenamefont {Reuter}, \citenamefont
  {Wieck},\ and\ \citenamefont {Bayer}}]{CrookerBayerSpinNoise2010}%
  \BibitemOpen
  \bibfield  {author} {\bibinfo {author} {\bibfnamefont {S.~A.}\ \bibnamefont
  {Crooker}}, \bibinfo {author} {\bibfnamefont {J.}~\bibnamefont {Brandt}},
  \bibinfo {author} {\bibfnamefont {C.}~\bibnamefont {Sandfort}}, \bibinfo
  {author} {\bibfnamefont {A.}~\bibnamefont {Greilich}}, \bibinfo {author}
  {\bibfnamefont {D.~R.}\ \bibnamefont {Yakovlev}}, \bibinfo {author}
  {\bibfnamefont {D.}~\bibnamefont {Reuter}}, \bibinfo {author} {\bibfnamefont
  {A.~D.}\ \bibnamefont {Wieck}}, \ and\ \bibinfo {author} {\bibfnamefont
  {M.}~\bibnamefont {Bayer}},\ }\href@noop {} {\bibfield  {journal} {\bibinfo
  {journal} {Phys. Rev. Lett.}\ }\textbf {\bibinfo {volume} {104}},\ \bibinfo
  {pages} {036601} (\bibinfo {year} {2010})}\BibitemShut {NoStop}%
\bibitem [{\citenamefont {Dahbashi}\ \emph {et~al.}(2012)\citenamefont
  {Dahbashi}, \citenamefont {H\"ubner}, \citenamefont {Berski}, \citenamefont
  {Wiegand}, \citenamefont {Marie}, \citenamefont {Pierz}, \citenamefont
  {Schumacher},\ and\ \citenamefont {Oestreich}}]{Dahbashi2012}%
  \BibitemOpen
  \bibfield  {author} {\bibinfo {author} {\bibfnamefont {R.}~\bibnamefont
  {Dahbashi}}, \bibinfo {author} {\bibfnamefont {J.}~\bibnamefont {H\"ubner}},
  \bibinfo {author} {\bibfnamefont {F.}~\bibnamefont {Berski}}, \bibinfo
  {author} {\bibfnamefont {J.}~\bibnamefont {Wiegand}}, \bibinfo {author}
  {\bibfnamefont {X.}~\bibnamefont {Marie}}, \bibinfo {author} {\bibfnamefont
  {K.}~\bibnamefont {Pierz}}, \bibinfo {author} {\bibfnamefont {H.~W.}\
  \bibnamefont {Schumacher}}, \ and\ \bibinfo {author} {\bibfnamefont
  {M.}~\bibnamefont {Oestreich}},\ }\href@noop {} {\bibfield  {journal}
  {\bibinfo  {journal} {Appl. Phys. Lett.}\ }\textbf {\bibinfo {volume}
  {100}},\ \bibinfo {pages} {031906} (\bibinfo {year} {2012})}\BibitemShut
  {NoStop}%
\bibitem [{\citenamefont {Li}\ \emph {et~al.}(2012)\citenamefont {Li},
  \citenamefont {Sinitsyn}, \citenamefont {Smith}, \citenamefont {Reuter},
  \citenamefont {Wieck}, \citenamefont {Yakovlev}, \citenamefont {Bayer},\ and\
  \citenamefont {Crooker}}]{LiBayer2012}%
  \BibitemOpen
  \bibfield  {author} {\bibinfo {author} {\bibfnamefont {Y.}~\bibnamefont
  {Li}}, \bibinfo {author} {\bibfnamefont {N.}~\bibnamefont {Sinitsyn}},
  \bibinfo {author} {\bibfnamefont {D.~L.}\ \bibnamefont {Smith}}, \bibinfo
  {author} {\bibfnamefont {D.}~\bibnamefont {Reuter}}, \bibinfo {author}
  {\bibfnamefont {A.~D.}\ \bibnamefont {Wieck}}, \bibinfo {author}
  {\bibfnamefont {D.~R.}\ \bibnamefont {Yakovlev}}, \bibinfo {author}
  {\bibfnamefont {M.}~\bibnamefont {Bayer}}, \ and\ \bibinfo {author}
  {\bibfnamefont {S.~A.}\ \bibnamefont {Crooker}},\ }\href {\doibase
  10.1103/PhysRevLett.108.186603} {\bibfield  {journal} {\bibinfo  {journal}
  {Phys. Rev. Lett.}\ }\textbf {\bibinfo {volume} {108}},\ \bibinfo {pages}
  {186603} (\bibinfo {year} {2012})}\BibitemShut {NoStop}%
\bibitem [{\citenamefont {Zapasskii}\ \emph {et~al.}(2013)\citenamefont
  {Zapasskii}, \citenamefont {Greilich}, \citenamefont {Crooker}, \citenamefont
  {Li}, \citenamefont {Kozlov}, \citenamefont {Yakovlev}, \citenamefont
  {Reuter}, \citenamefont {Wieck},\ and\ \citenamefont
  {Bayer}}]{ZapasskiiGreilichBayer2013}%
  \BibitemOpen
  \bibfield  {author} {\bibinfo {author} {\bibfnamefont {V.~S.}\ \bibnamefont
  {Zapasskii}}, \bibinfo {author} {\bibfnamefont {A.}~\bibnamefont {Greilich}},
  \bibinfo {author} {\bibfnamefont {S.~A.}\ \bibnamefont {Crooker}}, \bibinfo
  {author} {\bibfnamefont {Y.}~\bibnamefont {Li}}, \bibinfo {author}
  {\bibfnamefont {G.~G.}\ \bibnamefont {Kozlov}}, \bibinfo {author}
  {\bibfnamefont {D.~R.}\ \bibnamefont {Yakovlev}}, \bibinfo {author}
  {\bibfnamefont {D.}~\bibnamefont {Reuter}}, \bibinfo {author} {\bibfnamefont
  {A.~D.}\ \bibnamefont {Wieck}}, \ and\ \bibinfo {author} {\bibfnamefont
  {M.}~\bibnamefont {Bayer}},\ }\href {\doibase 10.1103/PhysRevLett.110.176601}
  {\bibfield  {journal} {\bibinfo  {journal} {Phys. Rev. Lett.}\ }\textbf
  {\bibinfo {volume} {110}},\ \bibinfo {pages} {176601} (\bibinfo {year}
  {2013})}\BibitemShut {NoStop}%
\bibitem [{\citenamefont {Ezer}\ and\ \citenamefont
  {Kosloff}(1984)}]{TalEzer-Kosloff-84}%
  \BibitemOpen
  \bibfield  {author} {\bibinfo {author} {\bibfnamefont {H.}\ \bibnamefont
  {Tal-Ezer}}\ and\ \bibinfo {author} {\bibfnamefont {R.}~\bibnamefont {Kosloff}},\
  }\href@noop {} {\bibfield  {journal} {\bibinfo  {journal} {J. Chem. Phys}\
  }\textbf {\bibinfo {volume} {81}},\ \bibinfo {pages} {3967} (\bibinfo {year}
  {1984})}\BibitemShut {NoStop}%
\bibitem [{\citenamefont {Kosloff}(1994)}]{Kosloff-94}%
  \BibitemOpen
  \bibfield  {author} {\bibinfo {author} {\bibfnamefont {R.}~\bibnamefont
  {Kosloff}},\ }\href@noop {} {\bibfield  {journal} {\bibinfo  {journal} {Ann.
  Rev. Phys. Chem.}\ }\textbf {\bibinfo {volume} {45}},\ \bibinfo {pages} {145}
  (\bibinfo {year} {1994})}\BibitemShut {NoStop}%
\bibitem [{\citenamefont {Wei\ss{}e}\ \emph {et~al.}(2006)\citenamefont
  {Wei\ss{}e}, \citenamefont {Wellein}, \citenamefont {Alvermann},\ and\
  \citenamefont {Fehske}}]{Fehske-RMP2006}%
  \BibitemOpen
  \bibfield  {author} {\bibinfo {author} {\bibfnamefont {A.}~\bibnamefont
  {Wei\ss{}e}}, \bibinfo {author} {\bibfnamefont {G.}~\bibnamefont {Wellein}},
  \bibinfo {author} {\bibfnamefont {A.}~\bibnamefont {Alvermann}}, \ and\
  \bibinfo {author} {\bibfnamefont {H.}~\bibnamefont {Fehske}},\ }\href
  {\doibase 10.1103/RevModPhys.78.275} {\bibfield  {journal} {\bibinfo
  {journal} {Rev. Mod. Phys.}\ }\textbf {\bibinfo {volume} {78}},\ \bibinfo
  {pages} {275} (\bibinfo {year} {2006})}\BibitemShut {NoStop}%
\bibitem [{\citenamefont {Dobrovitski}\ and\ \citenamefont
  {De~Raedt}(2003)}]{Dobrovitski2003}%
  \BibitemOpen
  \bibfield  {author} {\bibinfo {author} {\bibfnamefont {V.~V.}\ \bibnamefont
  {Dobrovitski}}\ and\ \bibinfo {author} {\bibfnamefont {H.~A.}\ \bibnamefont
  {De~Raedt}},\ }\href {\doibase 10.1103/PhysRevE.67.056702} {\bibfield
  {journal} {\bibinfo  {journal} {Phys. Rev. E}\ }\textbf {\bibinfo {volume}
  {67}},\ \bibinfo {pages} {056702} (\bibinfo {year} {2003})}\BibitemShut
  {NoStop}%
\bibitem [{\citenamefont {Zhang}\ \emph
  {et~al.}(2006{\natexlab{a}})\citenamefont {Zhang}, \citenamefont
  {Dobrovitski}, \citenamefont {Al-Hassanieh}, \citenamefont {Dagotto},\ and\
  \citenamefont {Harmon}}]{ZangHarmon2006}%
  \BibitemOpen
  \bibfield  {author} {\bibinfo {author} {\bibfnamefont {W.}~\bibnamefont
  {Zhang}}, \bibinfo {author} {\bibfnamefont {V.~V.}\ \bibnamefont
  {Dobrovitski}}, \bibinfo {author} {\bibfnamefont {K.~A.}\ \bibnamefont
  {Al-Hassanieh}}, \bibinfo {author} {\bibfnamefont {E.}~\bibnamefont
  {Dagotto}}, \ and\ \bibinfo {author} {\bibfnamefont {B.~N.}\ \bibnamefont
  {Harmon}},\ }\href {\doibase 10.1103/PhysRevB.74.205313} {\bibfield
  {journal} {\bibinfo  {journal} {Phys. Rev. B}\ }\textbf {\bibinfo {volume}
  {74}},\ \bibinfo {pages} {205313} (\bibinfo {year}
  {2006}{\natexlab{a}})}\BibitemShut {NoStop}%
\bibitem [{\citenamefont {Yuan}\ \emph {et~al.}(2008)\citenamefont {Yuan},
  \citenamefont {Katsnelson},\ and\ \citenamefont {De~Raedt}}]{Yuan2008}%
  \BibitemOpen
  \bibfield  {author} {\bibinfo {author} {\bibfnamefont {S.}~\bibnamefont
  {Yuan}}, \bibinfo {author} {\bibfnamefont {M.~I.}\ \bibnamefont
  {Katsnelson}}, \ and\ \bibinfo {author} {\bibfnamefont {H.}~\bibnamefont
  {De~Raedt}},\ }\href {\doibase 10.1103/PhysRevB.77.184301} {\bibfield
  {journal} {\bibinfo  {journal} {Phys. Rev. B}\ }\textbf {\bibinfo {volume}
  {77}},\ \bibinfo {pages} {184301} (\bibinfo {year} {2008})}\BibitemShut
  {NoStop}%
\bibitem [{\citenamefont {Yoshida}\ \emph {et~al.}(1990)\citenamefont
  {Yoshida}, \citenamefont {Whitaker},\ and\ \citenamefont
  {Oliveira}}]{YoshidaWithakerOliveira1990}%
  \BibitemOpen
  \bibfield  {author} {\bibinfo {author} {\bibfnamefont {M.}~\bibnamefont
  {Yoshida}}, \bibinfo {author} {\bibfnamefont {M.~A.}\ \bibnamefont
  {Whitaker}}, \ and\ \bibinfo {author} {\bibfnamefont {L.~N.}\ \bibnamefont
  {Oliveira}},\ }\href@noop {} {\bibfield  {journal} {\bibinfo  {journal}
  {Phys.~Rev.~B}\ }\textbf {\bibinfo {volume} {41}},\ \bibinfo {pages} {9403}
  (\bibinfo {year} {1990})}\BibitemShut {NoStop}%
\bibitem [{\citenamefont {Bulla}\ \emph {et~al.}(2008)\citenamefont {Bulla},
  \citenamefont {Costi},\ and\ \citenamefont
  {Pruschke}}]{BullaCostiPruschke2008}%
  \BibitemOpen
  \bibfield  {author} {\bibinfo {author} {\bibfnamefont {R.}~\bibnamefont
  {Bulla}}, \bibinfo {author} {\bibfnamefont {T.~A.}\ \bibnamefont {Costi}}, \
  and\ \bibinfo {author} {\bibfnamefont {T.}~\bibnamefont {Pruschke}},\
  }\href@noop {} {\bibfield  {journal} {\bibinfo  {journal} {Rev.~Mod.~Phys.}\
  }\textbf {\bibinfo {volume} {80}},\ \bibinfo {pages} {395} (\bibinfo {year}
  {2008})}\BibitemShut {NoStop}%
\bibitem [{\citenamefont {Anders}\ and\ \citenamefont
  {Schiller}(2005)}]{AndersSchiller2005}%
  \BibitemOpen
  \bibfield  {author} {\bibinfo {author} {\bibfnamefont {F.~B.}\ \bibnamefont
  {Anders}}\ and\ \bibinfo {author} {\bibfnamefont {A.}~\bibnamefont
  {Schiller}},\ }\href@noop {} {\bibfield  {journal} {\bibinfo  {journal}
  {Phys.~Rev.~Lett.}\ }\textbf {\bibinfo {volume} {95}},\ \bibinfo {pages}
  {196801} (\bibinfo {year} {2005})}\BibitemShut {NoStop}%
\bibitem [{\citenamefont {Anders}\ and\ \citenamefont
  {Schiller}(2006)}]{AndersSchiller2006}%
  \BibitemOpen
  \bibfield  {author} {\bibinfo {author} {\bibfnamefont {F.~B.}\ \bibnamefont
  {Anders}}\ and\ \bibinfo {author} {\bibfnamefont {A.}~\bibnamefont
  {Schiller}},\ }\href@noop {} {\bibfield  {journal} {\bibinfo  {journal}
  {Phys.~Rev.~B}\ }\textbf {\bibinfo {volume} {74}},\ \bibinfo {pages} {245113}
  (\bibinfo {year} {2006})}\BibitemShut {NoStop}%
\bibitem [{\citenamefont {Leggett}\ \emph {et~al.}(1987)\citenamefont
  {Leggett}, \citenamefont {Chakravarty}, \citenamefont {Dorsey},\ and\
  \citenamefont {Fisher}}]{Leggett1987}%
  \BibitemOpen
  \bibfield  {author} {\bibinfo {author} {\bibfnamefont {A.~J.}\ \bibnamefont
  {Leggett}}, \bibinfo {author} {\bibfnamefont {S.}~\bibnamefont
  {Chakravarty}}, \bibinfo {author} {\bibfnamefont {A.~T.}\ \bibnamefont
  {Dorsey}}, \ and\ \bibinfo {author} {\bibfnamefont {M.~P.~A.}\ \bibnamefont
  {Fisher}},\ }\href@noop {} {\bibfield  {journal} {\bibinfo  {journal} {Rev.
  Mod. Phys.}\ }\textbf {\bibinfo {volume} {59}},\ \bibinfo {pages} {1}
  (\bibinfo {year} {1987})}\BibitemShut {NoStop}%
\bibitem [{\citenamefont {Erlingsson}\ and\ \citenamefont
  {Nazarov}(2004)}]{ErlingssonNazarov2004}%
  \BibitemOpen
  \bibfield  {author} {\bibinfo {author} {\bibfnamefont {S.~I.}\ \bibnamefont
  {Erlingsson}}\ and\ \bibinfo {author} {\bibfnamefont {Y.~V.}\ \bibnamefont
  {Nazarov}},\ }\href {\doibase 10.1103/PhysRevB.70.205327} {\bibfield
  {journal} {\bibinfo  {journal} {Phys. Rev. B}\ }\textbf {\bibinfo {volume}
  {70}},\ \bibinfo {pages} {205327} (\bibinfo {year} {2004})}\BibitemShut
  {NoStop}%
\bibitem [{\citenamefont {Cucchietti}\ \emph {et~al.}(2005)\citenamefont
  {Cucchietti}, \citenamefont {Paz},\ and\ \citenamefont
  {Zurek}}]{Cucchietti2005}%
  \BibitemOpen
  \bibfield  {author} {\bibinfo {author} {\bibfnamefont {F.~M.}\ \bibnamefont
  {Cucchietti}}, \bibinfo {author} {\bibfnamefont {J.~P.}\ \bibnamefont {Paz}},
  \ and\ \bibinfo {author} {\bibfnamefont {W.~H.}\ \bibnamefont {Zurek}},\
  }\href {\doibase 10.1103/PhysRevA.72.052113} {\bibfield  {journal} {\bibinfo
  {journal} {Phys. Rev. A}\ }\textbf {\bibinfo {volume} {72}},\ \bibinfo
  {pages} {052113} (\bibinfo {year} {2005})}\BibitemShut {NoStop}%
\bibitem [{\citenamefont {Merkulov}\ \emph {et~al.}(2010)\citenamefont
  {Merkulov}, \citenamefont {Alvarez}, \citenamefont {Yakovlev},\ and\
  \citenamefont {Schulthess}}]{Merkulov2010}%
  \BibitemOpen
  \bibfield  {author} {\bibinfo {author} {\bibfnamefont {I.~A.}\ \bibnamefont
  {Merkulov}}, \bibinfo {author} {\bibfnamefont {G.}~\bibnamefont {Alvarez}},
  \bibinfo {author} {\bibfnamefont {D.~R.}\ \bibnamefont {Yakovlev}}, \ and\
  \bibinfo {author} {\bibfnamefont {T.~C.}\ \bibnamefont {Schulthess}},\ }\href
  {\doibase 10.1103/PhysRevB.81.115107} {\bibfield  {journal} {\bibinfo
  {journal} {Phys. Rev. B}\ }\textbf {\bibinfo {volume} {81}},\ \bibinfo
  {pages} {115107} (\bibinfo {year} {2010})}\BibitemShut {NoStop}%
\bibitem [{\citenamefont {Al-Hassanieh}\ \emph {et~al.}(2006)\citenamefont
  {Al-Hassanieh}, \citenamefont {Dobrovitski}, \citenamefont {Dagotto},\ and\
  \citenamefont {Harmon}}]{Al-Hassanieh2006}%
  \BibitemOpen
  \bibfield  {author} {\bibinfo {author} {\bibfnamefont {K.~A.}\ \bibnamefont
  {Al-Hassanieh}}, \bibinfo {author} {\bibfnamefont {V.~V.}\ \bibnamefont
  {Dobrovitski}}, \bibinfo {author} {\bibfnamefont {E.}~\bibnamefont
  {Dagotto}}, \ and\ \bibinfo {author} {\bibfnamefont {B.~N.}\ \bibnamefont
  {Harmon}},\ }\href {\doibase 10.1103/PhysRevLett.97.037204} {\bibfield
  {journal} {\bibinfo  {journal} {Phys. Rev. Lett.}\ }\textbf {\bibinfo
  {volume} {97}},\ \bibinfo {pages} {037204} (\bibinfo {year}
  {2006})}\BibitemShut {NoStop}%
\bibitem [{\citenamefont {Chen}\ \emph {et~al.}(2007)\citenamefont {Chen},
  \citenamefont {Bergman},\ and\ \citenamefont {Balents}}]{ChenBalents2007}%
  \BibitemOpen
  \bibfield  {author} {\bibinfo {author} {\bibfnamefont {G.}~\bibnamefont
  {Chen}}, \bibinfo {author} {\bibfnamefont {D.~L.}\ \bibnamefont {Bergman}}, \
  and\ \bibinfo {author} {\bibfnamefont {L.}~\bibnamefont {Balents}},\ }\href
  {\doibase 10.1103/PhysRevB.76.045312} {\bibfield  {journal} {\bibinfo
  {journal} {Phys. Rev. B}\ }\textbf {\bibinfo {volume} {76}},\ \bibinfo
  {pages} {045312} (\bibinfo {year} {2007})}\BibitemShut {NoStop}%
\bibitem [{\citenamefont {Sinitsyn}\ \emph {et~al.}(2012)\citenamefont
  {Sinitsyn}, \citenamefont {Li}, \citenamefont {Crooker}, \citenamefont
  {Saxena},\ and\ \citenamefont {Smith}}]{Sinitsyn2012}%
  \BibitemOpen
  \bibfield  {author} {\bibinfo {author} {\bibfnamefont {N.~A.}\ \bibnamefont
  {Sinitsyn}}, \bibinfo {author} {\bibfnamefont {Y.}~\bibnamefont {Li}},
  \bibinfo {author} {\bibfnamefont {S.~A.}\ \bibnamefont {Crooker}}, \bibinfo
  {author} {\bibfnamefont {A.}~\bibnamefont {Saxena}}, \ and\ \bibinfo {author}
  {\bibfnamefont {D.~L.}\ \bibnamefont {Smith}},\ }\href {\doibase
  10.1103/PhysRevLett.109.166605} {\bibfield  {journal} {\bibinfo  {journal}
  {Phys. Rev. Lett.}\ }\textbf {\bibinfo {volume} {109}},\ \bibinfo {pages}
  {166605} (\bibinfo {year} {2012})}\BibitemShut {NoStop}%
\bibitem [{\citenamefont {Glazov}\ and\ \citenamefont
  {Ivchenko}(2012)}]{Glazov2012}%
  \BibitemOpen
  \bibfield  {author} {\bibinfo {author} {\bibfnamefont {M.~M.}\ \bibnamefont
  {Glazov}}\ and\ \bibinfo {author} {\bibfnamefont {E.~L.}\ \bibnamefont
  {Ivchenko}},\ }\href {\doibase 10.1103/PhysRevB.86.115308} {\bibfield
  {journal} {\bibinfo  {journal} {Phys. Rev. B}\ }\textbf {\bibinfo {volume}
  {86}},\ \bibinfo {pages} {115308} (\bibinfo {year} {2012})}\BibitemShut
  {NoStop}%
\bibitem [{\citenamefont {Stanek}\ \emph {et~al.}(2013)\citenamefont {Stanek},
  \citenamefont {Raas},\ and\ \citenamefont {Uhrig}}]{StanekRaasUhrig2013}%
  \BibitemOpen
  \bibfield  {author} {\bibinfo {author} {\bibfnamefont {D.}~\bibnamefont
  {Stanek}}, \bibinfo {author} {\bibfnamefont {C.}~\bibnamefont {Raas}}, \ and\
  \bibinfo {author} {\bibfnamefont {G.~S.}\ \bibnamefont {Uhrig}},\ }\href
  {\doibase 10.1103/PhysRevB.88.155305} {\bibfield  {journal} {\bibinfo
  {journal} {Phys. Rev. B}\ }\textbf {\bibinfo {volume} {88}},\ \bibinfo
  {pages} {155305} (\bibinfo {year} {2013})}\BibitemShut {NoStop}%
\bibitem [{\citenamefont {White}(1992)}]{White92}%
  \BibitemOpen
  \bibfield  {author} {\bibinfo {author} {\bibfnamefont {S.~R.}~\bibnamefont
  {White}},\ }\href@noop {} {\bibfield  {journal} {\bibinfo  {journal}
  {Phys.~Rev.~Lett.}\ }\textbf {\bibinfo {volume} {69}},\ \bibinfo {pages}
  {2863} (\bibinfo {year} {1992})}\BibitemShut {NoStop}%
\bibitem [{\citenamefont {Schollw\"ock}(2005)}]{SchollwoeckDMRG2005}%
  \BibitemOpen
  \bibfield  {author} {\bibinfo {author} {\bibfnamefont {U.}~\bibnamefont
  {Schollw\"ock}},\ }\href@noop {} {\bibfield  {journal} {\bibinfo  {journal}
  {Rev. Mod. Phys.}\ }\textbf {\bibinfo {volume} {77}},\ \bibinfo {pages} {259}
  (\bibinfo {year} {2005})}\BibitemShut {NoStop}%
\bibitem [{\citenamefont {Schollw\"ock}(2011)}]{Schollwoeck2011}%
  \BibitemOpen
  \bibfield  {author} {\bibinfo {author} {\bibfnamefont {U.}~\bibnamefont
  {Schollw\"ock}},\ }\href {\doibase 10.1016/j.aop.2010.09.012} {\bibfield
  {journal} {\bibinfo  {journal} {Annals of Physics}\ }\textbf {\bibinfo
  {volume} {326}},\ \bibinfo {pages} {96 } (\bibinfo {year}
  {2011})}\BibitemShut {NoStop}%
\bibitem [{\citenamefont {Friedrich}(2006)}]{FriedrichPhD2006}%
  \BibitemOpen
  \bibfield  {author} {\bibinfo {author} {\bibfnamefont {A.}~\bibnamefont
  {Friedrich}},\ }\emph {\bibinfo {title} {Time-dependent Properties of
  one-dimensional Spin-Systems: a DMRG-Study}},\ \href@noop {} {Ph.D. thesis},\
  \bibinfo  {school} {RWTH Aachen} (\bibinfo {year} {2006})\BibitemShut
  {NoStop}%
\bibitem [{\citenamefont {Lee}\ \emph {et~al.}(2005)\citenamefont {Lee},
  \citenamefont {von Allmen}, \citenamefont {Oyafuso}, \citenamefont
  {Klimeck},\ and\ \citenamefont {Whaley}}]{LeeEtAl2005}%
  \BibitemOpen
  \bibfield  {author} {\bibinfo {author} {\bibfnamefont {S.}~\bibnamefont
  {Lee}}, \bibinfo {author} {\bibfnamefont {P.}~\bibnamefont {von Allmen}},
  \bibinfo {author} {\bibfnamefont {F.}~\bibnamefont {Oyafuso}}, \bibinfo
  {author} {\bibfnamefont {G.}~\bibnamefont {Klimeck}}, \ and\ \bibinfo
  {author} {\bibfnamefont {K.~B.}\ \bibnamefont {Whaley}},\ }\href {\doibase
  10.1063/1.1850605} {\bibfield  {journal} {\bibinfo  {journal} {J. Appl.
  Phys.}\ }\textbf {\bibinfo {volume} {97}},\ \bibinfo {pages} {043706}
  (\bibinfo {year} {2005})}\BibitemShut {NoStop}%
\bibitem [{\citenamefont {Abragam}(1961)}]{AbragamNMR1961}%
  \BibitemOpen
  \bibfield  {author} {\bibinfo {author} {\bibfnamefont {A.}~\bibnamefont
  {Abragam}},\ }\href@noop {} {\emph {\bibinfo {title} {The Principles of
  Nuclear Magnetism}}}\ (\bibinfo  {publisher} {Oxford U.P.},\ \bibinfo {year}
  {1961})\BibitemShut {NoStop}%
\bibitem [{\citenamefont {Abramowitz}\ and\ \citenamefont
  {Stegun}(1972)}]{AbramowitzStegun}%
  \BibitemOpen
  \bibinfo {editor} {\bibfnamefont {M.}~\bibnamefont {Abramowitz}}\ and\
  \bibinfo {editor} {\bibfnamefont {I.~A.}\ \bibnamefont {Stegun}},\ eds.,\
  \href@noop {} {\emph {\bibinfo {title} {Handbook of mathematical
  Functions}}}\ (\bibinfo  {publisher} {Dover Publications},\ \bibinfo {year}
  {1972})\BibitemShut {NoStop}%
\bibitem [{\citenamefont {Khaetskii}\ \emph {et~al.}(2002)\citenamefont
  {Khaetskii}, \citenamefont {Loss},\ and\ \citenamefont
  {Glazman}}]{KhaetskiiGlazman2002}%
  \BibitemOpen
  \bibfield  {author} {\bibinfo {author} {\bibfnamefont {A.~V.}\ \bibnamefont
  {Khaetskii}}, \bibinfo {author} {\bibfnamefont {D.}~\bibnamefont {Loss}}, \
  and\ \bibinfo {author} {\bibfnamefont {L.}~\bibnamefont {Glazman}},\
  }\href@noop {} {\bibfield  {journal} {\bibinfo  {journal} {Phys. Rev. Lett.}\
  }\textbf {\bibinfo {volume} {88}},\ \bibinfo {pages} {186802} (\bibinfo
  {year} {2002})}\BibitemShut {NoStop}%
\bibitem [{\citenamefont {Zhang}\ \emph
  {et~al.}(2006{\natexlab{b}})\citenamefont {Zhang}, \citenamefont
  {Dobrovitski}, \citenamefont {Al-Hassanieh}, \citenamefont {Dagotto},\ and\
  \citenamefont {Harmon}}]{Zhang2004}%
  \BibitemOpen
  \bibfield  {author} {\bibinfo {author} {\bibfnamefont {W.}~\bibnamefont
  {Zhang}}, \bibinfo {author} {\bibfnamefont {V.~V.}\ \bibnamefont
  {Dobrovitski}}, \bibinfo {author} {\bibfnamefont {K.~A.}\ \bibnamefont
  {Al-Hassanieh}}, \bibinfo {author} {\bibfnamefont {E.}~\bibnamefont
  {Dagotto}}, \ and\ \bibinfo {author} {\bibfnamefont {B.~N.}\ \bibnamefont
  {Harmon}},\ }\href@noop {} {\bibfield  {journal} {\bibinfo  {journal} {Phys.
  Rev. B}\ }\textbf {\bibinfo {volume} {74}},\ \bibinfo {pages} {205313}
  (\bibinfo {year} {2006}{\natexlab{b}})}\BibitemShut {NoStop}%
\bibitem [{\citenamefont {Koppens}\ \emph {et~al.}(2007)\citenamefont
  {Koppens}, \citenamefont {Klauser}, \citenamefont {Coish}, \citenamefont
  {Nowack}, \citenamefont {Kouwenhoven}, \citenamefont {Loss},\ and\
  \citenamefont {Vandersypen}}]{Koppens2007}%
  \BibitemOpen
  \bibfield  {author} {\bibinfo {author} {\bibfnamefont {F.~H.~L.}\
  \bibnamefont {Koppens}}, \bibinfo {author} {\bibfnamefont {D.}~\bibnamefont
  {Klauser}}, \bibinfo {author} {\bibfnamefont {W.~A.}\ \bibnamefont {Coish}},
  \bibinfo {author} {\bibfnamefont {K.~C.}\ \bibnamefont {Nowack}}, \bibinfo
  {author} {\bibfnamefont {L.~P.}\ \bibnamefont {Kouwenhoven}}, \bibinfo
  {author} {\bibfnamefont {D.}~\bibnamefont {Loss}}, \ and\ \bibinfo {author}
  {\bibfnamefont {L.~M.~K.}\ \bibnamefont {Vandersypen}},\ }\href {\doibase
  10.1103/PhysRevLett.99.106803} {\bibfield  {journal} {\bibinfo  {journal}
  {Phys. Rev. Lett.}\ }\textbf {\bibinfo {volume} {99}},\ \bibinfo {pages}
  {106803} (\bibinfo {year} {2007})}\BibitemShut {NoStop}%
\end{thebibliography}

%

\end{document}